\documentclass[journal]{IEEEtran}
\usepackage[applemac]{inputenc}
\newif\ifonecolumn

\usepackage{amsmath}
\usepackage{amssymb}
\usepackage{dsfont}
\usepackage{amsthm}
\usepackage{thmtools}
\usepackage{cases}
\usepackage{balance}
\usepackage[lined,boxed,commentsnumbered,linesnumbered, ruled]{algorithm2e}

\usepackage{psfrag}
\usepackage[capitalize]{cleveref}
\crefname{equation}{}{}
\usepackage{comment}
\usepackage{bm}

\newtheorem{lemma}{Lemma}
\newtheorem{theorem}{Theorem}

\newtheorem{remark}{Remark}
\newtheorem{proposition}{Proposition}
\newtheorem{conjecture}{Conjecture}

\newtheoremstyle{mydef}
	{3pt}		
	{3pt}		
	{}		
	{}		
	{\itshape}	
	{:}		
	{.5em}	
	{}		

\newtheorem{definition}{Definition}

\usepackage{cite} 

\usepackage[nolist, nohyperlinks]{acronym}
\renewcommand\acsfont{\normalfont}

\acrodefplural{t.u.}[t.u.]{time units}

\ifCLASSOPTIONcompsoc
	\usepackage[caption=false,font=normalsize,labelfont=sf,textfont=sf]{subfig}
\else
	\usepackage[caption=false,font=footnotesize]{subfig}
\fi

\renewcommand\thesubfigure{(\alph{subfigure})}

\usepackage{tikz}
\usetikzlibrary{calc,
shapes,
positioning,
snakes,
patterns}

\newcommand{\ccolora}{red}
\newcommand{\ccolorb}{blue}
\newcommand{\tcola}[1]{\textcolor{\ccolora}{#1}}
\newcommand{\tcolb}[1]{\textcolor{\ccolorb}{#1}}
\newcommand*\diff{\mathop{}\!\mathrm{d}}

\newcommand*{\Scale}[2][4]{\scalebox{#1}{$#2$}}%
\newcommand*{\Resize}[2]{\resizebox{#1}{!}{$#2$}}%
\newcommand{\ubar}[1]{\text{\b{$#1$}}}

\renewcommand{\IEEEQED}{\IEEEQEDopen} 

\newcommand{\F}{\mathcal{F}}
\newcommand{\C}{\mathcal{C}}
\newcommand{\E}{\mathcal{E}}

\newcommand{\ftilde}{\tilde{f}}
\newcommand{\gtilde}{\tilde{g}}

\newcommand{\bphi}{\boldsymbol{\phi}}
\newcommand{\x}{\boldsymbol{x}}
\newcommand{\y}{\boldsymbol{y}}
\newcommand{\f}{\boldsymbol{f}}

\newcommand{\Apsi}{\mathcal{A}^{(\ell)}_\psi}
\newcommand{\Amu}{\mathcal{A}^{(\ell)}_\mu}

\newcommand{\phif}{\phi(\mathsf{f})}
\newcommand{\phifh}{\hat{\phi}(\mathsf{f})}

\newcommand{\phifi}{\phi(\mathsf{f}_i)}
\newcommand{\phifhi}{\hat{\phi}(\mathsf{f}_i)}

\newcommand{\T}{^{\mathsf{T}}}            		
\newcommand{\tr}[1]{\mathrm{#1}}
\newcommand{\mc}[1]{\mathcal{#1}}
\newcommand{\mf}[1]{\mathsf{#1}}
\newcommand{\set}[1]{\{#1\}}
\newcommand{\cd}{\cdot}
\newcommand{\vd}{\vdots}
\newcommand{\ld}{\ldots}
\newcommand{\dd}{\ddots}
\newcommand{\ms}[1]{\mathds{#1}}
\newcommand{\ie}{i.e.,~}
\DeclareMathOperator{\Image}{Im}
\newcommand{\cf}{cf.~}

\newcommand{\bth}{\beta_{\rm th}}
\newcommand{\fmin}{f_{\rm min}}

\newcommand{\argmax}{\operatornamewithlimits{arg\, max}}
\newcommand{\deltaequals}{\stackrel{\Delta}{=}}
\newcommand{\calD}{\mathcal{D}}
\newcommand{\calE}{\mathcal{E}}
\newcommand{\calR}{\mathcal{R}}
\newcommand{\boldc}{\bm{c}}
\newcommand{\boldd}{\bm{d}}
\newcommand{\boldb}{\bm{b}}
\newcommand{\boldp}{\bm{p}}
\newcommand{\bolde}{\bm{e}}
\newcommand{\boldx}{\bm{x}}
\newcommand{\boldy}{\bm{y}}
\newcommand{\sync}{^{\mathrm{sync}}}
\newcommand{\dec}{^{\mathrm{dec}}}
\newcommand{\cc}{\mathsf{c}}
\newcommand{\ff}{\mathsf{f}}
\newcommand{\KK}{\mathsf{K}}
\newcommand{\bb}{\mathsf{b}}
\newcommand{\sfl}{\mathsf{l}}
\newcommand{\vv}{\mathsf{v}}
\newcommand{\VV}{\mathsf{V}}
\newcommand{\V}{\mathcal{V}}
\renewcommand{\P}{\mathcal{P}}

\newcommand{\new}[1]{#1}

\usepackage{pgfplots} 
\pgfplotsset{compat=newest} 
\pgfplotsset{plot coordinates/math parser=false} 
\newlength\figureheight 
\newlength\figurewidth 

\begin{document}
\title{Asymptotic Analysis and Spatial Coupling of Counter Braids}

\author{Eirik Rosnes,~\IEEEmembership{Senior Member,~IEEE} and Alexandre Graell i Amat,~\IEEEmembership{Senior Member,~IEEE}
\thanks{Parts of this paper have been presented at the IEEE Information Theory Workshop, Hobart, Australia, Nov. 2014, and at the IEEE Information Theory Workshop, Jeju Island, Korea, Oct. 2015.}
\thanks{E.\ Rosnes was supported by the Norwegian-Estonian Research Cooperation Programme (grant EMP133) and by the Research Council of Norway (grant 240985/F20). A.\ Graell i Amat was supported by the Swedish Research Council under grants \#2011-5961 and \#2016-04253.}
\thanks{E. Rosnes is with Simula@UiB, N-5020 Bergen, Norway (e-mail: eirikrosnes@simula.no).}
\thanks{A. Graell i Amat is with the Department of Electrical Engineering, Chalmers University of Technology, SE-41296 Gothenburg, Sweden (e-mail: alexandre.graell@chalmers.se).}
}

\maketitle


\begin{abstract}
A counter braid (CB) is a novel counter architecture introduced by Lu \emph{et al.} in 2007 for per-flow measurements on high-speed links which can be decoded with low complexity using message passing (MP). CBs achieve an asymptotic compression rate (under optimal decoding) that matches the entropy lower bound of the flow size distribution. In this paper, we apply the concept of spatial coupling to CBs to improve the performance of the original CBs and analyze the performance of the resulting spatially-coupled CBs (SC-CBs). We introduce an equivalent bipartite graph representation of CBs with identical iteration-by-iteration finite-length and asymptotic performance. Based on this equivalent representation, we then analyze the asymptotic performance of single-layer CBs and SC-CBs under the MP decoding algorithm proposed by Lu \emph{et al.}. In particular, we derive the potential threshold of the uncoupled system and show that it is equal to the area threshold. We also derive the Maxwell decoder for CBs and prove that the potential threshold is an upper bound on the Maxwell decoding threshold, which, in turn, is a lower bound on the maximum \emph{a posteriori} (MAP) decoding threshold. We then show that the area under the extended MP extrinsic information transfer curve (defined for the equivalent graph), computed for the expected residual CB graph  when a peeling decoder equivalent to the MP decoder stops, is equal to zero precisely at the area threshold. This, combined with the analysis of the Maxwell decoder and simulation results, leads us to the conjecture that the potential threshold is in fact equal to the Maxwell decoding threshold and hence a lower bound on the MAP decoding threshold. Interestingly, SC-CBs do not show the well-known phenomenon of threshold saturation of the MP decoding threshold to the potential threshold characteristic of spatially-coupled low-density parity-check codes and other coupled systems. However, SC-CBs yield better MP decoding thresholds than their uncoupled counterparts. Finally, we also consider SC-CBs as a compressed sensing scheme and show that low undersampling factors can be achieved. 

%
%
%
%
%
\end{abstract}


\section{Introduction}

Traffic measurement in large scale networks is key to network operators for network management, for example, in terms of pricing, billing, and diagnosing of network problems. Traffic measurement consists of measuring the size of network flows. A network flow is defined as a sequence of packets that traverse the network and that share a number of properties, e.g., the same source and destination. The size of a flow is the number of packets of a particular kind.
On high-speed links, the inter-arrival time between packets can be as short as $40$ nanoseconds for a $10$ Gbps link. Therefore, an exact estimation of the flow sizes is technologically challenging and even expensive to build,
since it requires large arrays of high-speed counters.
Thus, the design of efficient, low-complexity algorithms to measure the size of active flows in high-speed networks is of great practical interest. 

Recently, Lu \emph{et al.} proposed a novel counter architecture, inspired by sparse graph codes, for measuring network flow sizes, nicknamed Counter Braid (CB) \cite{lu08,lu08_1,lu07}. CBs address the problem of cheap high-speed memory-efficient approximate counting. In particular, they use less memory space than other approximate counting techniques, since the flow sizes are compressed \textit{on-the-fly}. CBs are asymptotically optimal (under some mild conditions) \cite{lu07}, i.e., the average number of bits needed to store the size of a flow tends to the information-theoretic limit (under maximum-likelihood (ML) decoding) when the number of flows goes to infinity. Furthermore, they are characterized by a layered structure which can be described by a graph. In \cite{lu07}, a low-complexity message passing (MP) decoding algorithm working on the resulting graph was proposed, where the messages exchanged within the graph are positive integers. In this sense, no reliability information is exchanged in the iterative process and the algorithm can be assimilated to hard-decision MP decoding algorithms for low-density parity-check (LDPC) codes. In general, good performance can be achieved with a small number of layers.

A single-layer CB can also be regarded as a compressed sensing (CS) scheme for nonnegative (integer) signals \cite{lu08_1}. CS establishes that sparse signals can be recovered from significantly fewer samples as compared to conventional Nyquist sampling. Indeed, the decoding of single-layer CBs can also be seen as a sparse signal recovery problem. In particular, if we set to zero the entries of minimum flow size in the vector of flow sizes, then it can be interpreted as a nonnegative $t$-sparse vector with $t$ nonzero (integer) entries. In this setup, the \emph{counters} of the CB architecture perform linear measurements of the sparse signal, i.e., each counter corresponds to a linear measurement in the CS context. In \cite{lu08_1}, the authors analyzed the threshold (when the number of flows goes to infinity) on the undersampling factor (the number of counters per flow, corresponding to the ratio between the number of measurements and the length of the sparse vector) of CBs above which recovery of the sparse vector is possible. It was shown that it is very close to the corresponding threshold by Donoho and Tanner \cite{don10} on the undersampling factor such that most signals of a given sparsity can be recovered exactly using $\ell_1$-norm minimization reconstruction, assuming random Gaussian measurement matrices and nonnegative real signals.



Spatial coupling of LDPC codes \cite{fel99,LentmaierTransITOct2010,kud11} has revealed as a powerful technique that improves the belief propagation decoding threshold of the spatially-coupled LDPC (SC-LDPC) code to the maximum \emph{a posteriori} (MAP) decoding threshold of the underlying block code ensemble, a phenomenon known as threshold saturation. The concept of spatial coupling is not exclusive of LDPC codes, and it applies to other classes of codes such as turbo-like codes \cite{MolLenGra14,Mol17} and to iterative hard-decision decoding (HDD) of spatially-coupled (SC) generalized LDPC codes \cite{Jian12,Hag16} as well as to other scenarios, such as relaying \cite{Sch14}, lossy compression \cite{Are12}, joint iterative decoding of LDPC codes on intersymbol-interference channels with erasure noise where decoding is performed on a large graph representing both the channel and the code constraints \cite{ngu12}, code-division multiple-access \cite{tak11}, and CS \cite{Don13}. In all these cases, threshold saturation to the so-called \emph{potential threshold} has been proven or observed numerically, while threshold saturation to the so-called \emph{condensation threshold}  was observed for lossy compression in \cite{are15}. In  \cite{has13}, it was observed that the \emph{survey propagation threshold} increases and saturates towards to the \emph{phase transition threshold} for SC constraint satisfaction problems, as well as saturation of the \emph{dynamic threshold} towards the condensation threshold. 

In this paper, we apply the concept of spatial coupling to CBs, and show that they yield improved MP decoding thresholds as compared to uncoupled CBs. Furthermore, we analyze the asymptotic performance of single-layer spatially-coupled CBs (SC-CBs). We derive an equivalent bipartite  graph representation of CBs, with identical iteration-by-iteration finite-length performance and asymptotic behavior to that of CBs decoded on the original bipartite graph. Based on this equivalent representation, we prove that the potential threshold, introduced by Yedla \emph{et al.} in \cite{yed13}, and the area threshold are equal. We further derive the Maxwell decoder \cite{mea08} for CBs and prove that the area threshold is an upper bound on the Maxwell decoding threshold, which, in turn, is a lower bound on the MAP decoding threshold. We also show that the area under the extended MP (EMP) extrinsic information transfer (EXIT) curve (defined for the equivalent graph), computed for the expected residual CB graph  when a peeling decoder equivalent to the MP decoder stops, is equal to zero precisely at the area threshold. This result, combined with an asymptotic analysis of the Maxwell decoder, leads us to formulate the conjecture that the potential threshold is equal to the Maxwell decoding threshold and hence a lower bound on the MAP decoding threshold. This conjecture is also supported by simulation results.

%
Interestingly, we observe that SC-CBs do not show threshold saturation of the MP decoding threshold to the potential threshold. Indeed, when coupling the original (or the equivalent) graph, there is a  remaining gap between the potential threshold (conjectured to be equal to the Maxwell decoding threshold) and the MP decoding threshold of SC-CBs even in the limit of large coupling chain length and smoothing parameter. The lack of threshold saturation seems to be fundamental and due to the fact that the flow node update rule for even and odd iterations is different. To the best of our knowledge, this is one of the rear coupled systems where threshold saturation does not occur. 

We also discuss the construction of multilayer SC-CBs and the extension of the density evolution (DE) analysis to this case. Finally, we discuss single-layer SC-CBs in the context of CS. We compare the threshold on the undersampling factor of SC-CBs with that of CBs \cite{lu08_1} and with the threshold in \cite{don10},  
and we show that lower thresholds can be achieved. 

The remainder of the paper is organized as follows. In Section~\ref{sec:CounterBraids}, we briefly review CBs and the corresponding MP decoding algorithm. In Section~\ref{sec:EquivalentGraph}, we introduce an equivalent bipartite graph representation of CBs. In Section~\ref{sec:AsymptoticAnalysis}, we give an asymptotic analysis of single-layer CBs, based on the equivalent graph representation of Section~\ref{sec:EquivalentGraph}, and show that the area threshold and the potential threshold are equal. In Section~\ref{sec:Maxwell}, we derive the Maxwell decoder for CBs and formulate the conjecture that the potential threshold is a lower bound on the MAP decoding threshold. Section~\ref{sec:SC-CBs} introduces SC-CBs, and Section~\ref{sec:NumericalResults} gives numerical results on their asymptotic and finite-length performance. A discussion on the lack of threshold saturation is provided in Section~\ref{sec:discussion}. Finally, the connection with CS is discussed in Section~\ref{sec:CS}, and Section~\ref{sec:Conclusion} draws some conclusions.






%


\section{Counter Braids}
\label{sec:CounterBraids}

A CB is a counter architecture consisting of $L \geq 1$ layers. At layer $l =1,\ldots,L$, it has $m_l$ counters of depth $d_l$ bits, with $m_i<m_j$ for $i>j$. The number of distinct flows to be counted is denoted by $m_0$.

The $l$-th layer of a CB can be represented by a bipartite graph $\mathcal{G}_l = \mathcal{G}_l(\F_l \cup \C_l,\E_l)$, where $\F_l$ (of size $m_{l-1}$) denotes the set of flow nodes, and $\C_l$  (of size $m_{l}$) denotes the set of counter nodes. The set of edges of the graph is denoted by $\E_l$. The $l$-th layer is connected to the $(l-1)$-th layer by a bijective mapping $\Xi_l(\mathsf{f}) = \mathsf{c}$ on the set of flow nodes of the $l$-th layer, where $\mathsf{f} \in \F_l$ and $\mathsf{c} \in \C_{l-1}$, $l=2,\ldots,L$. For notational convenience, the neighborhood of a node $\mathsf{a}$ (either a counter node $\mathsf{c}$ or a flow node $\mathsf{f}$) is denoted by $\Gamma(\mathsf{a})$.

When a flow is encountered (for instance, on a high-speed link), all connected counter nodes of the first layer are incremented modulo $2^{d_1}$. If a counter node $\mathsf{c}$ of the first layer overflows, all connected counter nodes of the second layer (formally the counter nodes in the set $\Gamma(\Xi^{-1}_2(\mathsf{c}))$), are incremented modulo $2^{d_2}$. Furthermore, if a counter node in the second layer overflows, all connected counter nodes of the third layer are incremented modulo $2^{d_3}$. This process is repeated for each level until we reach the final layer $L$. We denote by $\phi(\mathsf{c})$ the final value of a counter node $\mathsf{c}$ prior to decoding, and by $\hat{\phi}(\mathsf{f})$ the estimated value (after decoding) of a flow node $\mathsf{f}$.  The corresponding actual flow size is denoted by ${\phi}(\mathsf{f})$. 

\begin{figure}[!t]
\centering
\ifonecolumn
\includegraphics[width=0.75\columnwidth,height=5.35cm]{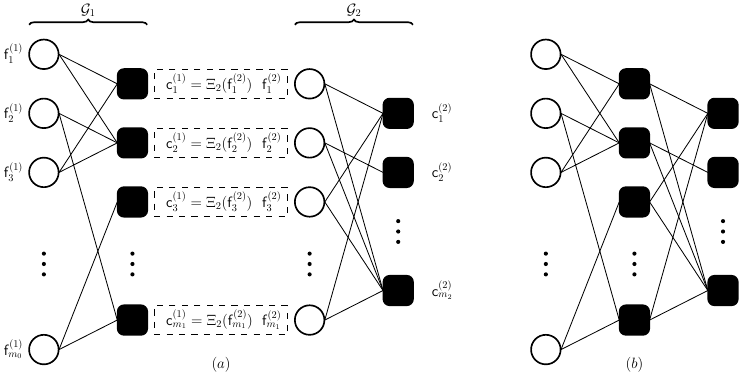}
\else
\includegraphics[width=\columnwidth,height=4.35cm]{Figure2}
\fi
\vspace{-4ex}
\caption{The graph of a two-layer CB. $(a)$ The bipartite graphs $\mathcal{G}_1$ and $\mathcal{G}_2$; $(b)$ the combined graph.}
\label{fig:cb}
\vspace{-2ex}
\end{figure}

An example of a two-layer CB is shown in Fig.~\ref{fig:cb}, where flow nodes are represented by empty circles and counter nodes by filled squares. Fig.~\ref{fig:cb}$(a)$ shows the bipartite graphs $\mathcal{G}_1$ and $\mathcal{G}_2$ of the two-layer CB, while Fig.~\ref{fig:cb}$(b)$ depicts an equivalent graph where a flow node of the second layer and its corresponding counter node of the first layer in Fig.~\ref{fig:cb}$(a)$ are represented using a single \textit{combined} counter node due to the bijective mapping $\Xi_2(\cdot)$.

In the following, we will assume that the bipartite graphs $\mathcal{G}_l$, $l=1,\ldots,L$, are left-regular, i.e., $|\Gamma(\mathsf{f})| = k_l$, for some integer $k_l \geq 2$, and for all $\mathsf{f} \in \F_l$. The assignment of graph sockets of flow nodes to graph sockets of counter nodes, i.e., the connections in the graph, is done in a random fashion. This means that, asymptotically as the number of flow nodes tends to infinity (while $\frac{|\C_l|}{|\F_l|}$ is kept fixed), the distribution of the fraction of edges connected to a given 
counter node approaches a Poisson distribution \cite{lu08}. We can now formally define the flow and counter node degree distributions. For counter nodes, we assume the asymptotic Poisson distribution. Let 
\begin{align*}
L_l(z)  = z^{k_l}, \quad\quad\quad\quad k_l \geq 2, 
\end{align*}
and 
\begin{align*}  
R_l(z) = \sum_{i=0}^{\infty} R^{(l)}_i z^i = \sum_{i=0}^{\infty} \frac{\mathrm{e}^{-\gamma_l} (\gamma_l z) ^{i}}{i!}
\end{align*}
denote the node-perspective flow node  and  counter node degree distributions of the $l$-th layer, respectively.  Here, $\gamma_l = \frac{m_{l-1}k_l}{m_l}$ is the average counter node degree. The corresponding edge-perspective  degree distributions are 
\begin{align}
\lambda_l(z) &= \frac{L_l'(z)}{L_l'(1)} = z^{k_l-1} \notag 
\end{align}
and
\begin{align}
\rho_l(z) &= \frac{R_l'(z)}{R_l'(1)} = \sum_{i=0}^{\infty} \frac{\mathrm{e}^{-\gamma_l} (\gamma_l z)^{i}}{i!} =  R_l(z), \notag 
\end{align}
respectively. Finally, we denote by $\beta_l = \frac{m_{l}}{m_{l-1}}$ the fraction of the number of counter and flow nodes of the $l$-th layer.

\subsection{Message Passing Decoder} \label{sec:message_passing}

CBs can be decoded using an MP decoding algorithm \cite{lu08} on the corresponding graph. 
In this section, we present the low-complexity MP decoding algorithm proposed in \cite{lu08}. We consider first the sum-product (SP) algorithm to decode CBs and motivate the derivation of the algorithm in \cite{lu08}. The derivation was not discussed in \cite{lu08} (or in \cite{lu08_1,lu07,cha10}), but may help in gaining some insight into the algorithm.


For simplicity, we consider a single-layer CB, and denote its set of edges by $\E=\E_1$. The optimality criterion behind the SP algorithm for CBs is the \emph{symbol-wise} MAP, 
\begin{align}
\label{eq:MAPrule}
\hat{\phi}_{\rm MAP}(\ff_i)=\argmax_{\phif} \text{Pr}(\phifi=\phif|\phi(\bm{\mathsf{c}})),
\end{align}
where $\phi(\bm{\mathsf{c}})=(\phi(\mathsf{c}_1),\ldots,\phi(\mathsf{c}_{m_1}))$.

Implementing the MAP decoder to solve \eqref{eq:MAPrule} directly is computationally infeasible. The SP algorithm attempts to solve \eqref{eq:MAPrule} approximately operating on the factor graph of the CB. Let $\psi_{\cc \rightarrow \ff }^{\rm SP, (\ell)}(a)$ denote the message sent from counter node $\cc$ to flow node $\ff$ at iteration $\ell$ of the SP algorithm, corresponding to the probability that the flow $\ff$ is of size $a$, i.e., $\phi(\ff)=a$. Note that $a$ is a nonnegative integer that runs from $f_{\rm min}$ to $\phi(\cc)- (|\Gamma(\cc)| -1) f_{\rm min}$, where $f_{\rm min}$ is the minimum flow size (from the flow size distribution) of a flow node. Furthermore, let $\mu_{\ff \rightarrow \cc}^{\rm SP, (\ell)}(a)$ denote the message sent from flow node $\ff$ to counter node $\cc$ at iteration $\ell$, corresponding to the probability that the flow $\ff$ is of size $a$. The update rules for $\psi_{\cc \rightarrow \ff }^{\rm SP, (\ell)}(a)$ and $\mu_{\ff \rightarrow \cc}^{\rm SP, (\ell)}(a)$ can be written as \cite{ksc01}
\begin{equation} \label{eq:counter_to_flow}
\psi_{\cc \rightarrow \ff  }^{\rm SP, (\ell)}(a) =  \sum_{ \substack{( a_{\mathsf{f}'} \colon \mathsf{f}' \in \Gamma(\mathsf{c}) \setminus  \mathsf{f})  \colon  a_{\mathsf{f}'} \geq f_{\rm min} \\ \text{and} \sum_{\mathsf{f}' \in \Gamma(\mathsf{c}) \setminus  \mathsf{f} } a_{\mathsf{f}'}=\phi(\mathsf{c})-a}}     \prod_{\mathsf{f}' \in \Gamma(\mathsf{c} )\setminus \mathsf{f} }  
\mu_{\ff' \rightarrow \cc}^{\rm SP, (\ell -1)}( a_{\mathsf{f}'})
\end{equation}
and
\begin{equation} \label{eq:flow_to_counter}
\mu_{\ff \rightarrow \cc}^{\rm SP, (\ell)}( a) =   \prod_{\cc'  \in \Gamma(\ff) \setminus \cc }  \psi_{\cc' \rightarrow \ff  }^{\rm SP, (\ell)}(a),
\end{equation}
respectively. More precisely, the message $\psi_{\cc \rightarrow \ff }^{\rm SP, (\ell)}(a)$ represents the probability that $\phi(\ff)=a$ given the incoming messages $\mu_{\ff' \rightarrow \cc}^{\rm SP, (\ell-1)}( a_{\mathsf{f}'})$ from all adjacent edges to counter node $\cc$ (except the edge from $\ff$) and the constraint of the counter node itself.

Replacing the outer summation in \cref{eq:counter_to_flow} with a maximum gives the \emph{max-product} algorithm (or, equivalently, the \emph{min-sum} algorithm in the logarithmic domain),

\ifonecolumn
\begin{align} \label{eq:counter_to_flow_max_prod}
\psi_{\cc \rightarrow \ff  }^{\rm MaxP, (\ell)}(a) 
=  \max_{ \substack{( a_{\mathsf{f}'} \colon \mathsf{f}' \in \Gamma(\mathsf{c}) \setminus  \mathsf{f})  \colon  a_{\mathsf{f}'} \geq f_{\rm min} \\ \text{and} \sum_{\mathsf{f}' \in \Gamma(\mathsf{c}) \setminus \mathsf{f} } a_{\mathsf{f}'}=\phi(\mathsf{c})-a}}     \prod_{\mathsf{f}' \in \Gamma(\mathsf{c} )\setminus  \mathsf{f} }  
\mu_{\ff' \rightarrow \cc}^{\rm MaxP, (\ell-1)}( a_{\mathsf{f}'}).
\end{align}
\else
\begin{align} \label{eq:counter_to_flow_max_prod}
&\psi_{\cc \rightarrow \ff  }^{\rm MaxP, (\ell)}(a) \nonumber\\
&=  \max_{ \substack{( a_{\mathsf{f}'} \colon \mathsf{f}' \in \Gamma(\mathsf{c}) \setminus  \mathsf{f})  \colon  a_{\mathsf{f}'} \geq f_{\rm min} \\ \text{and} \sum_{\mathsf{f}' \in \Gamma(\mathsf{c}) \setminus \mathsf{f} } a_{\mathsf{f}'}=\phi(\mathsf{c})-a}}     \prod_{\mathsf{f}' \in \Gamma(\mathsf{c} )\setminus  \mathsf{f} }  
\mu_{\ff' \rightarrow \cc}^{\rm MaxP, (\ell-1)}( a_{\mathsf{f}'})
\end{align}
and
\begin{equation} \label{eq:flow_to_counter_max_prod}
\mu_{\ff \rightarrow \cc}^{\rm MaxP, (\ell)}( a) =   \prod_{\cc'  \in \Gamma(\ff) \setminus \cc }  \psi_{\cc' \rightarrow \ff  }^{\rm MaxP, (\ell)}(a).
\end{equation}
\fi

Both the SP algorithm and the max-product algorithm are \emph{soft-decision} decoding algorithms which entail a high complexity.\footnote{The update rules in \cref{eq:counter_to_flow} and \cref{eq:counter_to_flow_max_prod} can be implemented in each counter node using the BCJR algorithm working on a trellis where each trellis section would be a complete bipartite graph with at most $\phi(\cc)- |\Gamma(\cc)| f_{\rm min} + 1$ states.} To reduce decoding complexity, one may consider HDD.  Assuming a uniform flow size distribution (i.e., not exploiting any information about the flow size distribution) the max-product algorithm above boils down to the low-complexity MP algorithm proposed in \cite{lu08}, which indeed can be seen as an HDD algorithm. In the following, we only give the update rules of the MP algorithm in \cite{lu08}. Its derivation from   \cref{eq:counter_to_flow_max_prod} and \cref{eq:flow_to_counter_max_prod} is given in Appendix~\ref{app:MPdecodingDerivation}.

We define the messages exchanged in the MP decoding algorithm as follows. Let $\mu^{(\ell)}_{\mathsf{f} \rightarrow \mathsf{c}} \in \mathbb{N}$, where $\mathbb{N}$ is the set of natural numbers and $(\mathsf{f}, \mathsf{c}) \in \E$, denote the message sent from flow node $\mathsf{f}$ to counter node $\mathsf{c}$ during the $\ell$-th iteration of the MP decoding algorithm. Likewise, let $\psi^{(\ell)}_{\mathsf{c} \rightarrow \mathsf{f}} \in \mathbb{N}$, $(\mathsf{f}, \mathsf{c}) \in \E$, denote the message sent from counter node $\mathsf{c}$ to flow node $\mathsf{f}$ during the $\ell$-th iteration of the algorithm. The counter node and flow node update rules (for $\ell=1,\ldots,\ell_{\max}$) are given as follows\cite{lu08},
\begin{align}
\psi^{(\ell)}_{\mathsf{c} \rightarrow \mathsf{f}} &= \max \left\{ \phi(\mathsf{c}) - \sum_{ \mathsf{f}' \in \Gamma(\mathsf{c}) \setminus \ff} \mu^{(\ell-1)}_{\mathsf{f}' \rightarrow \cc},\;f_{\rm min} \right\}, \label{eq:Bp1} \\
\mu^{(\ell)}_{\mathsf{f} \rightarrow \mathsf{c}} &= \begin{cases}
\min_{\mathsf{c}' \in \Gamma(\mathsf{f}) \setminus \mathsf{c}} \psi^{(\ell)}_{\mathsf{c}' \rightarrow \ff}, & \text{if $\ell$ is odd} \\
\max_{\mathsf{c}' \in \Gamma(\mathsf{f}) \setminus \mathsf{c}} \psi^{(\ell)}_{\mathsf{c}' \rightarrow \ff}, & \text{if $\ell$ is even} \end{cases}, \label{eq:Bp2}
\end{align}
where $\ell_{\max}$ is the maximum number of iterations and $\mu^{(0)}_{\mathsf{f} \rightarrow \mathsf{c}} = f_{\rm min}$ for all $(\mathsf{f}, \mathsf{c}) \in \E$. The final estimate of the flow sizes of the flow nodes is according to
\begin{displaymath}
\hat{\phi}(\mathsf{f}) = \begin{cases}
\min_{\mathsf{c} \in \Gamma(\mathsf{f})} \psi^{(\ell_{\max})}_{\mathsf{c} \rightarrow \mathsf{f}}, & \text{if $\ell_{\max}$ is odd} \\
\max_{\mathsf{c} \in \Gamma(\mathsf{f})} \psi^{(\ell_{\max})}_{\mathsf{c} \rightarrow \mathsf{f}}, & \text{if $\ell_{\max}$ is even} \end{cases}. \notag
\end{displaymath}


The MP algorithm in \cref{eq:Bp1}--\cref{eq:Bp2} can be seen as an HDD algorithm where only integer values  are exchanged between flow and counter nodes. Despite this, as  shown in \cite[Th.~1]{lu08}, the output of the MP decoder converges to the exact flow size vector when the underlying CB graph is a tree.

For a multi-layer CB, decoding proceeds starting with the right-most layer. After decoding the $l$-th layer, represented by the bipartite graph $\mathcal{G}_l$, the counter nodes of the $(l-1)$-th layer are updated (for all $\mathsf{f} \in \F_{l}$) based on the mapping $\Xi_{l}(\cdot)$ as
\begin{displaymath}
\phi(\Xi_{l}(\mathsf{f})) \leftarrow \hat{\phi}(\mathsf{f}) \cdot 2^{d_{l-1}} + \phi( \Xi_{l}(\mathsf{f})).
\end{displaymath}
Decoding proceeds layer-by-layer until the first layer is decoded.

\subsection{Peeling Decoder}
In this subsection, we introduce a peeling decoder version of the MP decoder mentioned above. The reason for doing this is that we need the concept of a residual graph and residual degree distributions in \cref{sec:Maxwell}.

Note that the MP decoder with the update rules in (\ref{eq:Bp1})--(\ref{eq:Bp2}) will always \emph{stop}, i.e., at some point the flow size estimates of the flow nodes will be the same at iterations $\ell$ and $\ell-2$ for some $\ell$. This is due to the monotonicity property of the messages, i.e., for each edge $(\mathsf{f}, \mathsf{c}) \in \mathcal{E}$, the messages $\mu_{\mathsf{f} \rightarrow \mathsf{c}}^{(2\ell)}$, $\ell \geq 0$, are monotonically nondecreasing lower bounds on $\phi(\mathsf{f})$, and the messages $\mu_{\mathsf{f} \rightarrow \mathsf{c}}^{(2\ell+1)}$, $\ell \geq 0$, are monotonically nonincreasing upper bounds on $\phi(\mathsf{f})$. This can be verified by induction \cite{cha10}. When the decoder stops, either the flow size estimates for a flow node at iterations $\ell$ and $\ell-1$ are the same (i.e., we have convergence (upper and lower bounds are the same and thus we have the correct value for the flow size)), or we have oscillation (i.e., the estimates for iterations $\ell$ and $\ell-2$ are the same, but the estimates for iterations $\ell$  and $\ell-1$ are not the same).

The MP decoder can be turned into a peeling decoder in the following way. Run the MP decoder and in addition apply the following two peeling rules:\footnote{Note that we can peel all neigboring flow nodes of a counter node $\mathsf{c}$ if its value $\phi(\mathsf{c})$ is equal to $f_{\rm min}$ times its degree $\Gamma(\mathsf{c})$, i.e., if $\phi(\mathsf{c}) = f_{\rm min} \cdot \Gamma(\mathsf{c})$. This follows from the second peeling rule.}

\begin{enumerate}
\item In case $|\Gamma(\mathsf{c})|=1$ for a counter node $\mathsf{c}$, then remove $\mathsf{c}$ and the connected flow node $\mathsf{f} \in \Gamma(\mathsf{c})$ and all its adjacent edges from the graph. Decrease the values of the counter nodes of $\Gamma(\mathsf{f}) \setminus \mathsf{c}$ by the value of $\mathsf{c}$.
\item For odd iterations, if a message from a counter node $\mathsf{c}$ to a flow node $\mathsf{f}$ is equal to $f_{\rm min}$, then remove the flow node $\mathsf{f}$ and all its adjacent edges from the graph. Decrease the values of the counter nodes of $\Gamma(\mathsf{f})$ by $f_{\rm min}$. 
\end{enumerate}

The MP and the peeling decoders are \emph{equivalent}, in the sense that the set of \emph{converged} flow nodes for the MP decoder, i.e., the set of flow nodes for which the flow size estimates for even and odd iterations are the same,  is equal to the set of \emph{peeled} flow nodes for the peeling decoder, i.e., the set of flow nodes that have been removed from the graph when running the peeling decoder.

\subsection{A Note on Notation}
In the rest of the paper, except for \cref{sec:ExtMoreLayers}, we assume a single-layer system and frequently omit, for notational convenience, the  subscript/superscript $l$. Also, we assume that the depth of the counters of the single-layer CB, $d_1$, is large enough so that they do not overflow. Note that if the counters overflow, one should consider multilayer CBs. This is discussed in \cref{sec:ExtMoreLayers}.



\section{Equivalent Bipartite Graph Representation of Counter Braids}
\label{sec:EquivalentGraph}

In this section, we construct an \textit{equivalent} bipartite graph representation of a single-layer CB, i.e., of the graph $\mathcal{G}$, and a corresponding MP decoding algorithm, which gives identical iteration-by-iteration finite-length and asymptotic performance to that of the original graph. The new graph has the same flow nodes as the original graph and the same number of counter nodes, but each counter node represents now a tree of depth three grown from the corresponding counter node in the original graph (a so-called computation tree of depth three,  see \cite[p.~27]{wib96}), and the update rules are different. 
This graph transformation enables us to bypass the fact that to apply the potential function framework of \cite{yed13}, a bipartite graph with left and right update rules that do not change over the iterations is required. 
Note that for the original bipartite graph, this property does not hold. 

The computation tree of depth three comes from the fact that, to obtain a bipartite graph with update rules that do not change with iterations, we combine two successive iterations (on the original graph) into a single one. Thus, the message from counter nodes to flow nodes of the transformed graph corresponds to the sequence (for the original graph) counter nodes -- flow nodes -- counter nodes -- flow nodes, which leads to a computation tree of depth three.


For simplicity, we will first consider the example graph $\mathcal{G}$ shown in Fig.~\ref{fig:example_graph}$(a)$ for the construction of the equivalent graph in \cref{sec:constr}, before presenting the general construction algorithm (Algorithm~\ref{alg:equivalent_graph}). The MP decoder for the equivalent graph is described in \cref{sec:message_passing_equiv} for the example graph $\tilde{\mathcal{G}}$ shown in Fig.~\ref{fig:example_graph}$(c)$, from which the general case follows in a straightforward manner.

\subsection{Construction} \label{sec:constr}

\begin{figure}[!t]
\centering
\ifonecolumn
\includegraphics[width=0.75\columnwidth]{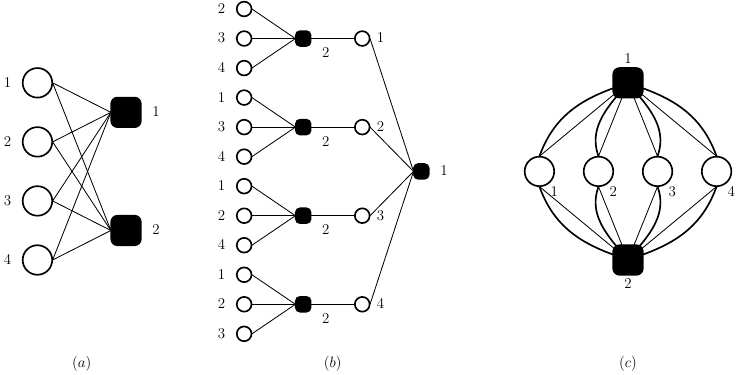}
\else
\includegraphics[width=\columnwidth]{Figure1}
\fi
\caption{$(a)$ The graph $\mathcal{G}$ of an example CB. $(b)$ The tree $\mathcal{T}(\mathsf{c}_1)$ of depth three grown from the counter node $\mathsf{c}_1$ (for notational convenience the counter node is labeled with the integer $1$ instead of $\mathsf{c}_1$ in the figure) from the original graph $\mathcal{G}$ in $(a)$. $(c)$ Equivalent bipartite graph representation $\tilde{\mathcal{G}}$ of the example CB from $(a)$.}
\label{fig:example_graph}
\end{figure}

The equivalent graph, denoted by $\tilde{\mathcal{G}}$, is constructed as follows for the example graph $\mathcal{G}$ shown in Fig.~\ref{fig:example_graph}$(a)$. First, for each counter node $\mathsf{c} \in \C$ in the graph in Fig.~\ref{fig:example_graph}$(a)$, we build a (computation) tree $\mathcal{T}(\mathsf{c})$ of depth three \cite[p.~27]{wib96}, as illustrated in Fig.~\ref{fig:example_graph}$(b)$ for the counter node $\mathsf{c}_1$ (for notational convenience this counter node is labeled with the integer $1$ instead of $\mathsf{c}_1$ in the figure). Then, for each counter node $\mathsf{c} \in \C$ we make a new counter node $\tilde{\mathsf{c}}$ in the new bipartite graph $\tilde{\mathcal{G}}$ (see Fig.~\ref{fig:example_graph}$(c)$). Furthermore, in the new graph, we make a copy $\tilde{\mathsf{f}}$ of each flow node $\mathsf{f} \in \F$ in the original graph. Counter nodes and flow nodes in the new graph are connected using two types of edges, type-$1$ and type-$2$, as explained in the following. Consider the counter node $\mathsf{c}_1$ in Fig.~\ref{fig:example_graph}$(a)$. For this counter node, we make a new counter node $\tilde{\mathsf{c}}_1$ in the equivalent graph in Fig.~\ref{fig:example_graph}$(c)$. Then, we connect $\tilde{\mathsf{c}}_1$ to a flow node $\tilde{\mathsf{f}}_i$ with a type-$2$ edge if $\mathsf{f}_i$ is a leaf in the tree of $\mathsf{c}_1$. Furthermore, we connect $\tilde{\mathsf{c}}_1$ to a flow node $\tilde{\mathsf{f}}_i$ with a type-$1$ edge if $\mathsf{f}_i$ is a flow node at depth one in the tree of $\mathsf{c}_1$. In this way there is an edge for each path in $\mathcal{T}(\mathsf{c}_1)$ from $\mathsf{c}_1$ to a flow node. When the path has length three, the edge is a type-$2$ edge, and when the path has length one, the edge is a type-$1$ edge.
\begin{algorithm}[t] \label{alg:equivalent_graph}
\SetKwInOut{Input}{Input}
\SetKwInOut{Output}{Output}
\SetKwComment{Comment}{$\triangleright$\ }{}
\DontPrintSemicolon

\Input{Original CB graph $\mathcal{G} =\left(\F \cup \C, \E \right)$}
\Output{Equivalent CB graph $\mathcal{\tilde{G}} = \left(\tilde{\F} \cup  \tilde{\C}, \tilde{\E} \right)$}
$\tilde{\F} \leftarrow \emptyset$ \Comment*[r]{$\tilde{\F}$ is a set} 
$\tilde{\C} \leftarrow \emptyset$ \Comment*[r]{$\tilde{\C}$ is a set} 
$\tilde{\E} \leftarrow \emptyset$ \Comment*[r]{$\tilde{\E}$ is a multiset} 
\For{$\ff \in \F$}{
$\tilde{\F} \leftarrow \ff$\\
}
\For{$\cc \in \C$}{
                $\tilde{\C} \leftarrow \cc$\\
	        \For{$\ff \in \F_1(\cc) \cup \F_3(\cc) $}{
	        $\tilde{\E} \leftarrow (\cc,\ff)$ with label equal to the length of $\P(\cc,\ff)$%
	        }
	        }
	        \KwRet{$\left(\tilde{\F} \cup  \tilde{\C}, \tilde{\E} \right)$}
	\caption{Construction of the equivalent graph}\label{alg:equivalentgraph}
	
\end{algorithm}

In Fig.~\ref{fig:example_graph}$(c)$, type-$1$ edges are represented by nonbold edges and correspond each to a distinct path of length one from $\mathsf{c}_1$ or $\mathsf{c}_2$ to a flow node at depth one in $\mathcal{T}(\mathsf{c}_1)$ or $\mathcal{T}(\mathsf{c}_2)$, respectively.  
On the other hand, bold edges represent edge bundles of type-$2$ edges. For the example in Fig.~\ref{fig:example_graph}$(c)$, an edge bundle between counter node $\tilde{\mathsf{c}}_1$ and flow node $\tilde{\mathsf{f}}_i$, $i=1,2,3,4$,  consists of three type-$2$ edges corresponding to the three distinct paths of length three in the tree $\mathcal{T}(\mathsf{c}_1)$ between $\mathsf{c}_1$ and flow node $\mathsf{f}_i$, at depth three. For notational convenience, these flow nodes are labeled with the integers $i$ instead of $\tilde{\mathsf{f}}_i$ in Fig.~\ref{fig:example_graph}$(c)$ and instead of $\mathsf{f}_i$ in Figs.~\ref{fig:example_graph}$(a)$ and \ref{fig:example_graph}$(b)$. Likewise, $\tilde{\mathsf{c}}_1$ is labeled by the integer $1$ in Fig.~\ref{fig:example_graph}$(c)$. Note that for a given bipartite graph $\mathcal{G}$, the number of type-$2$ edges in each edge bundle of the equivalent bipartite graph $\tilde{\mathcal{G}}$ depends on the connectivity of $\mathcal{G}$. Also, in the asymptotic limit there will be no duplicate flow nodes in the trees $\mathcal{T}(\mathsf{c})$ for any counter node $\mathsf{c} \in \mathcal{C}$ and the equivalent graph will be locally tree-like.

For the general case, let $\mathcal{T}(\cc)=\left(\V(\cc),\E(\cc) \right) $ denote the (computation) tree for node $\cc$, where $\V(\cc) = \cc \cup \F_1(\cc) \cup \C_2(\cc) \cup \F_3(\cc) $,  $\F_1(\cc)$ is the set of neighboring flow nodes of $\cc$ (distance-$1$ nodes from $\cc$), $\C_2(\cc)$ is the set of distance-$2$ nodes from $\cc$ (all counter nodes), and $\F_3(\cc)$ is the set of distance-$3$ nodes from $\cc$ (all flow nodes). The set of edges of the computation tree is denoted by $\E(\cc)$. For notational convenience, the (unique) path between two nodes $\vv_1 \in \V(\cc)$ and $\vv_2 \in \V(\cc)$ is denoted by ${\P}(\vv_1,\vv_2)$. The construction of the equivalent graph  $\mathcal{\tilde{G}} = (\tilde{\F} \cup  \tilde{\C}, \tilde{\E})$, which is a graph with parallel edges (or a multigraph),  can be described by Algorithm~\ref{alg:equivalent_graph}. For a multigraph, the edge set $\tilde{\E}$ is a multiset, i.e., unlike a set, it allows multiple instances of the elements of the multiset. If the label of an edge in $\tilde{\mathcal{G}}$ is $1$, then the edge is a type-$1$ edge. Otherwise, it is a type-$2$ edge. Since the original CB graph $\mathcal G$ does not contain parallel edges, there are no parallel type-$1$ edges in $\mathcal{\tilde{G}}$. However, there might be parallel type-$2$ edges, and these can be grouped together in edge bundles, where each edge bundle contains all parallel edges between a given pair of endpoints.


\subsection{Message Passing Decoding} \label{sec:message_passing_equiv}
We have the following update rules for the new graph. For the flow nodes $\tilde{\mathsf{f}} \in \tilde{\F}$ we use the same update rule as in (\ref{eq:Bp2}) for type-$1$ edges, using the maximum only, over all incoming messages on all connected edges (except the one on which the outgoing message is transmitted). For the type-$2$ edges in an edge bundle, the update rule is a bit different. Note that each edge bundle is in one-to-one correspondence with paths of length three from the root node in the tree to the leaf nodes. Thus, we can label each type-$2$ edge in the bundle from flow node $\tilde{\mathsf{f}}$ to counter node $\tilde{\mathsf{c}}$ with the counter node in the new graph that corresponds to the parent counter node of the corresponding leaf node $\mathsf{f}$ in the tree. Then, the output message on a type-$2$ edge in the bundle will be the same as the output message on the type-$1$ edge of $\tilde{\mathcal{G}}$ that connects flow node $\tilde{\mathsf{f}}$ to the counter node that labels the type-$2$ edge in the bundle which is being considered. Equivalently, we can use the update rule in (\ref{eq:Bp2}), using the maximum only, over all incoming messages on all connected edges except the particular type-$1$ edge connected to the counter node that labels the type-$2$ edge in the bundle under consideration.

For the counter nodes $\tilde{\mathsf{c}} \in \tilde{\C}$ in the new graph, the update rule is a bit more complicated and the corresponding tree $\mathcal{T}(\mathsf{c})$ must be used. The procedure consists of three steps as follows:

\begin{enumerate}
\item First, the outgoing message on an edge from a flow node $\tilde{\mathsf{f}}$ to a counter node $\tilde{\mathsf{c}}$ in the new graph is assigned as an outgoing message to the corresponding edge emanating from the corresponding node $\mathsf{f}$ in the tree $\mathcal{T}(\mathsf{c})$. Then, run the update rule in (\ref{eq:Bp1}) for all counter nodes in the tree $\mathcal{T}(\mathsf{c})$ using all incoming messages  (except the one on which the outgoing message is transmitted). Outgoing messages are sent to all connected flow nodes at depth one in the  tree $\mathcal{T}(\mathsf{c})$ (i.e., upward towards the root node).
\item Secondly, run the update rule in (\ref{eq:Bp2}) for all flow nodes at depth one in the tree $\mathcal{T}(\mathsf{c})$ (using the minimum only) using all incoming messages  (except the one on which the outgoing message is transmitted). Outgoing messages are sent only to the root node $\mathsf{c}$ in the tree $\mathcal{T}(\mathsf{c})$ (i.e., upward to the root node).
\item Finally, apply the update rule in (\ref{eq:Bp1}) on the root node using all incoming messages  (except the one on which the outgoing message is transmitted). Outgoing messages to all connected flow nodes at depth one are computed and assigned to the corresponding type-$1$ edges in the new graph, i.e., a message from the root node $\mathsf{c}$ to a flow node $\mathsf{f}$ at depth one is assigned as an outgoing message on a type-$1$ edge from counter node $\tilde{\mathsf{c}}$ to flow node $\tilde{\mathsf{f}}$ in the new graph $\tilde{\mathcal{G}}$.
\end{enumerate} 
Note that the three-step procedure above only assigns outgoing messages to type-$1$ edges connected to $\tilde{\mathsf{c}}$ in the new graph. The outgoing messages on the type-$2$ edges in the edge bundles are set to $-\infty$. 

Decoding on the equivalent graph $\tilde{\mathcal{G}}$ is performed as in the standard flooding schedule, i.e., all counter nodes in $\tilde{\mathcal{G}}$  should be processed as outlined above before processing all flow nodes. On the other hand, encoding can be done using the equivalent graph as well by ignoring the type-$2$ edges and using the encoding method of the original graph, i.e., the value of a counter node is the sum modulo $2^{d_1}$ of the values of the connected flow nodes (considering only type-$1$ edges). Notice  that for any CB graph $ \mathcal{G}$ we can construct an equivalent graph $\tilde{\mathcal{G}}$ using the construction above, and viceversa (removing the edge bundles in $\tilde{\mathcal{G}}$).




Let $\hat{\phi}_{\mathcal{G}}^{(\ell)}(\ff)$ denote the estimated flow size at flow node $\ff \in \mathcal{G}$ at iteration $\ell$ under MP decoding on the original CB graph $\mathcal{G}$ (see \cref{sec:message_passing}). The corresponding estimated flow size at iteration $\ell$ and flow node $\tilde{\ff} \in \tilde{\mathcal{G}}$ (corresponding to flow node $\ff$ in the original graph $\mathcal{G}$) under MP decoding on the equivalent graph $\tilde{\mathcal{G}}$ (as described above) is denoted by  $\hat{\phi}_{\tilde{\mathcal{G}}}^{(\ell)}(\tilde{\ff})$. Based on the above discussion, the proposition below follows by inspection.

\begin{proposition} \label{prop:equivalent}
For any given CB graph $\mathcal{G}$ and iteration number $\ell \geq 1$ (with respect to the original graph $\mathcal{G}$), 
\begin{displaymath}
\hat{\phi}_{\mathcal{G}}^{(2\ell)}(\ff)=\hat{\phi}_{\tilde{\mathcal{G}}}^{(\ell)}(\tilde{\ff}),\; \forall \ff \in \mathcal{G},
\end{displaymath}
i.e., the two graphs are \emph{iteration-by-iteration} equivalent.
\end{proposition}

\section{Asymptotic Analysis of Single-Layer Counter Braids}
\label{sec:AsymptoticAnalysis}


Consider a single-layer CB with $m_0$ flow nodes and $m_1$ counter nodes of depth $d_1$. When $m_0\rightarrow \infty$, the asymptotic performance of CBs can be analyzed by means of  DE. Denote by $x^{(\ell)}$ and $y^{(\ell)}$ the error probability of an outgoing message from a flow node and a counter node, respectively, at the $\ell$-th iteration. The flow node and counter node DE updates at iteration $\ell$ are described by the equations \cite{lu08,lu08_1}
\begin{align}
\label{eq:DECounter1}
x^{(\ell)} = \ftilde\left(y^{(\ell)}; \epsilon\right),~~~~
y^{(\ell)} = \gtilde \left(x^{(\ell-1)}\right),
\end{align}
where
\begin{align}
\label{eq:DECounter1a}
 \ftilde(y;\epsilon) &= \begin{cases}
y^{k-1},  & \text{if $\ell$ is odd} \\
\epsilon \cdot y^{k-1}, & \text{if $\ell$ is even}\end{cases},\\
\label{eq:DECounter1b}
\gtilde(x) & = 1-\rho\left(1-x\right),
\end{align}
and $\epsilon$ is the probability of observing a flow of size strictly larger than $f_{\rm min}$. Note that $\epsilon$ depends on the flow size distribution, denoted by $p_{\Phi}$.

In the following, we denote by ${\cal{X}}=[0,1]$ and ${\cal{Y}}=[0,1]$ the set of possible values for $x$ and $y$, respectively, and by ${\cal{E}}$ the set of possible values for $\epsilon$.

For a given flow size distribution, or more precisely, for  a given $\epsilon$, a relevant parameter for the performance of single-layer CBs is the number of counters per flow, $\beta=\frac{m_1}{m_0}=\frac{k}{\gamma}$. The average number of bits needed to represent a flow is therefore $\beta d_1$, hence $\beta$ is directly related to the compression rate. In particular, we are interested in the minimum value of $\beta$ so that decoding is successful, 
\begin{align*}
\beta_{\rm MP} = \beta_{\rm MP}(\epsilon)\triangleq\inf\left\{\beta>0 \; |\; x^{(\infty)}(\beta,\epsilon)=0\right\}.
\end{align*}
Alternatively, we can analyze the asymptotic behavior of CBs by fixing $\beta$ and finding the maximum value of $\epsilon$ such that decoding is successful, 
\begin{align*}
\epsilon_{\rm MP} = \epsilon_{\rm MP}(\beta)\triangleq\sup\left\{\epsilon\in\mathcal{E} \; |\;  x^{(\infty)}(\beta,\epsilon)=0\right\},
\end{align*}
since for a fixed $k$ and any $\epsilon$ it follows that $\epsilon_{\rm MP}(\beta_{\rm MP}(\epsilon))=\beta_{\rm MP}^{-1}(\beta_{\rm MP}(\epsilon))=\epsilon$.
 In this case, $\epsilon_{\rm MP}$ has a similar meaning as the MP decoding threshold for LDPC codes over the binary erasure channel (BEC), where $\epsilon$ can now be interpreted as the channel parameter.


Combining two successive iterations into a single one, 
we can rewrite the DE in an equivalent form as
\begin{align}
\label{eq:DECounter2}
x^{(2\ell)} = f\left(y^{(2\ell)}; \epsilon\right),~~~~
y^{(2\ell)} = g \left(x^{(2\ell-2)}\right),
\end{align}
where
\begin{align}
f(y;\epsilon) &= \epsilon \cdot y^{k-1}, \label{eq:f} \\
g(x) &= 1-\rho\left(1-\left(1-\rho\left(1-x\right)\right)^{k-1}\right). \label{eq:g}
\end{align}

Clearly, the DE recursions in (\ref{eq:DECounter1})--(\ref{eq:DECounter1b}) and (\ref{eq:DECounter2})--(\ref{eq:g}) give the same MP decoding threshold. 
Furthermore, it is easy to verify that the DE in (\ref{eq:DECounter2})--(\ref{eq:g}) corresponds to the DE of the equivalent graph introduced in the previous section. Therefore, from this and \cref{prop:equivalent}, we can analyze CBs based on the equivalent graph and the DE recursion in (\ref{eq:DECounter2})--(\ref{eq:g}). 





\begin{lemma} \label{th:1}
The functions $f(y;\epsilon)$ (for a fixed value of $\epsilon > 0$) and $g(x)$, $k \geq 2$, are strictly increasing in $y$ and $x$, respectively.
\end{lemma}

\begin{IEEEproof} 
From the definition above it follows directly that $f(y;\epsilon)$ is strictly increasing in $y$ for a fixed value of $\epsilon>0$. Also, since $\rho(x)$ is strictly increasing in $x$ and both $1-\rho(x)$ and $\rho(1-x)$ are strictly decreasing in $x$, it follows directly that $g(x)$ is strictly increasing in $x$.
\end{IEEEproof}

As a result of Lemma~\ref{th:1}, it follows that the DE recursion in (\ref{eq:DECounter2})--(\ref{eq:g}) (and thus the DE recursion in (\ref{eq:DECounter1})--(\ref{eq:DECounter1b})) for uncoupled CBs converges to a fixed-point. The fixed-point DE equation for $x = x^{(\infty)}(\beta,\epsilon)$ is
\begin{align}
\label{eq:DECounter2FixedPoint}
x = f(g(x);\epsilon).
\end{align}
Decoding is successful if the fixed-point is $x^{(\infty)}(\beta,\epsilon)=0$. 

We will need the definition of an \emph{admissible system} \cite{yed13}.

\begin{definition}[Admissible System] \label{lem:prop}
An admissible system is a system where the functions $f(y;\epsilon)$ and $g(x)$ satisfy the following properties.
\begin{enumerate}
\item The first derivative of $f(y;\epsilon)$ exists and is continuous on $\cal{Y}\times\E$, the first derivative of $g(x)$ exists and is continuous on $\cal{X}$,
\item $f(y;\epsilon)$ is nondecreasing in both $y$ and $\epsilon$, 
\item $g(x)$ is strictly increasing in $x$, and
\item the second derivative of $g(x)$ exists and is continuous on $\cal{X}$.
\end{enumerate}
\end{definition}

In addition, an admissible system is said to be \emph{proper} if the derivative of $h(x;\epsilon) \triangleq f(g(x);\epsilon)$ with respect to $\epsilon$ is strictly positive for all $(x,\epsilon) \in (0,1] \times \mathcal{E}$ \cite{yed13}.

\begin{lemma} \label{lem:proper_admissible}
The system defined in  (\ref{eq:f})--(\ref{eq:g}) is a proper admissible system.
\end{lemma}

\begin{IEEEproof}
It is easy to show that the DE updates in (\ref{eq:f})--(\ref{eq:g}) satisfy Properties~1 to 4 in \cref{lem:prop}. Properties 2 and 3 have already been proven in \cref{th:1}. Now, the derivative of $h(x;\epsilon)$ with respect to $\epsilon$ is
\begin{displaymath}
\left(1-\rho\left(1-\left(1-\rho\left(1-x\right)\right)^{k-1}\right)\right)^{k-1},
\end{displaymath}
and the result follows from the fact that $\rho(x)$ for $x \in (0,1]$ and $k \geq 2$ is strictly positive.
\end{IEEEproof}



\subsection{Extended Message Passing Extrinsic Information Transfer Curve}

The EMP EXIT curve of a single-layer CB is given in parametric form by
\begin{equation} \label{eq:EBP_EXIT}
(\epsilon,h^{\rm EMP}) = (\epsilon(x), (1-\rho(1-(1-\rho(1-x))^{k-1}))^{k}),
\end{equation}
where
\begin{equation} \notag 
\epsilon(x)= \frac{x}{(1-\rho(1-(1-\rho(1-x))^{k-1}))^{k-1}}
\end{equation}
is the solution of (\ref{eq:DECounter2FixedPoint}) for $\epsilon$,
and $x \in {\cal{X}}$. The curve is a trace of all fixed-points of the DE recursion in (\ref{eq:DECounter2FixedPoint}) (both stable and unstable).

We have the following theorem related to the EMP EXIT curve, adapting \cite[Th.~8]{mea08} to the recursion in (\ref{eq:f})--(\ref{eq:g}).

\ifonecolumn
\begin{theorem} \label{th:area}
The EMP EXIT curve satisfies
\begin{displaymath}
\int_{0}^1 h^{\rm EMP}(x)\, \diff \epsilon(x) = 
1 - \rho(1-(1-\mathrm{e}^{-\gamma})^{k-1}) + k \left( \rho(1-(1-\mathrm{e}^{-\gamma})^{k-1})  
-  \int_{0}^{1} \rho(1-(1-\rho(1-z))^{k-1})\, \diff z \right).
\end{displaymath}
\end{theorem}
\else
\begin{theorem} \label{th:area}
The EMP EXIT curve satisfies
\begin{displaymath}
\begin{split}
&\int_{0}^1 h^{\rm EMP}(x)\, \diff \epsilon(x) = \\
&1 - \rho(1-(1-\mathrm{e}^{-\gamma})^{k-1}) + k \left( \rho(1-(1-\mathrm{e}^{-\gamma})^{k-1}) - \right. \\
& \left. \int_{0}^{1} \rho(1-(1-\rho(1-z))^{k-1})\, \diff z \right).
\end{split}
\end{displaymath}
\end{theorem}
\fi

\begin{IEEEproof}
The result follows by applying integration by parts twice as in the  proof of \cite[Th.~8]{mea08}. Details are omitted for brevity.
\end{IEEEproof}

Note that, contrary to the case of standard (both regular and irregular) LDPC code ensembles on the BEC, the area is \emph{not} equal to the rate (in number of counters per flow) and the curve does \emph{not} start at the $(1,1)$ point for $x=1$. In fact, the EMP EXIT curve starts (for $x=1$) at the point $(1 / ( 1 - \rho(1-(1-\mathrm{e}^{-\gamma})^{k-1}))^{k-1},  ( 1 - \rho(1-(1-\mathrm{e}^{-\gamma})^{k-1}))^{k})$, where the first coordinate is larger than $1$ and the second coordinate is smaller than $1$. Also, as $x$ approaches $0$, when $k \geq 3$, the curves approach the point $(\infty, 0)$ (as could also be the case for LDPC code ensembles). 



\begin{definition}[Area Threshold] \label{def:MAP}
Let $(\epsilon(x^*), h^{\rm EMP}(x^*))$ be a point on the EMP EXIT curve $h^{\rm EMP}$ of a single-layer CB such that 
\begin{displaymath}
\int_{x^*}^1 h^{\rm EMP}(x)\, \diff \epsilon(x) = \int_{0}^1 h^{\rm EMP}(x)\, \diff \epsilon(x)
\end{displaymath}
and assume that there exist no $x' \in (x^*,1]$ such that $\epsilon(x') = \epsilon(x^*)$. Then, the area threshold, denoted by $\bar{\epsilon}$, is defined as $\bar{\epsilon} = \epsilon(x^*)$. 
\end{definition}


\begin{lemma}
For $k=2$ the area threshold is equal to $\bar{\epsilon}=1/\gamma^2$.
\end{lemma}

\begin{IEEEproof}
For $k=2$, $\epsilon(x) = \frac{x}{1-\rho(\rho(1-x))}$ is an increasing function of $x$. Thus, $x^*=0$, and the area threshold becomes
\ifonecolumn
\begin{displaymath}
\bar{\epsilon} = \lim_{x \rightarrow 0} \frac{x}{1-\rho(\rho(1-x))} = \frac{1}{\rho'(\rho(1-0))\rho'(1-0)} 
= \frac{1}{\gamma\rho(\rho(1))\gamma\rho(1)} = \frac{1}{\gamma^2},
\end{displaymath}
\else
\begin{displaymath}
\begin{split}
\bar{\epsilon} = \lim_{x \rightarrow 0} \frac{x}{1-\rho(\rho(1-x))} &= \frac{1}{\rho'(\rho(1-0))\rho'(1-0)} \\
&= \frac{1}{\gamma\rho(\rho(1))\gamma\rho(1)} = \frac{1}{\gamma^2},
\end{split}
\end{displaymath}
\fi
where $\rho'(x)$ denotes the derivative of $\rho(x)$ with respect to $x$, and the result follows.
\end{IEEEproof}

Note that this threshold corresponds to the \emph{stability} threshold from the theory of LDPC codes on the BEC.

In Section~\ref{sec:Maxwell}, we will prove that the area threshold $\bar{\epsilon}$ is an upper bound on the Maxwell decoding threshold. 

\subsection{Potential Function, Potential Threshold, and Area Threshold}

Since the DE recursion in (\ref{eq:DECounter2})--(\ref{eq:g})   
describes an admissible system (see Definition~\ref{lem:prop}), we can define a corresponding potential function as in \cite{yed13}. 

\begin{definition}
The potential function $U(x;\epsilon)$ of the system defined by the functions $f$ and $g$ from (\ref{eq:f}) and (\ref{eq:g}), respectively, is given by
\ifonecolumn
\begin{displaymath}
U(x;\epsilon) \triangleq x g(x) - \int_{0}^x g(z)\,\diff z - \int_{0}^{g(x)} f(z;\epsilon)\,\diff z 
= x g(x) - \int_{0}^x g(z)\,\diff z - \frac{\epsilon}{k} g(x)^k.
\end{displaymath}
\else
\begin{displaymath}
\begin{split}
U(x;\epsilon) &\triangleq x g(x) - \int_{0}^x g(z)\,\diff z - \int_{0}^{g(x)} f(z;\epsilon)\,\diff z \\
&= x g(x) - \int_{0}^x g(z)\,\diff z - \frac{\epsilon}{k} g(x)^k.
\end{split}
\end{displaymath}
\fi
\end{definition}

Following \cite[Def.~32]{yed13}, we make the following definitions,
\ifonecolumn
\begin{align}
\Psi(\epsilon) \triangleq \min_{x \in \cal{X}} U(x;\epsilon),\;
X^*(\epsilon) \triangleq \{x \in {\cal{X}}\; |\; U(x;\epsilon) = \Psi(\epsilon) \}, \text{ and } 
\bar{x}^*(\epsilon) \triangleq \max X^*(\epsilon). \notag
\end{align}
\else
\begin{align}
\Psi(\epsilon) &\triangleq \min_{x \in \cal{X}} U(x;\epsilon), \notag \\
X^*(\epsilon) &\triangleq \{x \in {\cal{X}}\; |\; U(x;\epsilon) = \Psi(\epsilon) \}, \text{ and} \notag \\
\bar{x}^*(\epsilon) &\triangleq \max X^*(\epsilon). \notag
\end{align}
\fi
%
Now, we can define the \emph{potential threshold} $\epsilon^*_{\rm p}$ as \cite[Def.~35]{yed13}
\begin{equation} \label{eq:area_thres}
\epsilon^*_{\rm p} \triangleq \sup \left\{\epsilon \in {\cal{E}}\; |\;  \bar{x}^*(\epsilon) = 0\right\}.
\end{equation}
%
%
\begin{definition} \label{def:fixed_point}
The \emph{fixed-point potential} $Q(x)$ is given by \cite[Def.~41]{yed13}
\ifonecolumn
\begin{displaymath}
Q(x) \triangleq U(x;\epsilon(x)) = x g(x) - \int_{0}^x g(z)\,\diff z - \frac{1}{k} \cdot \frac{x}{g(x)^{k-1}} \cdot g(x)^k 
= \left(1 -\frac{1}{k} \right) x g(x) -  \int_{0}^x g(z)\,\diff z.
\end{displaymath}
\else
\begin{displaymath}
\begin{split}
Q(x) &\triangleq U(x;\epsilon(x)) \\
&= x g(x) - \int_{0}^x g(z)\,\diff z - \frac{1}{k} \cdot \frac{x}{g(x)^{k-1}} \cdot g(x)^k \\
&= \left(1 -\frac{1}{k} \right) x g(x) -  \int_{0}^x g(z)\,\diff z.
\end{split}
\end{displaymath}
\fi
\end{definition}

\begin{definition} \label{def:trial}
The \emph{trial entropy} is given by
\begin{equation}  \notag 
P(x) \triangleq \int_{0}^{x} h^{\rm EMP}(z) \epsilon'(z)\,\diff z = \int_{0}^{x} g(z)^k \epsilon'(z)\,\diff z.
\end{equation}
\end{definition}

The following theorem shows that the area threshold and the potential threshold are equal, $\bar{\epsilon}=\epsilon^*_{\rm p}$.
\begin{theorem} \label{th:area_potential_thres_same}
The area threshold from Definition~\ref{def:MAP} is equal to the potential threshold from (\ref{eq:area_thres}).
\end{theorem}

\begin{IEEEproof}
It follows that
\ifonecolumn
\begin{displaymath}
g(z)^k \epsilon'(z) = g(z)^k \left( \frac{g(z)^{k-1} - (k-1) g(z)^{k-2} g'(z) z}{g(z)^{2(k-1)}} \right)
= g(z) - (k-1) z g'(z).
\end{displaymath}
\else
\begin{displaymath}
\begin{split}
g(z)^k \epsilon'(z) &= g(z)^k \left( \frac{g(z)^{k-1} - (k-1) g(z)^{k-2} g'(z) z}{g(z)^{2(k-1)}} \right)\\
&= g(z) - (k-1) z g'(z).
\end{split}
\end{displaymath}
\fi
Thus,
\ifonecolumn
\begin{align} 
P(x) &= \int_{0}^x g(z)\,\diff z - (k-1) \int_{0}^x z g'(z)\,\diff z 
= \int_{0}^x g(z)\,\diff z - (k-1) \left( z g(z)\Big{|}_{0}^x - \int_{0}^x g(z)\,\diff z \right) \nonumber\\
&= k \int_{0}^x g(z)\,\diff z + (1-k) x g(x) \nonumber\\
&= -k Q(x).\label{eq:PpropQ}
\end{align}
\else
\begin{align} 
P(x) &= \int_{0}^x g(z)\,\diff z - (k-1) \int_{0}^x z g'(z)\,\diff z \nonumber\\
&= \int_{0}^x g(z)\,\diff z - (k-1) \left( z g(z)\Big{|}_{0}^x - \int_{0}^x g(z)\,\diff z \right) \nonumber\\
&= k \int_{0}^x g(z)\,\diff z + (1-k) x g(x) \nonumber\\
&= -k Q(x).\label{eq:PpropQ}
\end{align}
\fi
%
%
%
%
We have
\ifonecolumn
\begin{align}
\int_{x}^{1} h^{\rm EMP}(z)\,\diff \epsilon(z) 
= \int_{0}^{1} h^{\rm EMP}(z)\,\diff \epsilon(z)
- \int_{0}^{x} h^{\rm EMP}(z)\,\diff \epsilon(z) 
\overset{(a)}{=} \int_{0}^{1} h^{\rm EMP}(z)\,\diff \epsilon(z) - P(x), \label{eq:EBPrelationTrial}
\end{align}
\else
\begin{align}
\int_{x}^{1} h^{\rm EMP}(z)\,\diff \epsilon(z) &
= \int_{0}^{1} h^{\rm EMP}(z)\,\diff \epsilon(z) \notag \\
&\;\;\;\;- \int_{0}^{x} h^{\rm EMP}(z)\,\diff \epsilon(z) \notag \\
&\overset{(a)}{=} \int_{0}^{1} h^{\rm EMP}(z)\,\diff \epsilon(z) - P(x), \label{eq:EBPrelationTrial}
\end{align}
\fi
where $(a)$ follows from the definition of the trial entropy in Definition~\ref{def:trial}. Now, using Definition~\ref{def:MAP} it follows that $P(x^*)=0$. Using (\ref{eq:PpropQ}) this implies that $Q(x^*)=0$. From \cite[Lem.~46]{yed13}, for a proper admissible system, which is the case here as shown in \cref{lem:proper_admissible}, the corresponding value $\bar{\epsilon} = \epsilon(x^*)$ is indeed equal to the potential threshold, and the result follows. 
\end{IEEEproof}

In the next section, we derive the Maxwell decoder for CBs and prove that the area threshold $\bar{\epsilon}$ is an upper bound on the Maxwell decoding threshold (formally defined below), which, in turn, is a lower bound on the MAP decoding threshold (which is also formally defined below). We then give a conjecture, supported by simulation results, that the area threshold is in fact equal to the Maxwell decoding threshold, and thus a lower bound on the MAP decoding threshold.

\section{Maxwell Decoder} \label{sec:Maxwell}

Similar to LDPC codes, a Maxwell decoder \cite{mea08} can be constructed for CBs, and it can be analyzed asymptotically using DE on the equivalent graph representation introduced in Section~\ref{sec:EquivalentGraph}. For LDPC codes, the Maxwell decoder is a MAP decoder, thus the MAP decoding threshold is equal to the Maxwell decoding threshold.  However, it is important to notice that, in general, for CBs the Maxwell decoder is \emph{not} a MAP decoder, since the flow size distribution is typically nonuniform. Thus, the Maxwell decoding threshold (on $\epsilon$) is in general a lower bound on the MAP decoding threshold. 


The MAP EXIT curve  of a single-layer CB  is defined as 
\begin{displaymath}
h^{\rm MAP}(\epsilon) = \limsup_{m_{0} \to \infty} \mathbb{E}\left[\frac{1}{m_{0}} \sum_{\mathsf{f} \in \F} \mathbb{H}(\phi(\mathsf{f}) | \{\phi(\mathsf{c}): \mathsf{c} \in \C\}) \right],
\end{displaymath}
where $\mathbb{H}(\cdot|\cdot)$ denotes conditional entropy and the expectation $\mathbb{E}[\cdot]$ is taken with respect to the ensemble of bipartite graphs $\mathcal{G}$ of the underlying  CBs. The MAP decoding threshold, denoted by $\epsilon_{\rm MAP}$, is defined as
\begin{align*}
\epsilon_{\rm MAP} \triangleq \sup \left\{\epsilon \in {\cal{E}}\; |\;  h^{\rm MAP}(\epsilon)= 0\right\}.
\end{align*}
The Maxwell decoding threshold (and EXIT chart) can be defined in an analogue manner under the constraint of a uniform flow size distribution.

\subsection{Maxwell Decoder on the Original Graph}

The Maxwell decoder is an MP decoder with guesses, inspired from statistical mechanics and the theory of phase transitions, which was outlined in \cite{mea08} in the context of LDPC codes. The concept of an MP decoder with guesses was considered in various works prior to \cite{mea08}, with the more practical motivation of improving the performance of the standard MP  decoder for finite-length codes \cite{nik04,ROS07}. The motivation in \cite{mea08}, on the contrary, was an asymptotic analysis.

The Maxwell decoder works in the following way. First, standard MP decoding as described in Section~\ref{sec:message_passing} is run until the algorithm stops, i.e., there is no further progress from one iteration to the next one. Then, the value of an \emph{unknown} (to be defined below) flow node is \emph{guessed} and all its outgoing messages (more precisely, the second component of the messages, see below for details) are set to '${\rm g}$'. Moreover, the label of the flow node is changed from \emph{unknown} to \emph{guessed}. Decoding is performed again until there is no further progress. If there are still \emph{unknown} flow nodes when the algorithm stops, a second \emph{unknown} flow node is \emph{guessed}, its label is changed to \emph{guessed} and, finally, all its outgoing messages (the second component of the messages) are set to '${\rm g}$'. This process continues until there is no more \emph{unknown} flow nodes to guess. In the end we end up with a system of linear equations for the values of the \emph{guessed} flow nodes that can be solved. The system of equations may or may not have a unique solution. In case there is no unique solution, ML decoding will also fail.

Let $\boldsymbol{\psi}^{(\ell), \rm M}_{\mathsf{c} \rightarrow \mathsf{f}}$ denote the message from counter node $\mathsf{c}$ to flow node $\mathsf{f}$ at iteration $\ell \geq 1$ of the Maxwell decoder. Likewise, let $\boldsymbol{\mu}^{(\ell),\rm M}_{\mathsf{f} \rightarrow \mathsf{c}}$ denote the corresponding message from flow node $\mathsf{f}$ to counter node $\mathsf{c}$ at iteration $\ell \geq 0$. The messages in the Maxwell decoder are $2$-tuples defined as
\begin{align*}
\boldsymbol{\psi}^{(\ell), \rm M}_{\mathsf{c} \rightarrow \mathsf{f}} &= (\psi^{(\ell), \rm M}_{1,\mathsf{c} \rightarrow \mathsf{f}}, \psi^{(\ell), \rm M}_{2,\mathsf{c} \rightarrow \mathsf{f}}),  \\
\boldsymbol{\mu}^{(\ell), \rm M}_{\mathsf{f} \rightarrow \mathsf{c}} &= (\mu^{(\ell), \rm M}_{1,\mathsf{f} \rightarrow \mathsf{c}}, \mu^{(\ell), \rm M}_{2,\mathsf{f} \rightarrow \mathsf{c}}).
\end{align*}
The first component of the messages is updated according to the update rules of the standard MP decoder described in Section~\ref{sec:message_passing}. 
 $t$ for brevity.
The second component of the messages takes values in the set $\{0,\ast,{\rm g}\}$. The meaning of a $0$-message is that the value of the flow node from which this message emanates is \emph{known}, the meaning of the $\ast$-message is that the value of the flow node is \emph{unknown} and, finally, the meaning of a $\rm g$-message is that the value of the flow node has been guessed or that the value of the flow node can be expressed as a linear combination of the values of other flow nodes that all have been guessed. Operationally, we can think of a $\rm g$-message emanating from a flow node $\mathsf{f}$ as a shorthand notation for a nonempty list $\{\mathsf{f}_{j_1},\dots,\mathsf{f}_{j_l}, \mathsf{b}_{j_1},\dots,\mathsf{b}_{j_l}, \KK \}$, where $\{\mathsf{f}_{j_1},\dots,\mathsf{f}_{j_l}\}$ is a list of flow nodes whose values have been guessed, $\{\mathsf{b}_{j_1},\dots,\mathsf{b}_{j_l}\}$, with $\mathsf{b}_{j_i} \in \mathbb{Z}$, where $\mathbb{Z}$ is the set of integers, is a corresponding list of integer coefficients, and $\KK$ is an integer that indicates that $\phi(\mathsf{f})$ can be expressed as $\phi(\mathsf{f})=\KK - \mathsf{b}_{j_1} \phi(\mathsf{f}_{j_1}) - \cdots - \mathsf{b}_{j_l} \phi(\mathsf{f}_{j_l})$.

For the second component of the counter-to-flow node messages we have the following update rule, for $\ell=1,\dotsc,\ell_{\rm max}$, 
\begin{align} \label{eq:counter-to-flow}
\psi^{(\ell), \rm M}_{2,\mathsf{c} \rightarrow \mathsf{f}} &= \begin{cases}
0, & \text{if $\forall \mathsf{f}' \in \Gamma(\mathsf{c}) \setminus \ff,\; \mu^{(\ell-1),\rm M}_{2,\mathsf{f}' \rightarrow \cc} = 0$} \\
\ast, & \text{if $\exists \mathsf{f}' \in \Gamma(\mathsf{c}) \setminus \ff,\; \mu^{(\ell-1), \rm M}_{2,\mathsf{f}' \rightarrow \cc} = \ast$} \\
{\rm g}, & \text{if $\forall \mathsf{f}' \in \Gamma(\mathsf{c}) \setminus \ff,\; \mu^{(\ell-1), \rm M}_{2,\mathsf{f}' \rightarrow \cc} \neq \ast$ and} \\
 & \text{\;\;\;\,$\exists \mathsf{f}' \in \Gamma(\mathsf{c}) \setminus \ff,\; \mu^{(\ell-1), \rm M}_{2,\mathsf{f}' \rightarrow \cc} = {\rm g}$} \end{cases}.
%
\end{align}
Initially, we set $\mu^{(0), \rm M}_{2,\mathsf{f} \rightarrow \mathsf{c}} = \ast$ for all $(\mathsf{f},\mathsf{c}) \in \mathcal{E}$. The last rule in (\ref{eq:counter-to-flow}) states that we will get a ${\rm g}$-message if for all connected flow nodes (except the one to which the outgoing message is transmitted) the value is either known, has been guessed, or can be expressed as a linear combination of the values of guessed flow nodes, and at least one connected flow node sends a ${\rm g}$-message. Since the outgoing message from a counter node is the difference of the value of the counter node and the sum of the values of all its neighboring flow nodes (except the one to which the outgoing message is transmitted) (see (\ref{eq:Bp1})), it follows that the outgoing message can also be expressed as a linear combination of guessed flow node values. 
Operationally, we have several lists entering the counter node $\cc$. 
The outgoing list of flow nodes will be the \emph{union} of the incoming lists, where the coefficient of flow nodes that occur a multiple number of times in the incoming lists is replaced by the corresponding sum of coefficients multiplied by $-1$. If the sum is equal to zero, the flow node is eliminated from the outgoing list.
%
Also, the integer $\KK$ of the outgoing list is $\phi(\cc)$ minus the sum of all $\KK$'s of the incoming lists.

For the flow-to-counter node messages, the first component is again updated according to the update rule of the standard MP decoder described in Section~\ref{sec:message_passing}. For the second component,  we have the following update rule, for $\ell=1,\dotsc,\ell_{\rm max}$,
\begin{align} \label{eq:flow-to-counter}
\mu^{(\ell), \rm M}_{2,\mathsf{f} \rightarrow \mathsf{c}} &= \begin{cases}
0, & \text{if $\exists \mathsf{c}' \in \Gamma(\mathsf{f}) \setminus \mathsf{c},\; \psi^{(\ell), \rm M}_{2,\mathsf{c}' \rightarrow \ff} = 0$ or} \\
&\text{\;\;\;\,$\mu^{(\ell), \rm M}_{1,\mathsf{f} \rightarrow \mathsf{c}} = \mu^{(\ell-1), \rm M}_{1,\mathsf{f} \rightarrow \mathsf{c}}$}\\
\ast, & \text{if $\forall \mathsf{c}' \in \Gamma(\mathsf{f}) \setminus \mathsf{c},\; \psi^{(\ell), \rm M}_{2,\mathsf{c}' \rightarrow \ff} = \ast$} \\
{\rm g}, & \text{if $\forall \mathsf{c}' \in \Gamma(\mathsf{f}) \setminus \mathsf{c},\; \psi^{(\ell), \rm M}_{2,\mathsf{c}' \rightarrow \ff} \neq 0$ and} \\
 & \text{\;\;\;\,$\exists \mathsf{c}' \in \Gamma(\mathsf{f}) \setminus \mathsf{c},\; \psi^{(\ell), \rm M}_{2,\mathsf{c}' \rightarrow \ff} = {\rm g}$} \end{cases},
\end{align}
where we assume that the first component of the messages, i.e., $\mu^{(\ell), \rm M}_{1,\mathsf{f} \rightarrow \mathsf{c}}$ is updated first.

\subsection{Maxwell Decoder on the Equivalent Graph}

The Maxwell decoder can also be implemented on the equivalent graph introduced in \cref{sec:EquivalentGraph}. In this case, the first component of the messages is updated according to the update rules of the MP decoder on the equivalent graph, as described in \cref{sec:message_passing_equiv}. The update rules for the second component are as follows.

For the flow nodes $\tilde{\mathsf{f}} \in \tilde{\F}$ in the equivalent graph, we use the same update rule as in (\ref{eq:flow-to-counter}) for type-$1$ edges over all incoming messages on all connected edges (except the one on which the outgoing message is transmitted). On the other hand, for type-$2$ edges in an edge bundle, the update rule is a bit different. Similar to the MP decoder (see \cref{sec:EquivalentGraph}), we can use the same update rule as for type-$1$ edges, i.e., the update rule in (\ref{eq:flow-to-counter}), applied over all incoming messages on all connected edges except the particular type-$1$ edge connected to the counter node that labels the type-$2$ edge in the bundle under consideration.

For the counter nodes $\tilde{\mathsf{c}} \in \tilde{\C}$ in the equivalent graph, the update rule follows a three-step procedure, similar to that of the MP decoder described in \cref{sec:message_passing_equiv}. In particular, the three-step procedure is the same as for the MP decoder, with the difference that the update rule in (\ref{eq:counter-to-flow}), the update rule in (\ref{eq:flow-to-counter}), and the update rule in (\ref{eq:counter-to-flow}) are used in steps 1, 2, and 3, respectively, of the three-step procedure described in \cref{sec:message_passing_equiv} (instead of (\ref{eq:Bp1}), (\ref{eq:Bp2}), and (\ref{eq:Bp1}), respectively, for the MP decoder). Furthermore, as for the MP decoder on the equivalent graph, this three-step procedure only assigns outgoing messages on the type-$1$ edges connected to $\tilde{\mathsf{c}}$ in the equivalent graph. The second component of the outgoing messages on the type-$2$ edges in the edge bundles is set to $*$.

Now, let $\hat{\phi}_{\mathcal{G}}^{(\ell),\rm M}(\ff)$ denote the estimated flow size at flow node $\ff \in \mathcal{G}$ at iteration $\ell$ under Maxwell  decoding on the original CB graph $\mathcal{G}$ and $\sfl_{\mathcal{G}}^{(\ell),\rm M}(\ff)$ its label (``guessed'', ``unknown'', or ``known''). The corresponding estimated flow size at iteration $\ell$ and flow node $\tilde{\ff} \in \tilde{\mathcal{G}}$ (corresponding to flow node $\ff$ in the original graph $\mathcal{G}$) under Maxwell decoding on the equivalent graph $\tilde{\mathcal{G}}$  is denoted by  $\hat{\phi}_{\tilde{\mathcal{G}}}^{(\ell), \rm M}(\tilde{\ff})$. Its label is denoted by $\sfl_{\tilde{\mathcal{G}}}^{(\ell), \rm M}(\tilde{\ff})$.  Based on the above discussion, the proposition below (which is analogous to \cref{prop:equivalent} for the MP decoder) follows by inspection.

\begin{proposition} \label{prop:equivalent_maxwell}
For any given CB graph $\mathcal{G}$ and iteration number $\ell \geq 1$ (with respect to the original graph $\mathcal{G}$),
\begin{displaymath}
\sfl_{\mathcal{G}}^{(2\ell), \rm M}(\ff)=\sfl_{\tilde{\mathcal{G}}}^{(\ell), \rm M}(\tilde{\ff}),\; \forall \ff \in \mathcal{G},
\end{displaymath}
and for all $\ff  \in \mathcal{G}$ that are labeled as ``known'',
\begin{displaymath}
\hat{\phi}_{\mathcal{G}}^{(2\ell), \rm M}(\ff)=\hat{\phi}_{\tilde{\mathcal{G}}}^{(\ell), \rm M}(\tilde{\ff}),
\end{displaymath}
i.e., the two graphs are \emph{iteration-by-iteration} equivalent.
\end{proposition}


We have the following result.

\begin{theorem} \label{th:MAPa}
The area threshold from Definition~\ref{def:MAP}, $\bar{\epsilon}$, is an upper bound on the Maxwell decoding threshold.
\end{theorem}
\begin{IEEEproof}
See Appendix~\ref{app:maxwell_decoder}.
\end{IEEEproof}




\begin{remark}The proof of \cref{th:MAPa} in Appendix~\ref{app:maxwell_decoder} is based on DE of the Maxwell decoder as outlined above. As in \cref{sec:AsymptoticAnalysis}, two consecutive iterations are combined into a single one, i.e., we perform the analysis on the equivalent graph of \cref{sec:EquivalentGraph}. From \cref{prop:equivalent}, \cref{prop:equivalent_maxwell}, and the equivalence of the DE on the two graphs, it follows directly that the upper bound from \cref{th:MAPa} applies also to the original CB graph.\end{remark}

Below, we provide an analogue to \cite[Th.~3.121]{modern_coding_theory} for the special case of $\epsilon=\bar{\epsilon}$ for the first statement of the theorem. The  theorem below, together with \cref{th:MAPa} and the simulation results from \cref{sec:NumericalResults} of the Maxwell decoder, lead to the conjecture that the area threshold is in fact equal to the Maxwell decoding threshold.

\begin{theorem} \label{th:maxwell_threshold}
The EMP EXIT curve (defined for the equivalent graph as in (\ref{eq:EBP_EXIT})) for the expected residual CB graph when the peeling decoder stops is given in parametric form by
\begin{displaymath}
(\tilde{\epsilon},\tilde{h}^{\rm EMP}) = (\tilde{\epsilon}(z;x), (1-\tilde{\rho}(1-z;x))^{k}),
\end{displaymath}
where
\begin{equation} \notag 
\tilde{\epsilon}(z;x)= \frac{z}{(1-\tilde{\rho}(1-z;x))^{k-1}} \text{ and } \tilde{\rho}(z;x) = 1 -\frac{g(x-zx)}{g(x)},
\end{equation}
and where $x=x(\epsilon)$ is the largest fixed-point of the DE recursion in (\ref{eq:DECounter2FixedPoint}) for a given $\epsilon$. Furthermore, for $\epsilon = \bar{\epsilon}$ (the area threshold from Definition~\ref{def:MAP}), the area $\int_{0}^1 \tilde{h}^{\rm EMP}(z;x) \, \diff \tilde{\epsilon}(z;x)$  is equal to zero.
\end{theorem}

\begin{IEEEproof}
See Appendix~\ref{proof:theorem2}.
\end{IEEEproof}

\begin{conjecture} \label{th:MAP}
The area threshold from Definition~\ref{def:MAP}, $\bar{\epsilon}$, is equal to the Maxwell decoding threshold and thus a lower bound on the MAP decoding threshold. 
\end{conjecture}

Note that ML decoding of CBs is equivalent to solving a system of linear equations (with binary coefficient matrix) over the set of positive integers larger than or equal to $f_{\rm min}$. This problem resembles the well-known \emph{integer knapsack problem}, which is known to be NP-hard. It can be shown that having a polynomial-time algorithm for the decoding of CBs implies having a polynomial-time algorithm for the integer knapsack problem. Thus, ML decoding of CBs is an NP-hard problem. This is in contrast to ML decoding of LDPC codes on the BEC which can be done in polynomial time, since it is equivalent to solving a system of linear equations over the binary field. Note that the ML decoding problem  (which is equivalent in terms of error performance to the Maxwell decoding problem) of CBs can be cast as a linear integer program as follows,
\ifonecolumn
\begin{equation} \label{eq:LIP}
\begin{array}{rl}
\text{maximize} &  \sum_{\mathsf{f} \in \mathcal{F}} \phi(\mathsf{f}) \\
\text{s.\,t.} & \sum_{\mathsf{f} \in \Gamma(\mathsf{c})} \phi(\mathsf{f}) = \phi(\mathsf{c}),\, \forall \mathsf{c} \in \mathcal{C} \text{ and } 
\phi(\mathsf{f}) \geq f_{\rm min},\, \forall \mathsf{f} \in \mathcal{F}
\end{array}.
\end{equation}
\else
\begin{equation} \label{eq:LIP}
\begin{array}{rl}
\text{maximize} &  \sum_{\mathsf{f} \in \mathcal{F}} \phi(\mathsf{f}) \\
\text{s.\,t.} & \sum_{\mathsf{f} \in \Gamma(\mathsf{c})} \phi(\mathsf{f}) = \phi(\mathsf{c}),\, \forall \mathsf{c} \in \mathcal{C} \text{ and } \\
&\phi(\mathsf{f}) \geq f_{\rm min},\, \forall \mathsf{f} \in \mathcal{F}
\end{array}.
\end{equation}
\fi
Thus, when simulating the Maxwell decoder in \cref{sec:NumericalResults}, we do not implement it using the guessing framework above (which is very useful for asymptotic analysis using DE as shown in Appendix~\ref{app:maxwell_decoder}), but instead solve the linear integer program in (\ref{eq:LIP}) using the commercial Gurobi software \cite{Gurobi600}.

\ifonecolumn
\else
\begin{figure*}[!t]
\normalsize 
\begin{equation} \label{eq:sc_density_evo}
\xi_i(k,\gamma,N,w,\bm{x}^{(2\ell-2)}) = 
 \sum_{g=1}^{N} A_{g,i} \left[ \sum_{h=1}^{M} A_{g,h} \left[ 1- \rho \left(1 - \sum_{p=1}^{N} A_{p,h} \left\{ \sum_{q=1}^{M} A_{p,q} \left[ 1-\rho(1-x_{q}^{(2\ell-2)}) \right] \right\}^{k-1} \right) \right]  \right]^{k-1}. 
\end{equation}
\begin{equation} \label{eq:sc_density_evo_1}
\tilde{\xi_i}(k,\gamma,N,w,\bm{x}^{(2\ell-2)}) = 
 \sum_{g=1}^{N} A_{g,i} \left[ \sum_{h=1}^{M} A_{g,h} \left[ 1- \rho \left(1 - \sum_{p=1}^{N} A_{p,h} \left\{ \sum_{q=1}^{M} A_{p,q} \left[ 1-\rho(1-x_{q}^{(2\ell-2)}) \right] \right\}^{k-1} \right) \right]  \right]^{k}. 
\end{equation}
 \hrulefill
\vspace*{-2mm}
\end{figure*}
\fi

\section{Spatially-Coupled Counter Braids} \label{sec:SC-CBs}


We consider the ensemble $(\lambda,\rho,N,w)$ of single-layer SC-CBs  (coupling the original bipartite graph), where $N$ is the coupling chain length and $w$, $1\le w\le N+1$, is a smoothing parameter \cite{kud11}. The corresponding ensemble of SC graphs is denoted by $\mathcal{G}_{\rm c}(\lambda,\rho,N,w)$. The ensemble is constructed as follows. A collection of $N$ flow-node groups are placed at positions $\{1,2,\ldots,N\}$, each containing $\kappa$ nodes of degree $k$, such that $\kappa=\frac{m_0}{N}$ (the total number of flow nodes, i.e., the number of distinct flows to be counted, is $m_0$). Furthermore, a collection of $M=N+w-1$ counter-node groups are placed at positions $\{1,2,\ldots,M\}$, each containing $\frac{\kappa R_{j} L'(1)}{R'(1)}=\frac{\kappa R_{j} k}{\gamma}$ nodes of degree $j$ (we implicitly assume that $N$ is chosen such that $\frac{\kappa R_{j} k}{\gamma}$ and also $\frac{m_0}{N}$ are integers).


The $\kappa k$ edge sockets in each group of flow and counter nodes are partitioned into $w$ equally-sized subgroups (assuming that $\frac{\kappa k}{w}$ is an integer) using a uniform random interleaver. We denote by $\mathcal{P}_{n,i}^{(\mathsf{f})}$ and $\mathcal{P}_{n,i}^{(\mathsf{c})}$ the set of flow and counter node sockets, respectively, in the $i$-th subgroup, $i=0,1,\ldots,w-1$, at position $n$, where $n=1,\ldots,N$ for flow node sockets and $n=1,\ldots,M$ for counter node sockets. The SC ensemble is constructed by adding edges that connect the sockets in $\mathcal{P}_{n,i}^{(\mathsf{f})}$ to the sockets in $\mathcal{P}_{n+i,i}^{(\mathsf{c})}$. Different graphs are obtained by different socket associations. Note that this construction leaves some sockets of the counter-node groups at the boundary unconnected and these are removed. In the following, 
we will denote  
 the coupled ensemble $\mathcal{G}_{\rm c}(\lambda,\rho,N,w)$  by the alternative notation  $\mathcal{G}_{\rm c}(k,\gamma,N,w)$. 

\subsection{Density Evolution Recursion}

Denote by $x_i^{(\ell)}$, $i=1,\ldots,M$, the error probability of an outgoing message from a flow node at position $i$ at iteration $\ell$. Note that since  there are no flow-node groups at positions $i > N$, initially $x_i^{(0)}=0$ for $N < i \leq M$.  Furthermore, let $\x^{(\ell)} = (x_{1}^{(\ell)},\ldots,x_{M}^{(\ell)})$.  Using the ensemble defined above, we get the recursion $x^{(2\ell)}_{i} = \epsilon \cdot \xi_i(k,\gamma,N,w,\bm{x}^{(2\ell-2)})$ where the function $\xi_i(k,\gamma,N,w,\bm{x}^{(2\ell-2)})$ is given in  (\ref{eq:sc_density_evo}) at the top of the page, and where $\bm{A} = \{A_{p,q}\}$ is the $N \times M$ matrix defined by
\begin{equation} \notag
A_{p,q} = [ \bm{A}]_{p,q} = \begin{cases}
\frac{1}{w}, & \text{if $1 \leq q-p+1 \leq w$} \\
0, & \text{otherwise} \end{cases}.
\end{equation}


Note that, contrary to SC-LDPC codes, for which the SC recursion contains two summations (an outer summation over $N$ terms and an inner summation over $M$ terms), the recursion for SC-CBs contains four summations\footnote{Due  to the band-structure of the matrix $\bm{A}$, there are only $w$ nonzero terms in all four summations.} (two summations over $N$ terms and two summations over $M$ terms), since the update rule for the flow nodes is different for odd and even iterations. As a result, SC-CBs do not fit within the general framework of coupled scalar recursions outlined in \cite{yed13}.

The MP decoding threshold of the coupled ensemble $\mathcal{G}_{\rm c}(k,\gamma,N,w)$ is defined (analogous to the uncoupled case) as ${\epsilon}_{\rm MP}^{\rm c} =  {\epsilon}_{\rm MP}^{\rm c}(k,\gamma,N,w)$, where
\begin{equation} \notag
%
{\epsilon}_{\rm MP}^{\rm c}(k,\gamma,N,w) 
\triangleq
\sup\left\{ \epsilon \in \mathcal{E}  \; |\;  \x^{(\infty)}(k,\gamma,\epsilon,N,w) = \bm{0} \right\}.
\end{equation}
Alternatively, we can analyze the asymptotic behavior of SC-CBs by fixing $\epsilon$ and finding the minimum value of $\beta$, denoted by ${\beta}_{\rm MP}^{\rm c} =  {\beta}_{\rm MP}^{\rm c}(k,\epsilon,N,w)$, such that decoding is successful,  where 
\begin{equation} \notag
{\beta}_{\rm MP}^{\rm c}(k,\epsilon,N,w) 
\triangleq
\inf\left\{ \beta > 0  \; |\;  \x^{(\infty)}(k,k/\beta,\epsilon,N,w) = \bm{0} \right\}.
\end{equation}

\ifonecolumn
\begin{figure*}[!t]
\normalsize 
\begin{equation} \label{eq:sc_density_evo}
\xi_i(k,\gamma,N,w,\bm{x}^{(2\ell-2)}) = 
\epsilon \sum_{g=1}^{N} A_{g,i} \left[ \sum_{h=1}^{M} A_{g,h} \left[ 1- \rho \left(1 - \sum_{p=1}^{N} A_{p,h} \left\{ \sum_{q=1}^{M} A_{p,q} \left[ 1-\rho(1-x_{q}^{(2\ell-2)}) \right] \right\}^{k-1} \right) \right]  \right]^{k-1}. 
\end{equation}
 \hrulefill
\vspace*{-2mm}
\end{figure*}
\fi




\begin{lemma} \label{lem:design_rate}
The design rate $\mathcal{R}^{\rm c}(k,\gamma,N,w,d)$ (in bits per flow) of the coupled ensemble $\mathcal{G}_{\rm c}(k,\gamma,N,w)$ with $w \leq N+1$ is
\ifonecolumn
\begin{equation} \notag
\mathcal{R}^{\rm c}(k,\gamma,N,w,d)
= \frac{d k\sum_{i=0}^{\infty} \frac{\mathrm{e}^{-\gamma} \gamma^{i}}{i!} \left[ N-w+1+2\sum_{j=1}^{w-1} \left(1- \left( \frac{j}{w} \right)^{i^+} \right)\right]}{\gamma N},
\end{equation}
\else
\begin{equation} \notag
\begin{split}
&\mathcal{R}^{\rm c}(k,\gamma,N,w,d)\\
&= \frac{d k\sum_{i=0}^{\infty} \frac{\mathrm{e}^{-\gamma} \gamma^{i}}{i!} \left[ N-w+1+2\sum_{j=1}^{w-1} \left(1- \left( \frac{j}{w} \right)^{i^+} \right)\right]}{\gamma N},
\end{split}
\end{equation}
\fi
where $i^+=i$ for $i > 0$, $i^+=1$ for $i = 0$, and $d$ is the depth (in number of bits) of the counters.
\end{lemma}

\begin{IEEEproof}
The proof is along the same lines as the proof of \cite[Lem.~3]{kud11} for the design rate of regular SC-LDPC code ensembles. Details are omitted for brevity.
\end{IEEEproof}

\begin{figure}[tbp]
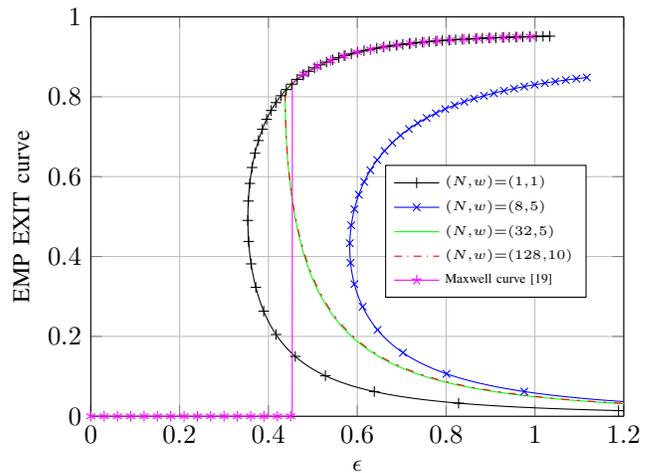

\centering
%
%
%
%
%
\definecolor{mycolor1}{rgb}{1.00000,0.00000,1.00000}%
%

%
\caption{EMP EXIT curves for CBs and SC-CBs with $k=3$ and $\gamma = 4.233585$. For the coupled ensembles, the EMP EXIT curve of the midpoint $i = \left\lfloor \frac{M}{2} \right\rfloor$ is given.}
\label{fig:ebp_exit_curve}
\end{figure}

For the  uncoupled ensemble  $\mathcal{G}_{\rm c}(k,\gamma,1,1)$, the design rate reduces to  $\frac{d k}{\gamma} = d \beta$. 
Not surprisingly, the coupling mechanism yields a rate increase that vanishes in the limit $N \to \infty$ (for $w$ fixed).

Finally, we define 
\begin{equation} \label{eq:beta_c}
\beta^{\rm c} =  \beta^{\rm c}(k,\gamma,N,w) \triangleq \frac{R^{\rm c}(k,\gamma,N,w,d)}{d}
\end{equation}
 for a coupled system (as in the uncoupled case) to be the fraction of the number of counter and flow nodes. 



\subsection{Extended Message Passing Extrinsic Information Transfer Curve}
As for the uncoupled system, we can define  EMP EXIT curves $(\epsilon(\bm{x}), h_{i}^{\rm EMP}(\bm{x}))$ for each position $i$, $i=1,\dots,M$,
for a SC system based on fixed-points of the coupled DE recursion $x^{(2\ell)}_{i} = \epsilon \cdot \xi_i(k,\gamma,N,w,\bm{x}^{(2\ell-2)})$, where $\xi_i(k,\gamma,N,w,\bm{x}^{(2\ell-2)})$ is given
in (\ref{eq:sc_density_evo}), and 
 where $\epsilon(\bm{x}) = \frac{x_i}{\xi_i(k,\gamma,N,w,\bm{x})}$, $h_{i}^{\rm EMP}(\bm{x}) = \tilde{\xi}_i(k,\gamma,N,w,\bm{x})$ (defined in (\ref{eq:sc_density_evo_1}) at the top of the page), and $\bm{x}$ is a fixed-point of the coupled recursion. 


As an example, in Fig.~\ref{fig:ebp_exit_curve} we plot the EMP EXIT curve $h_{\lfloor \frac{M}{2} \rfloor}^{\rm EMP}$ (as a function of $\epsilon$) for $k=3$ and $\gamma=4.233585$ for the pairs $(N,w) = (8,5)$, $(32,5)$, and $(128,10)$. As we can clearly see from the figure, SC-CBs exhibit improved MP decoding thresholds. Furthermore, the thresholds saturate to some fixed value when $N$ and $w$ become large. In fact, the curves for the $(k,\gamma,32,5)$ SC-CB (the green curve) and the $(k,\gamma,128,10)$ SC-CB (the red curve) are almost indistinguishable. We also plot the \emph{Maxwell curve} from the potential function framework of \cite{yed13} (i.e., $h^{\rm EMP}(\bar{x}^*(\epsilon))$ versus $\epsilon$ where $\bar{x}^*(\epsilon)$ is taken from \cite[Def.~28]{yed13}). 
%
%
Interestingly, there is a remaining gap between $\bar{\epsilon}$ (from Definition~\ref{def:MAP}) and the MP decoding thresholds of SC-CBs when $N$ and $w$ become large. This gap is discussed in \cref{sec:discussion} below.

\subsection{Extension to More Layers}
\label{sec:ExtMoreLayers}

We remark that SC-CBs with more than one layer can be easily constructed by applying the spatial coupling procedure described above to each layer. As for multilayer CBs, the decoding of multilayer SC-CBs is performed layer-by-layer (see Section~\ref{sec:message_passing}), from the inner-most layer to the outer-most layer. Furthermore, there are no iterations between layers. Decoding is successful if the decoding of each layer (starting from the inner-most layer) is successful. Accordingly, the corresponding DE is also performed layer-by-layer. 

Denote by $x^{(\ell)}_{i,l}$ the error probability of an outgoing message from a flow node at spatial position $i$ at layer $l$ at the $\ell$-th iteration, and let $\x^{(\ell)}_l=(x^{(\ell)}_{1,l}, \ldots, x^{(\ell)}_{M,l})$. Here, we assume that the coupling chain length $N$ and the smoothing parameter $w$ are the same for all layers. Also, denote by $\epsilon_l$ the probability of 
observing a flow at layer $l$ of size strictly larger than $f_{\min,l}$ ($\epsilon_1=\epsilon$). For layer one, $f_{\min,1}=\fmin$ (see Section~\ref{sec:message_passing}). On the other hand, for a layer $l>1$, $f_{\min,l}=1$, which corresponds to the case when a counter at layer $l-1$ overflows only once (the probability of this event is nonzero).
Furthermore, let ${\epsilon}_{\rm MP}^{{\rm c},l}$ be the coupled MP decoding threshold of layer $l$, i.e.,
\ifonecolumn
\begin{displaymath}
\epsilon_{\rm MP}^{{\rm c},l}=\epsilon_{\rm MP}^{{\rm c},l}(k_l,\gamma_l,N,w) 
\triangleq\sup\left\{\epsilon_l \in\mathcal{E} \; |\;  \x^{(\infty)}_l(k_l,\gamma_l,\epsilon_l,N,w)=\bm{0}\right\}.
\end{displaymath} 
\else
\begin{align*}
\begin{split}
\epsilon_{\rm MP}^{{\rm c},l}&=\epsilon_{\rm MP}^{{\rm c},l}(k_l,\gamma_l,N,w) \\
&\triangleq\sup\left\{\epsilon_l \in\mathcal{E} \; |\;  \x^{(\infty)}_l(k_l,\gamma_l,\epsilon_l,N,w)=\bm{0}\right\}.
\end{split}
\end{align*} 
\fi

It is important to note that the input flow size distribution $p_\Phi$ will induce a certain flow size distribution at layer $2$, and subsequently at the next layers. We denote by $p^{\rm i}_{\Phi,l}$ the flow size distribution at layer $l$ induced by $p_\Phi$. Asymptotically, for each layer, we can find the (induced) probability $\epsilon^{\rm i}_l$ of observing a flow size strictly larger than $f_{\min,l}$ from $p^{\rm i}_{\Phi,l}$. To show the dependency of $\epsilon^{\rm i}_l$ on $p_\Phi$ we will write $\epsilon_l^{\rm i}(p_\Phi)$.
The MP decoding threshold of the multilayer SC-CB can then be computed in two steps.
\begin{enumerate}
\item For each layer $l$, compute the corresponding MP decoding threshold ${\epsilon}_{\rm MP}^{{\rm c},l}$. 
\item The MP decoding threshold of the multilayer SC-CB is then
\ifonecolumn
\begin{align*}
\epsilon^{\rm c}_{\rm MP} = \epsilon^{\rm c}_{\rm MP}(\bm k, \boldsymbol{\gamma},N,w) 
\triangleq \sup\left\{\epsilon\in\mathcal{E} \; |\;  \x_1^{(\infty)}(k_1,\gamma_1,\epsilon, N,w)=\bm{0}, \epsilon_2^{\rm i}(p_\Phi)\le \epsilon_{\rm MP}^{{\rm c},2} ,\ldots, \epsilon_L^{\rm i}(p_\Phi) \le \epsilon_{\rm MP}^{{\rm c},L}\right\},
\end{align*}
\else
\begin{align*}
\epsilon^{\rm c}_{\rm MP} &= \epsilon^{\rm c}_{\rm MP}(\bm k, \boldsymbol{\gamma},N,w) \\
&\triangleq \sup\left\{\epsilon\in\mathcal{E} \; |\;  \x_1^{(\infty)}(k_1,\gamma_1,\epsilon, N,w)=\bm{0},\right.\\
&\;\;\;\;\;\;\;\;\;\;\;\;\; \left. \epsilon_2^{\rm i}(p_\Phi)\le \epsilon_{\rm MP}^{{\rm c},2} ,\ldots, \epsilon_L^{\rm i}(p_\Phi) \le \epsilon_{\rm MP}^{{\rm c},L}\right\},
\end{align*}
\fi
where $\boldsymbol{\gamma}=(\gamma_1,\ldots,\gamma_L)$ and $\bm k=(k_1,\ldots,k_L)$.
\end{enumerate}

For a given depth of the counters at layer $l$, $d_l$, the induced flow size distribution at layer $l+1$ can be easily computed. With some abuse of notation, for an arbitrary counter node $\cc^l$ at layer $l$, $l=1,\ldots,L$, let $\phi(\cc^l)=\sum_{\ff\in\Gamma(\cc^l)}\phi(\ff)$ (i.e., $\phi(\cc^l)$ ignores the fact that the counter node has finite depth). Also, let $\phi(\ff^l)$ be the size of an arbitrary flow node $\ff^l$ at layer $l$, $l=2,\ldots,L$.  We have
\begin{align*}
\mathrm{Pr}\left(\phi(\ff^l)=a\right)=\mathrm{Pr}\left(a2^{d_{l-1}}\le \phi(\cc^{l-1})<(a+1)2^{d_{l-1}}\right),
\end{align*}
and
\ifonecolumn
\begin{align*}
\mathrm{Pr}\left(\phi(\cc^l)=a\right)
=\sum_{b=1}^{\left\lfloor\frac{a}{f_{\min,l}}\right\rfloor}\mathrm{Pr}\left(\phi(\cc^l)=a \; |\; \deg(\cc^l)=b\right)\mathrm{Pr}\left(\deg(\cc^l)=b\right),
\end{align*}
\else
\begin{align*}
&\mathrm{Pr}\left(\phi(\cc^l)=a\right)\\
&=\sum_{b=1}^{\left\lfloor\frac{a}{f_{\min,l}}\right\rfloor}\mathrm{Pr}\left(\phi(\cc^l)=a \; |\; \deg(\cc^l)=b\right)\mathrm{Pr}\left(\deg(\cc^l)=b\right),
\end{align*}
\fi
where $\deg(\cc) = |\Gamma(\cc)|$ is the degree of counter node $\cc$. Note that $\mathrm{Pr}\left(\deg(\cc^l)=b\right)$ follows directly from the counter node degree distribution.

To compute $\mathrm{Pr}\left(\phi(\cc^l)=a \; |\; \deg(\cc^l)=b\right)$ we need to enumerate all order-$b$ integer partitions of $a$ (i.e., all different ways of writing $a$ as a sum of $b$ positive integers not considering the order of the summands) in which each summand is larger than or equal to $f_{\min,l}$. Let $\mathcal{A}(a,b)$ be the set of such integer partitions of $a$. Then,
\begin{align*}
\mathrm{Pr}\left(\phi(\cc^l)=a \; |\; \deg(\cc^l)=b\right)=\sum_{A\in\mathcal{A}(a,b)}\mathrm{Pr}(A),
\end{align*}
where the probability of partition $A \in \mathcal{A}$, $\mathrm{Pr}(A)$, follows directly from the induced flow node degree distribution $p_{\Phi,l}^{\rm i}$ at layer $l$. The induced probability $\epsilon^{\rm i}_l$ is obtained as
\ifonecolumn
\begin{align*}
\begin{split}
\epsilon^{\rm i}_l&=1-\mathrm{Pr}\left(\phi(\ff^l)=0\right)-\mathrm{Pr}\left(\phi(\ff^l)=1\right)\\
&= 1- \mathrm{Pr}\left( 0 \leq \phi(\cc^{l-1}) < 2^{d_{l-1}}\right) - \mathrm{Pr}\left( 2^{d_{l-1}} \leq \phi(\cc^{l-1}) < 2^{d_{l-1}+1}\right)\\
&=1-\mathrm{Pr}\left( 0 \leq \phi(\cc^{l-1}) < 2^{d_{l-1}+1}\right).
\end{split}
\end{align*}
\else
\begin{align*}
\begin{split}
\epsilon^{\rm i}_l&=1-\mathrm{Pr}\left(\phi(\ff^l)=0\right)-\mathrm{Pr}\left(\phi(\ff^l)=1\right)\\
&= 1- \mathrm{Pr}\left( 0 \leq \phi(\cc^{l-1}) < 2^{d_{l-1}}\right) \\
&\;\;\;\;\;\;\,- \mathrm{Pr}\left( 2^{d_{l-1}} \leq \phi(\cc^{l-1}) < 2^{d_{l-1}+1}\right)\\
&=1-\mathrm{Pr}\left( 0 \leq \phi(\cc^{l-1}) < 2^{d_{l-1}+1}\right).
\end{split}
\end{align*}
\fi
For the particular case of two layers, the induced probability $\epsilon^{\rm i}_2$ is easily obtained, since only the two first terms of the first induced probability distribution $p_{\Phi,2}^{\rm i}$ are needed.

Finally, we remark that the procedure above to derive the induced flow size distributions is general and applies also to multilayer CBs (by setting the coupling chain length to $N=1$ and the smoothing parameter to $w=1$).

\section{Numerical Results}
\label{sec:NumericalResults}

In this section, we present several numerical results that show threshold improvement, but lack of threshold saturation. In addition, we present simulation results for both the MP and the Maxwell decoder of CBs that show that, in fact, there is no threshold saturation.

We assume the power law  $\mathrm{Pr}(\phi(\mathsf{f}) > \eta) = \eta^{-\alpha}$, where $\eta$ is a positive integer, for the flow size distribution \cite{kum04}. Typically, flow size distributions from real Internet traces have $\alpha \approx 2$, while for distributions with a heavier tail a smaller value of $\alpha$ should be used \cite{kum04}.  Due to the nature of the power law distribution, $f_{\rm min}=2$ for all exponents $\alpha$. The particular value of $\epsilon$ is related to $\alpha$, since $\epsilon = 2^{-\alpha}$. As an example, $\alpha = 1.5$ gives $\epsilon = 2^{-1.5}$. We use this value for $\epsilon$ in the simulations below.

\subsection{Comparison of Area and Message Passing Decoding Thresholds}

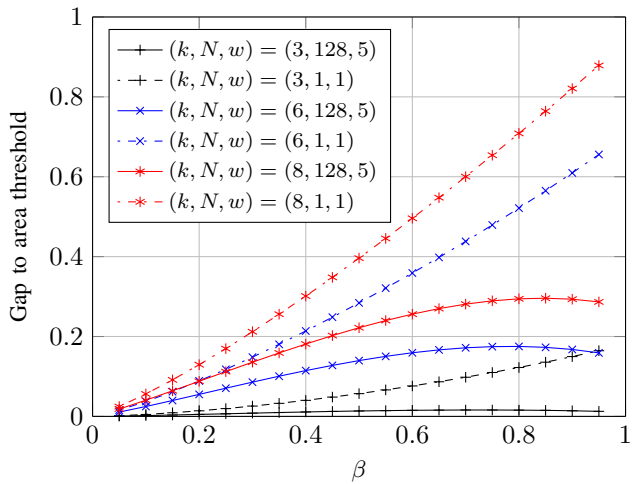
\begin{figure}[tbp]
\centering
%
%
\begin{tikzpicture}[font = \small]

\begin{axis}[%
width=0.8\columnwidth,
height=0.6\columnwidth,
at={(0.808889in,0.513333in)},
scale only axis,
separate axis lines,
every outer x axis line/.append style={black},
every x tick label/.append style={font=\color{black}},
xmin=0,
xmax=1,
xlabel={$\beta$},
xmajorgrids,
every outer y axis line/.append style={black},
every y tick label/.append style={font=\color{black}},
ymin=0,
ymax=1,
ylabel={Gap to area threshold},
ymajorgrids,
legend style={at={(0.03,0.97)},anchor=north west,legend cell align=left,align=left,draw=black}
]
\addplot [color=black,solid,mark=+,mark options={solid}]
  table[row sep=crcr]{%
0.05	0.000767\\
0.1	0.002026\\
0.15	0.003443\\
0.2	0.005019\\
0.25	0.00650899999999999\\
0.3	0.008147\\
0.35	0.00969199999999998\\
0.4	0.011079\\
0.45	0.012435\\
0.5	0.013577\\
0.55	0.014448\\
0.6	0.01523\\
0.65	0.01569\\
0.7	0.016008\\
0.75	0.016011\\
0.8	0.0156999999999999\\
0.85	0.015139\\
0.9	0.014096\\
0.95	0.012632\\
};
\addlegendentry{\footnotesize $ (k,N,w) = (3,128,5)$};

\addplot [color=black,dash pattern=on 1pt off 3pt on 3pt off 3pt,mark=+,mark options={solid}]
  table[row sep=crcr]{%
0.05	0.001699\\
0.1	0.004897\\
0.15	0.009066\\
0.2	0.013991\\
0.25	0.019548\\
0.3	0.025852\\
0.35	0.03266\\
0.4	0.040149\\
0.45	0.048323\\
0.5	0.057002\\
0.55	0.066367\\
0.6	0.076241\\
0.65	0.08675\\
0.7	0.098073\\
0.75	0.109919\\
0.8	0.122409\\
0.85	0.135725\\
0.9	0.149755\\
0.95	0.164441\\
};
\addlegendentry{\footnotesize $(k,N,w) = (3,1,1)$};

\addplot [color=blue,solid,mark=x,mark options={solid}]
  table[row sep=crcr]{%
0.05	0.010869\\
0.1	0.024762\\
0.15	0.03977\\
0.2	0.055127\\
0.25	0.070571\\
0.3	0.085784\\
0.35	0.100644\\
0.4	0.114659\\
0.45	0.127705\\
0.5	0.139725\\
0.55	0.150291\\
0.6	0.159347\\
0.65	0.166535\\
0.7	0.171795\\
0.75	0.174652\\
0.8	0.17517\\
0.85	0.172874\\
0.9	0.167591\\
0.95	0.15903\\
};
\addlegendentry{\footnotesize $(k,N,w) = (6,128,5)$};

\addplot [color=blue,dash pattern=on 1pt off 3pt on 3pt off 3pt,mark=x,mark options={solid}]
  table[row sep=crcr]{%
0.05	0.015742\\
0.1	0.037203\\
0.15	0.061902\\
0.2	0.088862\\
0.25	0.117824\\
0.3	0.14835\\
0.35	0.180437\\
0.4	0.213831\\
0.45	0.248411\\
0.5	0.284237\\
0.55	0.321121\\
0.6	0.359008\\
0.65	0.398017\\
0.7	0.438089\\
0.75	0.479227\\
0.8	0.521496\\
0.85	0.565017\\
0.9	0.609619\\
0.95	0.655729\\
};
\addlegendentry{\footnotesize $(k,N,w) = (6,1,1)$};

\addplot [color=red,solid,mark=asterisk,mark options={solid}]
  table[row sep=crcr]{%
0.05	0.018354\\
0.1	0.040536\\
0.15	0.06407\\
0.2	0.087948\\
0.25	0.112028\\
0.3	0.135753\\
0.35	0.158874\\
0.4	0.181138\\
0.45	0.202303\\
0.5	0.221949\\
0.55	0.239947\\
0.6	0.255998\\
0.65	0.269862\\
0.7	0.281055\\
0.75	0.289343\\
0.8	0.294362\\
0.85	0.295878\\
0.9	0.293412\\
0.95	0.286669\\
};
\addlegendentry{\footnotesize $ (k,N,w) = (8,128,5)$};

\addplot [color=red,dash pattern=on 1pt off 3pt on 3pt off 3pt,mark=asterisk,mark options={solid}]
  table[row sep=crcr]{%
0.05	0.024545\\
0.1	0.056207\\
0.15	0.091465\\
0.2	0.129459\\
0.25	0.169689\\
0.3	0.211957\\
0.35	0.255893\\
0.4	0.301246\\
0.45	0.348011\\
0.5	0.396248\\
0.55	0.445469\\
0.6	0.495974\\
0.65	0.54764\\
0.7	0.600225\\
0.75	0.653852\\
0.8	0.708517\\
0.85	0.764225\\
0.9	0.820856\\
0.95	0.878593\\
};
\addlegendentry{\footnotesize $(k,N,w) = (8,1,1)$};

\end{axis}
\end{tikzpicture}%
\caption{The difference $\bar{\epsilon}-{\epsilon}_{\rm MP}^{\rm c}$, where $\bar{\epsilon}$ is the area threshold from Definition~\ref{def:MAP} and ${\epsilon}_{\rm MP}^{\rm c}$ is the MP decoding threshold of SC-CBs, as a function  of $\beta$ for different values of the left-degree $k$ with $(N,w)=(128,5)$ and $(1,1)$ (uncoupled).}
\label{fig:relative_gap}
\vspace{-2ex}
\end{figure}

In Fig.~\ref{fig:relative_gap}, we plot the difference between the area threshold from Definition~\ref{def:MAP} (conjectured to be a lower bound on the MAP decoding threshold, see Conjecture~\ref{th:MAP}) and the MP decoding threshold of SC-CBs, $\bar{\epsilon}-{\epsilon}_{\rm MP}^{\rm c}$, as a function of $\beta=\frac{k}{\gamma}$ for different values of the left-degree $k$. The SC-CBs have parameters $(N,w)=(128,5)$. For comparison purposes, we also plot $\bar{\epsilon}-{\epsilon}_{\rm MP}$ for the uncoupled CBs ($(N,w)=(1,1)$). As we can observe from  the figure, there is a gap between the MP decoding threshold of SC-CBs  and the conjectured lower bound on the MAP decoding threshold.
Thus, surprisingly, threshold saturation does not occur for SC-CBs. However, the gap to the conjectured lower bound is significantly larger for uncoupled CBs, meaning that spatial coupling indeed improves performance. Note that the gap varies with $\beta$ 
and depends on $k$.

We remark that when $\beta$ becomes large (in the sense of approaching one), the area threshold $\bar{\epsilon}$ becomes larger than one. The minimum value of $\beta$ for which $\bar{\epsilon}$ is larger than one, referred to as the transition point, depends on $k$ (larger $k$ implies a lower transition point). From the perspective of the DE this is not really a problem (the recursion is a valid recursion also for values of $\epsilon$ above one), but of course there is no physical system corresponding to such values of $\epsilon$ (since $\epsilon$ is a probability). From a practical perspective, however, there is no need to design a system with a  value of $\beta$ above its transition point.

\subsection{Simulation Results}

\begin{figure}[tbp]
\centering
%
%
%
%
%
%
%

\begin{tikzpicture}[font=\small]

\begin{axis}[%
width=0.8\columnwidth,
height=0.6\columnwidth,
at={(0.808889in,0.513333in)},
scale only axis,
separate axis lines,
every outer x axis line/.append style={black},
every x tick label/.append style={font=\color{black}},
xmin=0.4,
xmax=1.1,
xlabel={$\beta$},
xmajorgrids,
every outer y axis line/.append style={black},
every y tick label/.append style={font=\color{black}},
ymode=log,
ymin=1e-06,
ymax=1,
yminorticks=true,
ylabel={SER},
ymajorgrids,
yminorgrids,
legend style={at={(0.03,0.03)},anchor=south west,legend cell align=left,align=left,draw=black}
]
\addplot [color=black,solid,mark=+,mark options={solid}]
  table[row sep=crcr]{%
%
%
0.800000 0.469639 \\
0.825000 0.421484 \\
0.850000 0.336461 \\
0.875000 0.293468 \\
0.900000 0.170969 \\
0.925000 0.069666 \\
0.950000 0.014120 \\
0.975000 0.003094 \\
1.000000 0.000164 \\
1.025000 0.000009 \\
};
\addlegendentry{$\scriptscriptstyle (N,w,m_0)= (1,1,1024)$};

\addplot [color=black,dash pattern=on 1pt off 3pt on 3pt off 3pt,mark=+,mark options={solid}]
  table[row sep=crcr]{%
%
%
0.800000 0.439875 \\
0.825000 0.408486 \\
0.850000 0.383143 \\
0.875000 0.275107 \\
0.900000 0.062784 \\
0.925000 0.002542 \\
0.950000 0.000010 \\
};
\addlegendentry{$\scriptscriptstyle (N,w,m_0)= (1,1,4096)$};

\addplot [color=blue,solid,mark=x,mark options={solid}]
  table[row sep=crcr]{%
%
%
0.650391 0.405164 \\
0.663574 0.376807 \\
0.676758 0.366477 \\
0.689941 0.236641 \\
0.703125 0.167449 \\
0.720703 0.059450 \\
0.733887 0.021035 \\
0.747070 0.006022 \\
0.760254 0.000796 \\
0.773438 0.000096 \\
0.791016 0.000005 \\
};
\addlegendentry{$\scriptscriptstyle (N,w,m_0)= (16,3,4096)$};

\addplot [color=blue,dash pattern=on 1pt off 3pt on 3pt off 3pt,mark=x,mark options={solid}]
  table[row sep=crcr]{%
%
%
0.647095 0.408662 \\
0.661377 0.265629 \\
0.675659 0.094292 \\
0.689941 0.004564 \\
0.703125 0.000131 \\
};
\addlegendentry{$\scriptscriptstyle (N,w,m_0)= (16,3,16384)$};

\addplot [color=red,solid,mark=asterisk,mark options={solid}]
  table[row sep=crcr]{%
%
%
0.575043 0.522840 \\
0.588135 0.516685 \\
0.601227 0.496101 \\
0.614319 0.427166 \\
0.627411 0.168452 \\
0.639496 0.018235 \\
0.652588 0.000146 \\
};
\addlegendentry{$\scriptscriptstyle (N,w,m_0)= (64,3,65536)$};

\addplot [color=green,solid,mark=o,mark options={solid}]
  table[row sep=crcr]{%
%
%
0.400000 0.369840\\
0.450000 0.198101 \\
0.500000 0.040796 \\
0.550000 0.002046 \\
};
\addlegendentry{$\scriptscriptstyle (N,w,m_0)= (1,1,100)$};

\addplot [color=black,solid,forget plot]
  table[row sep=crcr]{%
0.878951	1e-06\\
0.878951	1e-05\\
0.878951	0.0001\\
0.878951	0.001\\
0.878951	0.01\\
0.878951	0.1\\
0.878951	1\\
};
\addplot [color=black,solid,forget plot]
  table[row sep=crcr]{%
0.636362	1e-06\\
0.636362	1e-05\\
0.636362	0.0001\\
0.636362	0.001\\
0.636362	0.01\\
0.636362	0.1\\
0.636362	1\\
};
\addplot [color=black,solid,forget plot]
  table[row sep=crcr]{%
0.583332	1e-06\\
0.583332	1e-05\\
0.583332	0.0001\\
0.583332	0.001\\
0.583332	0.01\\
0.583332	0.1\\
0.583332	1\\
};
\addplot [color=black,solid,forget plot]
  table[row sep=crcr]{%
0.574493	1e-06\\
0.574493	1e-05\\
0.574493	0.0001\\
0.574493	0.001\\
0.574493	0.01\\
0.574493	0.1\\
0.574493	1\\
};
\addplot [color=black,solid,forget plot]
  table[row sep=crcr]{%
0.431	1e-06\\
0.431	1e-05\\
0.431	0.0001\\
0.431	0.001\\
0.431	0.01\\
0.431	0.1\\
0.431	1\\
};
\end{axis}
\end{tikzpicture}%
\caption{SER performance of CBs and SC-CBs as a function of $\beta$ for $k=6$ and $\epsilon = 2^{-1.5}$.}
\label{fig:simulation}
\end{figure}
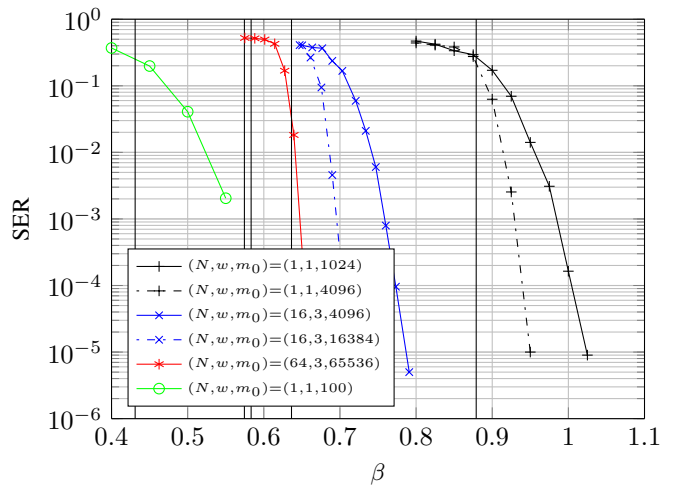

In Fig.~\ref{fig:simulation}, we give symbol error rate (SER) results for CBs decoded using the MP decoding algorithm and the Maxwell decoder, for $k=6$ and $\epsilon = 2^{-1.5}$, as a function of $\beta$.  In the figure we also plot the SER for several SC-CBs with parameters $(N,w,m_0) = (16,3,4096)$, $(16,3,16384)$, and $(64,3,65536)$, for the same $k$ and $\epsilon$. The Maxwell decoder has been implemented as described at the end of  \cref{sec:Maxwell}, and due to its high computational complexity (the problem is indeed NP-hard) only a short block length ($m_0=100$) has been considered (the green curve in Fig.~\ref{fig:simulation}). For MP decoding of CBs, we have considered the block lengths $m_0 = 1024$ and $4096$. The four black vertical lines show different thresholds as follows (in the order from left to right): The conjectured Maxwell decoding threshold (or the area threshold) and the coupled MP decoding thresholds $\beta^{\rm c}_{\rm MP}$ for $w=3$ and $N=128$, $64$, and $16$.

For both CBs and SC-CBs, we can clearly see how the performance improves as the block length increases (for a given coupling chain length and smoothing parameter) and approaches the corresponding MP decoding threshold. As expected, there is a larger gap to the MP decoding threshold for SC-CBs than for uncoupled CBs for a given fixed overall block length due to the fact that the block length at each position becomes shorter as the coupling chain length increases. SC-CBs achieve significantly better performance than uncoupled CBs. In particular, the threshold is improved from $\beta_{\rm MP}=0.88$ to $\beta^{\rm c}_{\rm MP}=0.57$ for coupling length $N=128$ and $w=3$. However, it is apparent from the figure that the MP decoding threshold for the coupled ensembles does not saturate to the area threshold (it remains a gap to the area threshold $\bar{\epsilon}=0.43$). The lack of threshold saturation is also supported by the simulation results of the Maxwell decoder (green curve). Despite of the short block length, the Maxwell decoder for the uncoupled CB performs significantly better than MP decoding of SC-CBs.


\subsection{Rate Gap to Entropy Lower Bound}
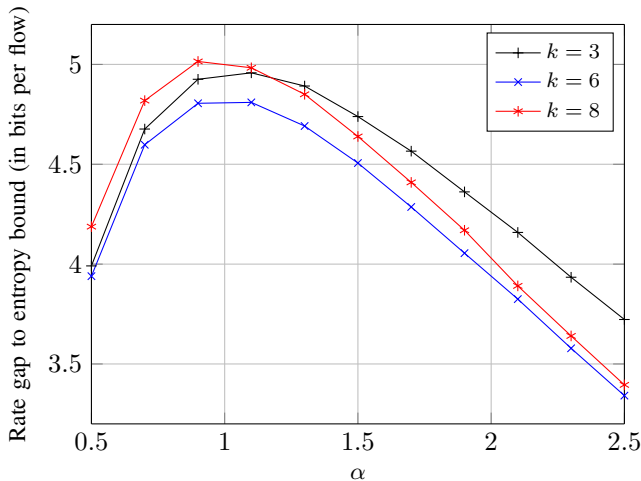
\begin{figure}[tbp]
\centering
%
%
\begin{tikzpicture}[font=\small]

\begin{axis}[%
width=0.8\columnwidth,
height=0.6\columnwidth,
at={(0.808889in,0.513333in)},
scale only axis,
separate axis lines,
every outer x axis line/.append style={black},
every x tick label/.append style={font=\color{black}},
xmin=0.5,
xmax=2.5,
xlabel={$\alpha$},
xmajorgrids,
every outer y axis line/.append style={black},
every y tick label/.append style={font=\color{black}},
ymin=3.2,
ymax=5.2,
ylabel={Rate gap to entropy bound (in bits per flow)},
ymajorgrids,
legend style={legend cell align=left,align=left,draw=black}
]
\addplot [color=black,solid,mark=+,mark options={solid}]
  table[row sep=crcr]{%
0.5	3.99102445359599\\
0.7	4.67634309902622\\
0.9	4.92551036798261\\
1.1	4.957862605214\\
1.3	4.89175545273087\\
1.5	4.73883572911071\\
1.7	4.56565259233277\\
1.9	4.36169161316312\\
2.1	4.15863193545282\\
2.3	3.93427079060252\\
2.5	3.72265457393932\\
};
\addlegendentry{\footnotesize $k=3$};

\addplot [color=blue,solid,mark=x,mark options={solid}]
  table[row sep=crcr]{%
0.5	3.93965697981357\\
0.7	4.59765067180456\\
0.9	4.80529482739464\\
1.1	4.80985753250398\\
1.3	4.69152406528204\\
1.5	4.506155674254\\
1.7	4.28603616518027\\
1.9	4.05486071189569\\
2.1	3.8248041504142\\
2.3	3.57844969825624\\
2.5	3.34148324990705\\
};
\addlegendentry{\footnotesize $k=6$};

\addplot [color=red,solid,mark=asterisk,mark options={solid}]
  table[row sep=crcr]{%
0.5	4.18763590566537\\
0.7	4.81875554298398\\
0.9	5.01480556198709\\
1.1	4.9830459335482\\
1.3	4.85037692549746\\
1.5	4.63913423223953\\
1.7	4.40925355888499\\
1.9	4.16952618949802\\
2.1	3.8909559991348\\
2.3	3.64040665993228\\
2.5	3.39529424150549\\
};
\addlegendentry{\footnotesize $k=8$};

\end{axis}
\end{tikzpicture}%
\caption{Rate gap of an optimized (over $\gamma$) single-layer SC-CB to the entropy lower bound of the flow size distribution (in bits per flow) as a function of $\alpha$  for different values of the flow node degree $k$. The number of spatial positions is fixed to $N=128$ and the smoothing parameter is fixed to $w=5$.} 
\label{fig:layer1_versus_alpha}
\end{figure}


In Fig.~\ref{fig:layer1_versus_alpha}, we depict the rate gap of an optimized (over $\gamma$) single-layer SC-CB to the entropy lower bound  of the flow size distribution (in bits per flow) as a function of $\alpha$ for different values of the flow node degree $k$.  The number of spatial positions is fixed to $N=128$ and the smoothing parameter is fixed to $w=5$, while $\alpha$ is sampled from $0.5$ to $2.5$ in steps of $0.2$. The depth $d$ of the counters has been chosen 
such that the average asymptotic  probability that a counter node overflows (computed from simulations for a large value of $m_0$) over the ensemble of possible CBs (for given values of $k$, $\gamma$, and $\alpha$)   is at most $10^{-4}$. Degree $k=6$ gives the lowest gap for all values of $\alpha$. 
Moreover, $k=8$ gives a larger gap than $k=3$ when $\alpha$ is small. The reason is that a larger depth $d$ is required  to maintain  the same average asymptotic overflow probability. The same observation was made in \cite{ros14} for $\alpha=1.5$. 

For $\alpha=1.5$, the optimized SC-CB with $(k,N,w)=(6,128,5)$ has a gap 
to the entropy lower bound of approximately $4.5$ bits per flow. For an uncoupled CB with $k=3$ (which gives the best performance, since the performance degrades with increasing $k$ as shown in \cite[Fig.~3]{ros14}) the corresponding rate gap (with the same overflow  probability) is approximately $5.3$ bits per flow,\footnote{The gap of approximately $5.3$ bits per flow to the entropy lower bound matches well with the results in \cite[Fig.~12]{lu08}. Note that in \cite[Fig.~12]{lu08} the entropy lower bound is shown to be slightly below $3$ bits per flow. However, this seems to be a misprint, as it is just below $2$ bits per flow for $\alpha = 1.5$. Moreover, the explicit overflow probability used when calculating the rate is not explicitly stated in  \cite{lu08}.} meaning that spatial coupling gives a gain of approximately $0.8$ bits per flow. 

The rather large gap to the entropy lower bound in Fig.~\ref{fig:layer1_versus_alpha} is due to having only a single-layer system. To increase performance, more layers are needed. As shown in Figs.~\ref{fig:relative_gap}, \ref{fig:simulation}, and \ref{fig:layer1_versus_alpha}, spatial coupling significantly improves performance of single-layer uncoupled CBs. Since the decoding of multilayer CBs proceeds layer-by-layer, we also expect multilayer SC-CBs to significantly improve the performance of their uncoupled counterparts: Spatial coupling will improve the decoding threshold for each layer and hence a better overall decoding threshold (see Section~\ref{sec:ExtMoreLayers}) is expected.

%
%

\section{Discussion on the Gap to the Area Threshold} 
\label{sec:discussion}

\begin{figure}[tbp]
\centering
%
%
\begin{tikzpicture}[font=\small]

\begin{axis}[%
width=0.8\columnwidth,
height=0.6\columnwidth,
at={(0.808889in,0.513333in)},
scale only axis,
separate axis lines,
every outer x axis line/.append style={black},
every x tick label/.append style={font=\color{black}},
xmin=3,
xmax=8,
xlabel={$k$},
xmajorgrids,
every outer y axis line/.append style={black},
every y tick label/.append style={font=\color{black}},
ymin=0,
ymax=0.7,
ylabel={Gap to area threshold},
ymajorgrids,
legend style={at={(0.03,0.97)},anchor=north west,legend cell align=left,align=left,draw=black}
]
\addplot [color=black,solid,mark=+,mark options={solid}]
  table[row sep=crcr]{%
3	0.00777900000000001\\
4	0.027816\\
5	0.051244\\
6	0.074326\\
7	0.095617\\
8	0.115683\\
};
\addlegendentry{$\scriptscriptstyle (N,w,\beta)=(128,5,1/4)$};

\addplot [color=black,dash pattern=on 1pt off 3pt on 3pt off 3pt,mark=+,mark options={solid}]
  table[row sep=crcr]{%
3	0.022303\\
4	0.057924\\
5	0.092384\\
6	0.123132\\
7	0.150203\\
8	0.174354\\
};
\addlegendentry{$\scriptscriptstyle (N,w,\beta)=(1,1,1/4)$};

\addplot [color=blue,solid,mark=x,mark options={solid}]
  table[row sep=crcr]{%
3	0.018589\\
4	0.062952\\
5	0.111274\\
6	0.157131\\
7	0.198948\\
8	0.237369\\
};
\addlegendentry{$\scriptscriptstyle (N,w,\beta)=(128,5,1/2)$};

\addplot [color=blue,dash pattern=on 1pt off 3pt on 3pt off 3pt,mark=x,mark options={solid}]
  table[row sep=crcr]{%
3	0.073448\\
4	0.164124\\
5	0.242695\\
6	0.309231\\
7	0.366257\\
8	0.416128\\
};
\addlegendentry{$\scriptscriptstyle (N,w,\beta)=(1,1,1/2)$};

\addplot [color=red,solid,mark=asterisk,mark options={solid}]
  table[row sep=crcr]{%
4	0.08171\\
6	0.200688\\
8	0.301042\\
};
\addlegendentry{$\scriptscriptstyle (N,w,\beta)=(128,5,2/3)$};

\addplot [color=red,dash pattern=on 1pt off 3pt on 3pt off 3pt,mark=asterisk,mark options={solid}]
  table[row sep=crcr]{%
4	0.258932\\
6	0.458916\\
8	0.601824\\
};
\addlegendentry{$\scriptscriptstyle (N,w,\beta)=(1,1,2/3)$};

\end{axis}
\end{tikzpicture}%
\caption{The difference $\bar{\epsilon}- {\epsilon}_{\rm MP}^{\rm c}$, where $\bar{\epsilon}$ is the area threshold from Definition~\ref{def:MAP} and ${\epsilon}_{\rm MP}^{\rm c}$ is the MP decoding threshold of \emph{regular} SC-CBs, as a function  of the left-degree $k$ for different values of $\beta$ with $(N,w)=(128,5)$ and $(1,1)$ (uncoupled).}
\label{fig:relative_gap_regular}
\end{figure}
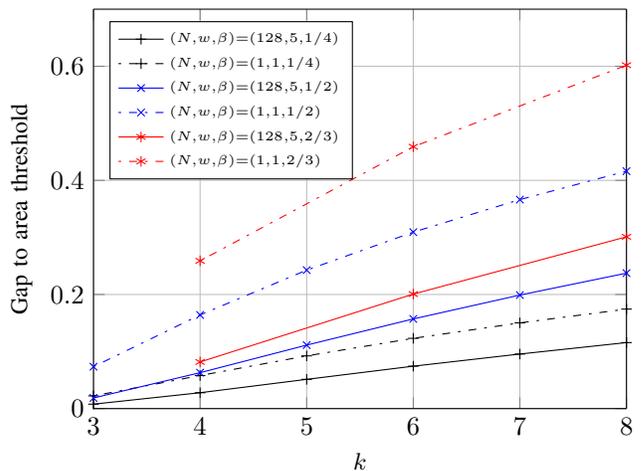

As shown in the previous section, we observe an improvement of the MP decoding threshold with spatial coupling, but no threshold saturation to the potential threshold by coupling the original bipartite graph (see Section~\ref{sec:SC-CBs}). The gap to the area threshold seems to be fundamental and is still present even for other more structured ways to do spatial coupling. For instance, the gap is still present with regular right degree distributions, i.e., for SC-CBs where the underlying uncoupled ensemble has a regular right degree distribution, as can be seen in Fig.~\ref{fig:relative_gap_regular}. Different curves correspond to different values of $\beta$. For instance, for the two blue curves  (the dash-dotted curve is for uncoupled CBs, while the solid curve is for SC-CBs), the left-degree $k$ increases from $3$ to $8$, while the right-degree is equal to $2k$, which results in $\beta=1/2$. Again, we observe a gap between the MP decoding threshold of $(k,\gamma,128,5)$ SC-CBs  and the area threshold. Note that the gap grows with $k$ and depends on $\beta$, 
as also shown in Fig.~\ref{fig:relative_gap}. 
Also, as shown in Fig.~\ref{fig:relative_gap}, the gap to area threshold is significantly larger for uncoupled CBs, meaning that spatial coupling indeed improves performance for regular CBs.

We remark that coupling the equivalent graph representation of Section~\ref{sec:EquivalentGraph} 
gives exactly the same coupled DE recursion $x^{(2\ell)}_{i} = \epsilon \cdot \xi_i(k,\gamma,N,w,\bm{x}^{(2\ell-2)})$, where $\xi_i(k,\gamma,N,w,\bm{x}^{(2\ell-2)})$ is given 
in (\ref{eq:sc_density_evo}). However, interestingly, if we consider the \textit{coupled version} of (\ref{eq:DECounter2})--(\ref{eq:g}), i.e., we substitute $y$ and $x$ in (\ref{eq:f}) and (\ref{eq:g}), respectively, by an average over spatial positions, we do indeed observe threshold saturation to the potential threshold. The DE recursion derived below corresponds to the coupled version of (\ref{eq:DECounter2})--(\ref{eq:g}), split into two steps.

Let $x_i^{(\ell)}$ and $y_i^{(\ell)}$, $i=1,\ldots,M$, denote the output message error probability at a flow and counter node, respectively, at the $\ell$-th iteration at coupling chain position $i$. As before, we initialize $x_i^{(0)}=0$ for $N < i \leq M$. 
Now, define the following DE  equations at iteration $\ell$,
\begin{align}
y_i^{(\ell)} &= \begin{cases}
1- \rho\left( 1- \frac{1}{w} \sum_{j=0}^{\min(i-1,w-1)} x_{i-j}^{(\ell-1)} \right), & \text{if $\ell$ is odd} \\[1.0ex]
1- \rho\left( 1- x_{i}^{(\ell-1)} \right), & \text{if $\ell$ is even} \end{cases},
\notag \\
x_i^{(\ell)} &= \begin{cases}
\left( y_i^{(\ell)} \right)^{k-1},  & \text{if $\ell$ is odd} \\[1.0ex]
\epsilon \left( \frac{1}{w} \sum_{j=0}^{\min(M-i,w-1)} y_{i+j}^{(\ell)} \right)^{k-1},  & \text{if $\ell$ is even} \end{cases}. \notag
\end{align}
With this modified DE, threshold saturation to the potential threshold, which (from Conjecture~\ref{th:MAP}) is a lower bound on the MAP decoding threshold, can be proved using the potential function framework by Yedla \emph{et al.} outlined in \cite{yed13}. This DE is characterized by an average only for odd and even iterations for the counter node and flow node updates, respectively. However, when coupling the original (or the equivalent) graph in the standard way, the average appears for all iterations (hence the four summations in (\ref{eq:sc_density_evo})). This effect, which is due to the fact that, as opposed to LDPC codes, the flow node update is different for odd and even iterations, seems to be the responsible for the lack of threshold saturation. 


A question that remains open is whether the DE equations above correspond to a physical system, i.e., whether threshold saturation can be achieved with an alternative coupling. 


The lack of threshold saturation has been observed in the past for other coupled systems. For example, SC-LDPC codes decoded using the corrected min-sum decoding algorithm show no threshold saturation, albeit some improvement to the threshold \cite{san14}. The lack of threshold saturation has also been observed from simulation of linear programming decoding of long SC codes using the alternating direction method of multipliers \cite{san14}.

 It is interesting to remark, however, that threshold saturation has been numerically observed in \cite{Jian12} for iterative HDD of SC generalized LDPC codes even with sub-optimal (bounded distance) component decoding, which shows that the lack of saturation cannot by  itself be explained by the hard-decision nature of the MP decoding algorithm. 
 In fact, a soft-decision SP decoding algorithm as outlined in (\ref{eq:counter_to_flow})--(\ref{eq:flow_to_counter}) could possibly perform better than the MP decoding algorithm studied  in this paper.  However, the complexity of such an algorithm is significantly higher.

\section{Connection With Compressed Sensing}
\label{sec:CS}

In this section, we explore the connection of CBs to CS as established in \cite{lu08_1}. 

We consider a single-layer SC-CB with $m_0$ flow nodes, $\mathsf{f}_1,\ldots,\mathsf{f}_{m_0}$, corresponding to the distinct flows whose sizes are to be measured, and $m_1$ counter nodes. We define the flow size vector $\bphi=(\phi_1,\dots,\phi_{m_0})$, of length $m_0$, where $\phi_i=\phi(\mathsf{f}_i)$ is the actual flow size of flow $i$. As in Section~\ref{sec:AsymptoticAnalysis}, the minimum flow size is $\fmin$, i.e., $\phi_i\ge \fmin~\forall i$. Now, if $f_{\rm min}=0$, the vector $\bphi$ has a number of zero entries and the rest are \emph{nonzero} positive integers. Thus, $\bphi$ can be interpreted as a \emph{nonnegative} (integer) signal vector. In this case, the $m_1$ counters can be also regarded as $m_1$ linear measurements of $\bphi$. If the number of nonzero entries in $\bphi$, denoted by $t$, is small (as compared to $m_0$), $\bphi$ is a $t$-sparse vector. Thus, the CB architecture can be regarded as a CS scheme with the important difference that for CBs, the signal is nonnegative. The MP decoding of the CB corresponds now to the recovery of the vector $\bphi$ based on the $m_1$ linear measurements.
We remark that, while for CBs $\bphi$ is a vector with nonnegative integer entries, the asymptotic analysis in Section~\ref{sec:AsymptoticAnalysis} depends only on the properties of the CB graph and $\epsilon$, and not on the exact value of $f_{\rm min}$ or the fact that the flow sizes are integers. Thus, our previous analysis is also valid for CS of nonnegative real signals.
\begin{figure}[!t]
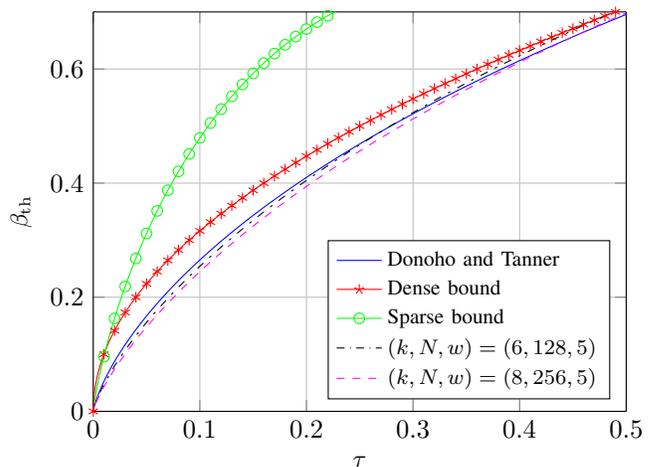

\centering
%
%
\definecolor{mycolor1}{rgb}{1.00000,0.00000,1.00000}%
%
\caption{Undersampling-sparsity phase transition trajectory (under MP decoding) of SC-CBs compared to two bounds from \cite{lu08_1} and the Donoho-Tanner phase transition trajectory for random Gaussian  measurement matrices with $\ell_1$-norm minimization reconstruction.}
\label{fig:UnderSamplSparsity}
\end{figure}

Let $\mathbb{R}_{\geq 0}$ denote the set of nonnegative real numbers. Given a vector $\bphi\in\mathbb{R}_{\geq 0}^{m_0}$, of size $m_0$, with $t$ nonzero entries, we define the \emph{sparsity ratio} $\tau=\frac{t}{m_0}$ as the probability that a given entry of the signal vector $\bphi$ is nonzero. Note that $\tau=\epsilon$, i.e., the probability of observing a flow size strictly larger than $f_{\rm min}=0$ (see Section~\ref{sec:AsymptoticAnalysis}). Furthermore, we define the \emph{undersampling ratio} (or factor) as the number of measurements over the signal dimension, $m_0$. This corresponds precisely to the parameter $\beta = m_1/m_0$ defined in Section~\ref{sec:AsymptoticAnalysis} for uncoupled CBs. For SC-CBs the undersampling ratio is equal to $\beta^{\rm c}$ (taking into account the rate increase due to edge effects (or termination) for SC-CBs) defined in (\ref{eq:beta_c}). We also define the undersampling threshold, $\bth$, as the lowest $\beta^{\rm c}$ (for a given sparsity ratio $\tau$) such that the error probability for recovering $t$-sparse vectors goes to zero as $m_0\rightarrow\infty$ and $t \rightarrow\infty$, while the ratio $\frac{t}{m_0}$ is fixed and equal to $\tau$.

In Fig.~\ref{fig:UnderSamplSparsity}, we plot $\bth$ as a function of $\tau$ for single-layer SC-CBs, with parameters $(k,N,w)=(6,128,5)$ and $(8,256,5)$. The curve $\bth(\tau)$ is commonly known as the phase transition curve in the CS literature. For comparison purposes, we also report in the figure the corresponding threshold $\bth$ for the two explicit constructions of CBs from \cite{lu08_1}, which we refer to as \emph{sparse bound} and \emph{dense bound} (see \cite{lu08_1}), respectively. The sparse bound is for a family of left-regular CBs with $\gamma= \ln(2) / \epsilon$, $k=\ln(1 / \epsilon) / \ln(2)$, and $\beta(\epsilon)= 1 / (\ln 2)^2 \cdot \epsilon \cdot \ln (1 / \epsilon)$ (see \cite[Th.~5]{lu08_1} for details). The dense bound is for a family of irregular CBs (found by making use of results from the optimization of LDPC code ensembles over the BEC) and has $\beta(\epsilon)=\sqrt{\epsilon}$ (see \cite[Th.~6]{lu08_1}). It is observed that SC-CBs yield a significant improvement of $\bth$.

In  Fig.~\ref{fig:UnderSamplSparsity}, we also compare the phase transition curve of SC-CBs to that of the ensemble of random Gaussian measurement matrices with $\ell_1$-norm minimization reconstruction by Donoho and Tanner \cite{don10}.  The curve is taken from \cite{don10} (see also \cite{urlTanner}, which contains the actual data in Matlab format). Surprisingly, SC-CBs give an improved phase transition trajectory in the sparse region. As compared to the $\ell_1$-norm minimization reconstruction algorithm in \cite{don10}, CS with CBs has the advantage of using sparse measurement matrices, i.e., measurements can be taken in $O(1)$ operations, and of a corresponding MP decoding algorithm of low complexity (linear in the signal dimension, $O(m_0)$), where $O(\cdot)$ is the standard $O$-notation.


Finally, we remark that iterative decoding for CS has also been considered in the literature in \cite{zha12,rav12,cha10,sar06,pha09,Don13,don09_1}. One such algorithm is the interval-passing algorithm by Chandar \emph{et al.} \cite{cha10} for binary measurements matrices, which was further generalized to nonbinary measurements matrices in \cite{rav12}. In fact, for CBs the interval-passing algorithm is equivalent to the MP decoding algorithm outlined in \cite{lu08} and presented here in \cref{sec:message_passing}.

\section{Conclusion}
\label{sec:Conclusion}

We proposed spatial coupling of CBs as a means of improving the MP decoding performance of (uncoupled) CBs. As expected, SC-CBs yield significantly better MP decoding thresholds than uncoupled CBs. However, surprisingly, our results suggest that the well-known phenomenon of threshold saturation does not occur for SC-CBs. We therefore analyzed the asymptotic performance of single-layer CBs. In particular, we introduced an equivalent graph representation of CBs with identical asymptotic and iteration-by-iteration finite-length performance to that of the original CB graph, which makes the analysis more amenable. Based on this equivalent representation, we showed that the area threshold and the potential threshold of CBs are equal. We also derived the Maxwell decoder for CBs and formulated the conjecture that the area threshold is equal to the Maxwell decoding threshold and therefore a lower bound on the MAP decoding threshold. There is a gap between the MP decoding threshold of SC-CBs and the conjectured MAP decoding threshold, showing that threshold saturation does not occur. Simulation results of the Maxwell decoder for finite length, which show better performance than MP decoding of SC-CBs for larger block lengths, support the conjecture. Finally, we also considered SC-CBs for CS and showed that they can yield very low undersampling thresholds.


\appendices


\section{Derivation of the Message Passing Decoding Algorithm in \cite{lu08}} \label{app:MPdecodingDerivation}

We derive the MP decoding algorithm proposed in \cite{lu08} starting from the flow node and counter node updates $\mu_{\ff \rightarrow \cc}^{{\sf MaxP},(\ell)}(a)$ and
 $\psi_{\cc \rightarrow \ff }^{{\sf MaxP},(\ell)}(a)$ of the max-product algorithm  in \cref{eq:flow_to_counter_max_prod} and  \cref{eq:counter_to_flow_max_prod}, respectively. We call the \emph{support} of a probability distribution $h(a)$ the set $\mathcal{A}$ of values for which $h(a)\neq 0$, i.e., $\mathcal{A}=\{a:h(a)\neq 0\}$. Note that for CBs $a$ is a nonnegative integer. Furthermore, it follows that the support $\Amu$ of $\mu_{\ff \rightarrow \cc}^{{\sf MaxP},(\ell)}(a)$ is the set of all integers in a given range $[L^{(\ell)}_{\ff \rightarrow \cc}, U^{(\ell)}_{\ff \rightarrow \cc}]$. Initially, $\Amu$ is the set of all integers in $[\fmin,\infty)$ (we know that the minimum flow size is $\fmin$ and there is no limit in the maximum flow size). Likewise, 
the support $\Apsi$
of $\psi_{\cc \rightarrow \ff }^{{\sf MaxP},(\ell)}(a)$ consists of all integers within a range $[L^{(\ell)}_{\cc \rightarrow \ff}, U^{(\ell)}_{\cc \rightarrow \ff}]$. Therefore, the supports $\Amu$ and $\Apsi$ are completely defined by the ranges $[L^{(\ell)}_{\ff \rightarrow \cc}, U^{(\ell)}_{\ff \rightarrow \cc}]$ and $[L^{(\ell)}_{\cc \rightarrow \ff}, U^{(\ell)}_{\cc \rightarrow \ff}]$, respectively, and as a consequence by the left and right limits of the ranges. Equivalently, we will write $L^{(\ell)}_{\ff \rightarrow \cc}\le a \le U^{(\ell)}_{\ff \rightarrow \cc}$ and $L^{(\ell)}_{\cc \rightarrow \ff} \le a \le U^{(\ell)}_{\cc \rightarrow \ff}$. We would like to remark that the supports $\Amu$ and $\Apsi$ are iteration-dependent. In particular, the limits of the corresponding ranges change with iterations. In the following, with some abuse of language, we will refer to the range on which a support is defined as the support itself.

Assuming that the decoder does not exploit any information about the flow size distribution, the message from flow node $\ff$ to counter node $\cc$ at iteration $\ell=0$ can be initialized as
\begin{equation*}
\mu_{\ff \rightarrow \cc}^{\sf{MaxP},(0)}( a) = \begin{cases}
\frac{1}{ \phi(\cc)- f_{\rm min} +1}, & \text{if $L^{(0)}_{\ff \rightarrow \cc} \leq  a \leq U^{(0)}_{\ff \rightarrow \cc}$} \\
0, & \text{otherwise}
\end{cases},
\end{equation*}
where $L^{(0)}_{\ff \rightarrow \cc} = f_{\rm min} $ and $U^{(0)}_{\ff \rightarrow \cc} =  \phi(\cc)$, i.e., the flow size distribution is assumed to be uniform over its support $[L^{(0)}_{\ff \rightarrow \cc},U^{(0)}_{\ff \rightarrow \cc}]$. For notational simplicity, we define $K_{\cc} \triangleq \frac{1}{ \phi(\cc)-  f_{\rm min} +1}$. 

Consider now the first iteration, $\ell=1$, of the max-product algorithm. The update rule  \cref{eq:counter_to_flow_max_prod} can now be rewritten as
\begin{equation*}
\psi_{\cc \rightarrow \ff  }^{\rm MaxP, (1)}(a) = \begin{cases}
K_{\cc}^{ |\Gamma(\cc)|-1}, & \text{if $L^{(1)}_{\cc \rightarrow \ff} \leq  a \leq U^{(1)}_{\cc \rightarrow \ff}$}  \\
0, & \text{otherwise} \end{cases},
\end{equation*}
where $L_{\cc \rightarrow \ff}^{(1)} = \max\{\phi(\cc)- \sum_{\ff' \in \Gamma(\cc) \setminus  \ff  }  U^{(0)}_{\ff' \rightarrow \cc},\fmin\}$, $U_{\cc \rightarrow \ff}^{(1)} = \max\{\phi(\cc)- \sum_{\ff' \in \Gamma(\cc) \setminus  \ff  }  L^{(0)}_{\ff' \rightarrow \cc},\fmin\}$, and the support of the distribution $\psi_{\cc \rightarrow \ff  }^{{\sf MaxP},(1)}(a)$ is $[L_{\cc \rightarrow \ff}^{(1)},U_{\cc \rightarrow \ff}^{(1)}]$. The values $L_{\cc \rightarrow \ff}^{(1)}$ and $U_{\cc \rightarrow \ff}^{(1)}$ come directly from the fact that the value of a counter node is equal to the sum of the values of its neighboring flow nodes. On the flow node side, the update rule in \cref{eq:flow_to_counter_max_prod} can be written as
\begin{equation*}
\mu_{\ff \rightarrow \cc}^{\rm MaxP, (1)}( a) = \begin{cases}
\left( K_{\cc}^{ |\Gamma(\cc)|-1} \right)^{ |\Gamma(\ff)| -1}, & \text{if $L_{\ff \rightarrow \cc}^{(1)} \leq  a \leq U_{\ff \rightarrow \cc}^{(1)}$ } \\
0, & \text{otherwise} \end{cases}
\end{equation*} 
where $L^{(1)}_{\ff \rightarrow \cc} =  \max_{\cc'  \in \Gamma(\ff) \setminus \cc } L_{\cc' \rightarrow \ff}^{(1)}$ and $U^{(1)}_{\ff \rightarrow \cc} =  \min_{\cc'  \in \Gamma(\ff) \setminus \cc } U_{\cc' \rightarrow \ff}^{(1)}$. The new support $[L^{(1)}_{\ff \rightarrow \cc},U_{\ff \rightarrow \cc}^{(1)}]$ comes directly from the fact that the update rule of \cref{eq:flow_to_counter_max_prod} is the product of the incoming messages. 

Consider now the second iteration from counter nodes to flow nodes. We get
\ifonecolumn
\begin{align*}
\psi_{\cc \rightarrow \ff  }^{\rm MaxP, (2)}(a) 
= \begin{cases}
\left( \left( K_{\cc}^{ |\Gamma(\cc)|-1} \right)^{ |\Gamma(\ff)| -1} \right)^{ |\Gamma(\cc)|-1}, & \text{if $L^{(2)}_{\cc \rightarrow \ff} \leq  a \leq U^{(2)}_{\cc \rightarrow \ff}$}  \\
0, & \text{otherwise} \end{cases},
\end{align*}
\else
\begin{align*}
&\psi_{\cc \rightarrow \ff  }^{\rm MaxP, (2)}(a) \nonumber\\
&= \begin{cases}
\left( \left( K_{\cc}^{ |\Gamma(\cc)|-1} \right)^{ |\Gamma(\ff)| -1} \right)^{ |\Gamma(\cc)|-1}, & \text{if $L^{(2)}_{\cc \rightarrow \ff} \leq  a \leq U^{(2)}_{\cc \rightarrow \ff}$}  \\
0, & \text{otherwise} \end{cases},
\end{align*}
\fi
where $L_{\cc \rightarrow \ff}^{(2)} = \max\{\phi(\cc)- \sum_{\ff' \in \Gamma(\cc) \setminus \ff }  U^{(1)}_{\ff' \rightarrow \cc},\fmin\}$ and $U_{\cc \rightarrow \ff}^{(2)} = \max\{\phi(\cc)- \sum_{\ff' \in \Gamma(\cc) \setminus  \ff  }  L^{(1)}_{\ff' \rightarrow \cc},\fmin\}$. 
Similarly, one can derive the messages  $\psi_{\cc \rightarrow \ff  }^{{\sf MaxP},(\ell)}(a)$ and $\mu_{\ff \rightarrow \cc}^{{\sf MaxP},(\ell)}(a)$  for a generic iteration $\ell$,
\begin{align}\label{eq:psiitl}
\psi_{\cc \rightarrow \ff  }^{\sf{MaxP},(\ell)}(a)
&= \begin{cases}
A^{(\ell)}, & \text{if $L^{(\ell)}_{\cc \rightarrow \ff} \leq  a \leq U^{(\ell)}_{\cc \rightarrow \ff}$}  \\ 
0, & \text{otherwise} \end{cases},\\
\mu_{\ff \rightarrow \cc}^{\sf{MaxP},(\ell)}( a) &= \begin{cases}
B^{(\ell)}, & \text{if $L_{\ff \rightarrow \cc}^{(\ell)} \leq  a \leq U_{\ff \rightarrow \cc}^{(\ell)}$ } \\
0, & \text{otherwise} \end{cases},
\label{eq:muitl}
\end{align} 
where $A^{(\ell)}$ and $B^{(\ell)}$ are constants and
\begin{align}
\label{eq:L1}
L_{\cc \rightarrow \ff}^{(\ell)} &= \max\left\{\phi(\cc)- \sum_{\ff' \in \Gamma(\cc) \setminus \ff }  U^{(\ell-1)}_{\ff' \rightarrow \cc},\fmin\right\},\\
\label{eq:L2}
U_{\cc \rightarrow \ff}^{(\ell)} &= \max\left\{\phi(\cc)- \sum_{\ff' \in \Gamma(\cc) \setminus  \ff  }  L^{(\ell-1)}_{\ff' \rightarrow \cc},\fmin\right\},\\
L^{(\ell)}_{\ff \rightarrow \cc} &=  \max_{\cc'  \in \Gamma(\ff) \setminus \cc } L_{\cc' \rightarrow \ff}^{(\ell)},\\
U^{(\ell)}_{\ff \rightarrow \cc} &=  \min_{\cc'  \in \Gamma(\ff) \setminus \cc } U_{\cc' \rightarrow \ff}^{(\ell)}.
\label{eq:L4}
\end{align}
 
It is important to observe that the messages exchanged between flow nodes and counter nodes in each iteration (see \cref{eq:psiitl}--\cref{eq:muitl}) are constant within a given support and $0$ otherwise. Therefore, it is enough to exchange the limits of the supports $L^{(\ell)}_{\ff \rightarrow \cc} , U^{(\ell)}_{\ff \rightarrow \cc}$ and $L_{\cc \rightarrow \ff}^{(\ell)}, U_{\cc \rightarrow \ff}^{(\ell)}$ at each iteration, and we can simplify the update rules of the max-product algorithm \cref{eq:psiitl}--\cref{eq:muitl} as
\begin{align}\label{eq:psiv}
\psi_{\cc \rightarrow \ff}^{\sf{MaxP},(\ell)} &=(L^{(\ell)}_{\cc \rightarrow \ff},U^{(\ell)}_{\cc \rightarrow \ff}),\\
\mu_{\ff \rightarrow \cc}^{\sf{MaxP},(\ell)} &=(L^{(\ell)}_{\ff \rightarrow \cc},U^{(\ell)}_{\ff \rightarrow \cc}),
\label{eq:muv}
\end{align}
i.e., each message is a two-dimensional vector. Note that these vectors convey all relevant information. Furthermore, it can be easily shown by induction that, for $\ell\ge1$,
\begin{align*}
L^{(2\ell-1)}_{\ff \rightarrow \cc} &= L^{(2\ell-2)}_{\ff \rightarrow \cc}, &U^{(2\ell)}_{\ff \rightarrow \cc} &= U^{(2\ell-1)}_{\ff \rightarrow \cc},\\
L_{\cc \rightarrow \ff}^{(2\ell-1)} &= L_{\cc \rightarrow \ff}^{(2\ell-2)}, &U_{\cc \rightarrow \ff}^{(2\ell)} &= U_{\cc \rightarrow \ff}^{(2\ell-1)},
\end{align*}
where $L_{\cc \rightarrow \ff}^{(0)} \triangleq \fmin$.
Thus, for odd iterations it is only necessary to exchange the upper limits $U^{(2\ell-1)}_{\ff \rightarrow \cc}$ and $U^{(2\ell-1)}_{\cc \rightarrow \ff}$. Similarly, for even iterations it is only necessary to exchange the lower limits $L^{(2\ell)}_{\ff \rightarrow \cc}$ and $L^{(2\ell)}_{\cc \rightarrow \ff}$. Therefore, we can rewrite \cref{eq:psiv} and \cref{eq:muv} as
\begin{align}\label{eq:UpdateRulesa}
\psi_{\cc \rightarrow \ff  }^{\sf{MaxP},(\ell)}
&= \begin{cases}
U_{\cc \rightarrow \ff}^{(\ell)}, & \text{if $\ell$ is odd}  \\ %
L_{\cc \rightarrow \ff}^{(\ell)}, & \text{if $\ell$ is even} \end{cases},\\
\mu_{\ff \rightarrow \cc}^{\sf{MaxP},(\ell)} &= \begin{cases}
U_{\ff \rightarrow \cc}^{(\ell)}, & \text{if $\ell$ is odd} \\
L_{\ff \rightarrow \cc}^{(\ell)}, & \text{if $\ell$ is even} \end{cases}.
\label{eq:UpdateRulesb}
\end{align} 
Finally, using \cref{eq:L1}--\cref{eq:L4} and observing that the structure of the update rules in \cref{eq:L1} and \cref{eq:L2} is the same, \cref{eq:UpdateRulesa} and \cref{eq:UpdateRulesb} can be rewritten as
\begin{align*}
\psi^{\sf{MaxP},(\ell)}_{\mathsf{c} \rightarrow \mathsf{f}} &= \max \left\{ \phi(\mathsf{c}) - \sum_{ \mathsf{f}' \in \Gamma(\mathsf{c}) \setminus \ff} \mu^{\sf{MaxP},(\ell-1)}_{\mathsf{f}' \rightarrow \cc},\;f_{\rm min} \right\},\\ 
\mu^{\sf{MaxP},(\ell)}_{\mathsf{f} \rightarrow \mathsf{c}} &= \begin{cases}
\min_{\mathsf{c}' \in \Gamma(\mathsf{f}) \setminus \mathsf{c}} \psi^{\sf{MaxP},(\ell)}_{\mathsf{c}' \rightarrow \ff}, & \text{if $\ell$ is odd} \\
\max_{\mathsf{c}' \in \Gamma(\mathsf{f}) \setminus \mathsf{c}} \psi^{\sf{MaxP},(\ell)}_{\mathsf{c}' \rightarrow \ff}, & \text{if $\ell$ is even} \end{cases}, 
\end{align*}
which is the MP decoding algorithm proposed in \cite{lu08}.

Notice  that if the summation within the counter nodes had been over some \emph{group}, i.e., the flow node values were elements of some group, then the support of the flow size distribution would have stayed the same throughout the iterations, and thus the MP algorithm would not work at all. Also, in this case, both the SP and the max-product algorithms would have been identical. The crucial point for reducing the support of the flow size distribution throughout the iterations of the algorithm is the nonexistence of additive inverses. Thus, in a pure lossy source coding setup, more sophisticated algorithms, for example the one in \cite{are15} (in this context for encoding), are required.

\section{Asymptotic Analysis of the Maxwell Decoder} \label{app:maxwell_decoder}

In this appendix, we perform an asymptotic analysis (using DE) of the Maxwell decoder introduced in \cref{sec:Maxwell} which ultimately results in the proof of \cref{th:MAPa}. 
The derivations parallel and extend those performed in \cite{mea08} for the Maxwell decoder of LDPC codes on the BEC, highlighting the main differences in the derivations. We assume in the following that the reader is already to some extent familiar with the proof given in \cite{mea08}.


\subsection{Density Evolution Analysis}

Let $x^{(\ell)}_{0}$, $x^{(\ell)}_{\ast}$, and $x^{(\ell)}_{\rm g}$ denote the probability that the second component of a message from a flow node to a counter node at iteration $\ell\geq 0$ is $0$, $\ast$, or $\rm g$, respectively, according to the update rules of the Maxwell decoder outlined in (\ref{eq:flow-to-counter}). Likewise, let $y^{(\ell)}_{0}$, $y^{(\ell)}_{\ast}$, and $y^{(\ell)}_{\rm g}$, $\ell\geq 1$,  denote the corresponding probabilities for the second component of a message from a counter node to a flow node, according to the update rules of the Maxwell decoder outlined in (\ref{eq:counter-to-flow}). The counter node DE updates are
\begin{equation} \label{eq:counter_update_maxwell}
\begin{split}
y_0^{(\ell)} &= \rho\left(x_0^{(\ell-1)}\right), \\
y_*^{(\ell)} &= 1-\rho\left(x_0^{(\ell-1)}+x_{\rm g}^{(\ell-1)}\right) = 1-\rho\left(1-x_*^{(\ell-1)}\right), \\
y_{\rm g}^{(\ell)} &= 1- y_0^{(\ell)}-y_*^{(\ell)}.
\end{split}
\end{equation}
On the other hand, the flow node DE updates for odd iterations are 
\begin{equation} \notag
\begin{split}
x_0^{(2\ell-1)} &= 1-\lambda\left(y_{\rm g}^{(2\ell-1)}+y_*^{(2\ell-1)}\right) = 1-\lambda\left(1-y_0^{(2\ell-1)}\right), \\
x_*^{(2\ell-1)} &= \lambda\left(y_*^{(2\ell-1)}\right),  \\
x_{\rm g}^{(2\ell-1)} &= 1-x_0^{(2\ell-1)}-x_*^{(2\ell-1)},
\end{split}
\end{equation}
and for even iterations
\begin{align}
x_0^{(2\ell)} &= 1-\epsilon\lambda\left(y_{\rm g}^{(2\ell)}+y_*^{(2\ell)}\right) = 1-\epsilon\lambda\left(1-y_0^{(2\ell)}\right),\notag \\
x_*^{(2\ell)} &= (1-\delta)\epsilon\lambda\left(y_*^{(2\ell)}\right), \notag \\
x_{\rm g}^{(2\ell)} &= 1-x_0^{(2\ell)}-x_*^{(2\ell)}, \notag
\end{align}
where $\delta$ represents the fraction of guesses performed so far. Combining odd and even iterations, we get
\ifonecolumn
\begin{align} 
x_0^{(2\ell)} &=  1-\epsilon\lambda\left(1-\rho\left(1-\lambda\left(1-\rho\left(x_0^{(2\ell-2)}\right)\right)\right)\right) 
= 1-\epsilon\lambda\left(g( 1- x_0^{(2\ell-2)})\right), \label{eq:DE_Maxwell_a}\\
x_*^{(2\ell)} &= (1-\delta)\epsilon\lambda\left(1-\rho\left(1-\lambda\left(1-\rho\left(1-x_*^{(2\ell-2)}\right)\right)\right)\right) 
= (1-\delta) \epsilon\lambda\left(g\left( x_*^{(2\ell-2)}\right)\right), \label{eq:DE_Maxwell_b} \\
x_{\rm g}^{(2\ell)} &= 1-x_0^{(2\ell)}-x_*^{(2\ell)}.  \label{eq:DE_Maxwell_c}
\end{align}
\else
\begin{align} 
x_0^{(2\ell)} &=  1-\epsilon\lambda\left(1-\rho\left(1-\lambda\left(1-\rho\left(x_0^{(2\ell-2)}\right)\right)\right)\right) \notag \\
&= 1-\epsilon\lambda\left(g( 1- x_0^{(2\ell-2)})\right), \label{eq:DE_Maxwell_a}\\
x_*^{(2\ell)} &= (1-\delta)\epsilon\lambda\left(1-\rho\left(1-\lambda\left(1-\rho\left(1-x_*^{(2\ell-2)}\right)\right)\right)\right) \notag \\
&= (1-\delta) \epsilon\lambda\left(g\left( x_*^{(2\ell-2)}\right)\right), \label{eq:DE_Maxwell_b} \\
x_{\rm g}^{(2\ell)} &= 1-x_0^{(2\ell)}-x_*^{(2\ell)}.  \label{eq:DE_Maxwell_c}
\end{align}
\fi
To settle the notation, when $\ell$ tends to infinity, we define
\begin{align*}
\x & \triangleq\left( x_0,x_*,x_{\rm g} \right)  \triangleq \left( x_0^{(\infty)},x_*^{(\infty)},x_{\rm g}^{(\infty)} \right),  \\
\y & \triangleq \left( y_0,y_*,y_{\rm g} \right) \triangleq \left( y_0^{(\infty)},y_*^{(\infty)},y_{\rm g}^{(\infty)} \right).
\end{align*}
Now, since  $\left( y_0,y_*,y_{\rm g} \right)$ is a function of $\left( x_0,x_*,x_{\rm g} \right)$ (see (\ref{eq:counter_update_maxwell})), any function of $(\x,\y)$ is indeed just a function of $\x$.

Note that the DE recursion in (\ref{eq:DE_Maxwell_a})--(\ref{eq:DE_Maxwell_c}) is for the Maxwell decoder on the equivalent graph introduced \cref{sec:EquivalentGraph}. However, according to \cref{prop:equivalent_maxwell}, the original and equivalent graphs are iteration-by-iteration equivalent for the Maxwell decoder. Thus, we can equivalently base our analysis on the  DE recursion in (\ref{eq:DE_Maxwell_a})--(\ref{eq:DE_Maxwell_c}).

\subsection{Analysis}	

For the Maxwell decoder we choose the following guessing strategy. Set $\delta=0$ and run the ordinary MP decoding algorithm for CBs (with the update rules in (\ref{eq:Bp1}) and (\ref{eq:Bp2})), until there is no further progress from one iteration round to the next. Then, for each unknown flow node, assign ${\rm g}$-messages as outgoing messages to all its adjacent edges with probability $\Delta \delta / (1-\delta)$, after which $\delta$ is incremented by $\Delta \delta$. Now, run the Maxwell decoder (with the update rules in  (\ref{eq:counter-to-flow})  and (\ref{eq:flow-to-counter})),  until it stops. This whole procedure is repeated until $\delta=1$, or until there are no more unknown flow nodes.

Let $\mathbb{G}$ be the total number of performed guesses  and $m_0  \Delta \mathbb{G}$ the number of newly guessed flow nodes when $\delta$ is incremented by $\Delta \delta$. In more detail, for each $i \in \{1,\dots,m_0\}$, $i$ is chosen independently with probability $\Delta \delta / (1-\delta)$. Then, if the $i$-th selected flow node is unknown, then assign ${\rm g}$-messages as outgoing messages to all its edges and increase the number of newly guessed flow nodes by one. It can be shown that \cite{mea08}
\begin{equation} \label{eq:deltaG}
\mathbb{E}[\Delta \mathbb{G}] = \epsilon L(y_*) \Delta \delta.
\end{equation}
As explained above, after $\delta$ has been incremented by $\Delta \delta$ and ${\rm g}$-messages have been assigned as outgoing messages to all edges connected to the $m_0 \Delta \mathbb{G}$ newly guessed flow nodes, the Maxwell decoder is run again until no further progress can be made.

For each iteration of the Maxwell decoder there could be flow nodes for which more than one of the incoming messages is a ${\rm g}$-message. As explained in \cref{sec:Maxwell}, a ${\rm g}$-message is just a nonempty list of flow nodes, a corresponding list of coeffcients, and an integer $\KK$,  giving an explicit resolution rule for the value of the flow node (the value of the flow node is equal to $\KK$ minus the sum of the values of the flow nodes in the list weighted by their respective coefficients). The fact that there are several incoming ${\rm g}$-messages imposes some linear conditions and the number of independent conditions (over all flow nodes) can be lower-bounded using \cite[Lem.~11]{mea08}. As an example, if there are two incoming ${\rm g}$-messages to a flow node $\mathsf{f}_5$ and they are represented by the lists $\{\mathsf{f}_1,\mathsf{f}_3,1,1,3 \}$ and $\{\mathsf{f}_1,\mathsf{f}_2,\mathsf{f}_4,1,1,1, 2 \}$, respectively, then we get the linear condition $3 - \phi(\mathsf{f}_1) - \phi(\mathsf{f}_3) = 2 - \phi(\mathsf{f}_1) - \phi(\mathsf{f}_2) -\phi(\mathsf{f}_4)$, which simplifies to $\phi(\mathsf{f}_3) = 1 +\phi(\mathsf{f}_2) +\phi(\mathsf{f}_4)$, and $\phi(\mathsf{f}_5) = 3-\phi(\mathsf{f}_1) - \phi(\mathsf{f}_3)$. Let ${\rm l}^{\rm g}_i$ be the number of incoming ${\rm g}$-messages at a flow node of index $i$ (i.e., $\mathsf{f}_i$), including the flow node itself if it has been guessed. Now, it is not so hard to see that the number of imposed conditions (over all flow nodes) is $\sum_{i =1}^{m_0} \left( {\rm l}^{\rm g}_i -1 \right)$. On the other hand, some of these conditions may also be dependent. Suppose that there are three incoming ${\rm g}$-messages to a counter node (from the flow nodes $\mathsf{f}_1$, $\mathsf{f}_2$, and $\mathsf{f}_3$, respectively) and that they are represented by the lists $\{\mathsf{f}_1,-1,0\}$, $\{\mathsf{f}_2,-1,0\}$, and  $\{\mathsf{f}_3,-1,0\}$, respectively, i.e., the $i$-th ${\rm g}$-message indicates that $\phi(\ff_i) = 0 - (-1 \cdot \phi(\ff_i)) = \phi(\ff_i)$, $i=1,2,3$.  Then, the ${\rm g}$-message from counter node $\cc$ to flow node $\mathsf{f}_1$ is $\{\mathsf{f}_2,\mathsf{f}_3,1,1,\phi(\cc)\}$, to flow node $\mathsf{f}_2$ is $\{\mathsf{f}_1,\mathsf{f}_3,1,1,\phi(\cc)\}$, and to flow node $\mathsf{f}_3$ is $\{\mathsf{f}_1,\mathsf{f}_2,1,1,\phi(\cc)\}$, from which we get the condition $\phi(\cc)=\phi(\mathsf{f}_1) +\phi(\mathsf{f}_2)+\phi(\mathsf{f}_3)$ at all three flow nodes. In particular, using the operational update rule described in the paragraph following (\ref{eq:counter-to-flow}), the list of flow nodes of the ${\rm g}$-message from $\cc$ to $\mathsf{f}_1$ is the \emph{union} of $\{\ff_2 \}$ and $\{\ff_3 \}$, the corresponding list of coefficients is the sum of the corresponding coefficients of the incoming lists multiplied by $-1$, i.e., $\{ -(-1), - (-1) \} = \{1,1\}$, while the integer $\KK$ of the outgoing list is $\phi(\cc)$ minus the sum of all $\KK$'s of the incoming lists, i.e., $\KK = \phi(\cc) - (0 + 0) = \phi(\cc)$, which results in the ${\rm g}$-message $\{\mathsf{f}_2,\mathsf{f}_3,1,1,\phi(\cc)\}$ from $\cc$ to $\mathsf{f}_1$. The other two ${\rm g}$-messages from  $\cc$ to $\mathsf{f}_2$ and from $\cc$ to $\mathsf{f}_3$, respectively, follow by symmetry.  Thus, the number of independent imposed linear conditions (over all flow nodes) is in general lower than the number of conditions from the formula $\sum_{i =1}^{m_0} \left( {\rm l}^{\rm g}_i -1 \right)$. The correction term can be upper-bounded (see \cite[Lem.~11]{mea08}) by $\sum_{\mathsf{c} \in \mathcal{C}_g} (\Gamma(\mathsf{c}) -1)$, where $\mathcal{C}_g$ is the subset of all counter nodes all of whose incoming messages are ${\rm g}$-messages. Note that this is a worst-case situation in which all incoming messages to a counter node are ${\rm g}$-messages (see the example above).

Now, we can compute the expected value of the lower bound from \cite[Lem.~11]{mea08} on the number of independent imposed conditions after any stage of the Maxwell decoder (right-hand side of \cite[Lemma~11, Eq.~(21)]{mea08}, excluding the term $\mathbb{G}$). We perform the same steps as in the analysis in \cite[Sec.~VI-G]{mea08}. 
%
%
%
%
In fact, the analysis of the Maxwell decoder carried out in \cite[Sec.~VI-G]{mea08}  is valid also in the context of CBs, except that  \cite[Eqs.\ (22) and (23)]{mea08} need to be properly modified following the DE equations in (\ref{eq:DE_Maxwell_a}) to (\ref{eq:DE_Maxwell_c}). In particular, the right-hand side of \cite[Eq.~(22)]{mea08} becomes
\ifonecolumn
\begin{equation} \label{eq:22}
\begin{split}
&\epsilon \cdot (1-\delta) \left\{ L'(y_* + y_{\rm g}) y_g  - L(y_*+y_{\rm g}) + L(y_*) \right\}
+ \epsilon \delta L'(y_* + y_{\rm g}) y_{\rm g} \\
&= k \left[ x_{*} \left(1-y_{*} \right) - \left( 1-x_{0} \right)y_0  \right] 
-\epsilon \cdot (1-\delta) \left[ L(1-y_0) - L(y_*) \right] 
+k (1-y_*) x_{\rm g},
\end{split}
\end{equation}
\else
\begin{equation} \label{eq:22}
\begin{split}
&\epsilon \cdot (1-\delta) \left\{ L'(y_* + y_{\rm g}) y_g  - L(y_*+y_{\rm g}) + L(y_*) \right\} \\
&+ \epsilon \delta L'(y_* + y_{\rm g}) y_{\rm g} \\
&= k \left[ x_{*} \left(1-y_{*} \right) - \left( 1-x_{0} \right)y_0  \right] \\
&\;\;\;\; -\epsilon \cdot (1-\delta) \left[ L(1-y_0) - L(y_*) \right] \\
&\;\;\;\; +k (1-y_*) x_{\rm g},
\end{split}
\end{equation}
\fi
and the right-hand side of \cite[Eq.~(23)]{mea08} becomes
\ifonecolumn
\begin{equation} \label{eq:23}
\frac{L'(1)}{\bar{R}'(1)} \left\{ \bar{R}'(1-x_*)x_{\rm g} - \bar{R}(1-x_*) + \bar{R}(1-x_* - x_{\rm g}) \right\} 
= \frac{k}{\bar{R}'(1)} \left[ -\bar{R}(1-x_*) + \bar{R}(x_0) \right] +k (1-y_*) x_{\rm g},
\end{equation}
\else
\begin{equation} \label{eq:23}
\begin{split}
&\frac{L'(1)}{\bar{R}'(1)} \left\{ \bar{R}'(1-x_*)x_{\rm g} - \bar{R}(1-x_*) + \bar{R}(1-x_* - x_{\rm g}) \right\} \\
&= \frac{k}{\bar{R}'(1)} \left[ -\bar{R}(1-x_*) + \bar{R}(x_0) \right] +k (1-y_*) x_{\rm g},
\end{split}
\end{equation}
\fi
where 
\begin{equation} \label{eq:Rbar}
\bar{R}(x) = \frac{\int_0^x \left(1-g(1-z) \right)\, \diff z}{ \int_0^1 \left(1-g(1-z) \right)\, \diff z}.
\end{equation}
As in \cite{mea08}, let the function $F(\x,\epsilon,\delta)$ denote the difference of (\ref{eq:22}) and (\ref{eq:23}). We get
\ifonecolumn
\begin{displaymath}
\begin{split}
F(\x,\epsilon,\delta) &\triangleq k \left[ x_{*} \left(1-y_{*} \right) - \left( 1-x_{0} \right)y_0  \right] 
-\epsilon \cdot (1-\delta) \left[ L(1-y_0) - L(y_*) \right] 
+ \frac{k}{\bar{R}'(1)} \left[ \bar{R}(1-x_*) - \bar{R}(x_0) \right] \\
&=  k  x_{*} \left(1-y_{*} \right) +\epsilon \cdot (1-\delta) L(y_*) + k \frac{\bar{R}(1-x_*)}{\bar{R}'(1)} 
+ \bar{F}(x_0,\epsilon,\delta),
\end{split}
\end{displaymath}
\else
\begin{displaymath}
\begin{split}
F(\x,\epsilon,\delta) &\triangleq k \left[ x_{*} \left(1-y_{*} \right) - \left( 1-x_{0} \right)y_0  \right] \\
&\;\;\;\; -\epsilon \cdot (1-\delta) \left[ L(1-y_0) - L(y_*) \right] \\
&\;\;\;\; + \frac{k}{\bar{R}'(1)} \left[ \bar{R}(1-x_*) - \bar{R}(x_0) \right] \\
&=  k  x_{*} \left(1-y_{*} \right) +\epsilon \cdot (1-\delta) L(y_*) + k \frac{\bar{R}(1-x_*)}{\bar{R}'(1)} \\
&\;\;\;\; + \bar{F}(x_0,\epsilon,\delta),
\end{split}
\end{displaymath}
\fi
where 
\begin{displaymath}
\bar{F}(x_0,\epsilon,\delta) \triangleq  -k(1-x_0) y_0 - \epsilon \cdot (1-\delta) L(1-y_0) - k \frac{\bar{R}(x_0)}{\bar{R}'(1)}
\end{displaymath}
depends only on $x_0$. The function $F(\x,\epsilon,\delta)$ is the expected value of the number of independent imposed conditions after any stage of the Maxwell decoder. Since $y_* = g( x_*)$ and $\epsilon \cdot (1-\delta) L(g(x_*)) = x_* g(x_*)$ (from (\ref{eq:DE_Maxwell_b})), we have
\ifonecolumn
\begin{displaymath}
\begin{split}
F(\x,\epsilon,\delta) &=  k  x_{*} \left(1-g(x_{*}) \right) +\epsilon \cdot (1-\delta) L(g(x_*))  
+ k \frac{\bar{R}(1-x_*)}{\bar{R}'(1)} + \bar{F}(x_0,\epsilon,\delta) \\
&=  k  x_{*} \left(1-g(x_{*}) \right) +x_* g(x_*) + k \frac{\bar{R}(1-x_*)}{\bar{R}'(1)} 
+\bar{F}(x_0,\epsilon,\delta) \\
&=  k  x_{*} \left(1-g(x_{*}) \right) +x_* g(x_*) 
+ k \int_0^{1-x_*} \left(1 - g\left(1-z\right) \right)\, \diff z +\bar{F}(x_0,\epsilon,\delta) \\
&=  k  x_{*} \left(1-g(x_{*}) \right) +x_* g(x_*) 
+ k \left[ \int_{0}^1 \left(1 - g\left(z\right) \right)\, \diff z - \int_{0}^{x_*} \left(1 - g\left(z\right) \right)\, \diff z \right] 
+\bar{F}(x_0,\epsilon,\delta) \\
&\overset{(a)}{=} P(x_*) + k \int_{0}^1 \left(1 - g\left(z\right) \right)\, \diff z +\bar{F}(x_0,\epsilon,\delta),
\end{split}
\end{displaymath}
\else
\begin{displaymath}
\begin{split}
F(\x,\epsilon,\delta) &=  k  x_{*} \left(1-g(x_{*}) \right) +\epsilon \cdot (1-\delta) L(g(x_*))  \\
&\;\;\;\; + k \frac{\bar{R}(1-x_*)}{\bar{R}'(1)} + \bar{F}(x_0,\epsilon,\delta) \\
&=  k  x_{*} \left(1-g(x_{*}) \right) +x_* g(x_*) + k \frac{\bar{R}(1-x_*)}{\bar{R}'(1)} \\
&\;\;\;\; +\bar{F}(x_0,\epsilon,\delta) \\
&=  k  x_{*} \left(1-g(x_{*}) \right) +x_* g(x_*) \\
&\;\;\;\;+ k \int_0^{1-x_*} \left(1 - g\left(1-z\right) \right)\, \diff z +\bar{F}(x_0,\epsilon,\delta) \\
&=  k  x_{*} \left(1-g(x_{*}) \right) +x_* g(x_*) \\
&\;\;\;\;+ k \left[ \int_{0}^1 \left(1 - g\left(z\right) \right)\, \diff z - \int_{0}^{x_*} \left(1 - g\left(z\right) \right)\, \diff z \right] \\
&\;\;\;\; +\bar{F}(x_0,\epsilon,\delta) \\
&\overset{(a)}{=} P(x_*) + k \int_{0}^1 \left(1 - g\left(z\right) \right)\, \diff z +\bar{F}(x_0,\epsilon,\delta),
\end{split}
\end{displaymath}
\fi
where $(a)$ follows from the definition of the trial entropy in Definition~\ref{def:trial}. Thus,
\begin{displaymath}
F(\x^1,\epsilon,\delta) - F(\x^2,\epsilon,\delta) = P(x_*^1) - P(x_*^2)
\end{displaymath}
when $\x^1=(x_0^1,x_*^1,x_{\rm g}^1)$ and $\x^2=(x_0^2,x_*^2,x_{\rm g}^2)$ have the same $x_0$-component, i.e., $x_0^1 = x_0^2$. This will become important in the next subsection in the proof of Theorem~\ref{th:MAPa}.

Now, if $\delta$ is incremented by $\Delta \delta$, the normalized (with respect to the number of flow nodes $m_0$) expected number of new independent conditions on the values of the newly guessed flow nodes, denoted by $\Delta \mathbb{C}$, can be upper-bounded by the difference
\begin{equation} \label{eq:delta_def}
\mathbb{E}[\Delta \mathbb{C}] \triangleq F(\x(\epsilon,\delta+ \Delta \delta),\epsilon,\delta+ \Delta \delta) - F(\x(\epsilon,\delta),\epsilon,\delta+ \Delta \delta).
\end{equation}
Following \cite[Sec.~VI-G]{mea08}, we can consider two separate cases:
\begin{itemize}
\item $\x(\epsilon,\delta)$ is continuous (in the second component)  in the interval $[\delta,\delta + \Delta \delta]$, from which it follows (using Taylor series expansion) that 
\begin{equation} \label{eq:deltaC1}
\mathbb{E}[\Delta \mathbb{C}] = O((\Delta \delta)^2),
\end{equation}
since it can be shown that the gradient is zero for $\x = \x(\epsilon,\delta+\Delta \delta)$.
\item There is a discontinuity point at $\delta = \delta_j$ where $\delta_j \in [\delta,\delta + \Delta \delta]$, in which case we get
\ifonecolumn
\begin{align} 
\mathbb{E}[\Delta \mathbb{C}] - O(\Delta \delta)= \Delta F_j &\triangleq F \left( \lim_{\delta \downarrow \delta_j} \x(\epsilon,\delta),\epsilon,\delta_j \right) -   F \left( \lim_{\delta \uparrow \delta_j} \x(\epsilon,\delta),\epsilon,\delta_j \right) \label{eq:deltaC2a} \\
&\overset{(a)}{=} P\left( \lim_{\delta \downarrow \delta_j} x_*(\epsilon,\delta) \right) - P\left( \lim_{\delta \uparrow \delta_j} x_*(\epsilon,\delta) \right) \label{eq:deltaC2b} \\
&\overset{(b)}{=}\int_{\lim_{\delta \uparrow \delta_j} x_*(\epsilon,\delta)}^{\lim_{\delta \downarrow \delta_j} x_*(\epsilon,\delta)} h^{\rm EMP}(z)\, \diff \epsilon(z), \label{eq:deltaC2c}
\end{align}
\else
\begin{align} 
&\mathbb{E}[\Delta \mathbb{C}] - O(\Delta \delta)= \Delta F_j \notag \\
&\triangleq F \left( \lim_{\delta \downarrow \delta_j} \x(\epsilon,\delta),\epsilon,\delta_j \right) -   F \left( \lim_{\delta \uparrow \delta_j} \x(\epsilon,\delta),\epsilon,\delta_j \right) \label{eq:deltaC2a} \\
&\overset{(a)}{=} P\left( \lim_{\delta \downarrow \delta_j} x_*(\epsilon,\delta) \right) - P\left( \lim_{\delta \uparrow \delta_j} x_*(\epsilon,\delta) \right) \label{eq:deltaC2b} \\
&\overset{(b)}{=}\int_{\lim_{\delta \uparrow \delta_j} x_*(\epsilon,\delta)}^{\lim_{\delta \downarrow \delta_j} x_*(\epsilon,\delta)} h^{\rm EMP}(z)\, \diff \epsilon(z), \label{eq:deltaC2c}
\end{align}
\fi
where $(a)$ follows since the $x_0$-component does not depend on $\delta$ 
(see (\ref{eq:DE_Maxwell_a})), 
and $(b)$ follows from (\ref{eq:EBPrelationTrial}).
\end{itemize}

%

\subsection{Proof of \cref{th:MAPa}}
\label{proof:theoremUpperBound}

We follow the guessing strategy explained above  in which for each stage the Maxwell decoder is stuck, $\delta$ in incremented by $\Delta \delta$ and then new unknown flow nodes are guessed with probability $\Delta \delta / (1-\delta)$.  Assume that at a given stage the Maxwell decoder has reached a fixed-point. The normalized (with respect to the number of flow nodes $m_0$) number of new guessed flow nodes at this stage is $\Delta \mathbb{G}_{\delta}$ (see (\ref{eq:deltaG})) and the normalized number of new imposed conditions is upper-bounded by $\Delta \mathbb{C}_{\delta}$ (see (\ref{eq:delta_def})). Note that we have made an explicit reference to the specific value of $\delta$ by including it as a subscript. We assume the procedure continues until $\delta=1$ from $\delta=0$ in small increments of $\Delta \delta$. When the algorithm stops, each assignment of integer values to the guessed flow nodes compatible with the imposed conditions at the end yields a valid flow node size configuration. Thus, we can lower bound the Maxwell EXIT curve as (see \cite[Lem.~11]{mea08})
\begin{align} 
h^{\rm Maxwell}(\epsilon) &\geq \sum_{\delta} \mathbb{E}[\Delta \mathbb{G}_{\delta}] - \sum_{\delta} \mathbb{E}[\Delta \mathbb{C}_{\delta}] \notag \\
&\overset{(a)}{=}  \int_{0}^1 \epsilon L(y_*(\epsilon,\delta))\, \diff \delta - \sum_{\delta_j} \Delta F_{j} + O(\Delta \delta) \label{eq:MAPlower1} \\
&\overset{(b)}{=} \int_{0}^\epsilon L(y(\epsilon'))\, \diff \epsilon' -  \sum_{\delta_j} \Delta F_{j} + O(\Delta \delta), \label{eq:MAPlower2}
\end{align}
where the summations in (\ref{eq:MAPlower1}) and (\ref{eq:MAPlower2}) are over all discontinuity points $\delta_j$ (see (\ref{eq:deltaC2a})). The  $(a)$ follows from (\ref{eq:deltaG}), (\ref{eq:deltaC1}), and  (\ref{eq:deltaC2a}), while $(b)$ follows from the fact that $y_*(\epsilon,\delta) = y(\epsilon \cdot (1-\delta))$, where $y(\epsilon')$ corresponds to a fixed-point of the standard MP decoder of a CB with a probability of observing a flow of size strictly larger than $f_{\rm min}$ equal to $\epsilon'$ (see (\ref{eq:DE_Maxwell_b})). 

Now, since $h^{\rm Maxwell}(\epsilon)$ does not depend of $\Delta \delta$, we can take the limit $\Delta \delta \to 0$ in (\ref{eq:MAPlower2}) and we get the lower bound
\ifonecolumn
\begin{align}
h^{\rm Maxwell}(\epsilon) &\geq \int_{0}^\epsilon L(y(\epsilon'))\, \diff \epsilon' -  \sum_{\delta_j} \Delta F_{j} \notag \\
&\overset{(a)}{=}\int_{0}^\epsilon L(y(\epsilon'))\, \diff \epsilon' 
- \sum_{\delta_j} \int_{\lim_{\delta \uparrow \delta_j} x_*(\epsilon,\delta)}^{\lim_{\delta \downarrow \delta_j} x_*(\epsilon,\delta)} h^{\rm EMP}(z)\, \diff \epsilon(z) \notag \\
&=\int_{0}^\epsilon L(y(\epsilon'))\, \diff \epsilon' - \int_{x^{\rm MP}}^0 h^{\rm EMP}(z)\, \diff \epsilon(z),  \label{eq:MAPlower_final}
\end{align}
\else
\begin{align}
h^{\rm Maxwell}(\epsilon) &\geq \int_{0}^\epsilon L(y(\epsilon'))\, \diff \epsilon' -  \sum_{\delta_j} \Delta F_{j} \notag \\
&\overset{(a)}{=}\int_{0}^\epsilon L(y(\epsilon'))\, \diff \epsilon' \notag \\
&\;\;\;\;- \sum_{\delta_j} \int_{\lim_{\delta \uparrow \delta_j} x_*(\epsilon,\delta)}^{\lim_{\delta \downarrow \delta_j} x_*(\epsilon,\delta)} h^{\rm EMP}(z)\, \diff \epsilon(z) \notag \\
&=\int_{0}^\epsilon L(y(\epsilon'))\, \diff \epsilon' - \int_{x^{\rm MP}}^0 h^{\rm EMP}(z)\, \diff \epsilon(z),  \label{eq:MAPlower_final}
\end{align}
\fi
where $x^{\rm MP}$ corresponds to the MP decoding threshold, i.e., $\epsilon(x^{\rm MP})=\epsilon_{\rm MP}$, and $(a)$ follows from (\ref{eq:deltaC2a})--(\ref{eq:deltaC2c}). 
Now, the first term in (\ref{eq:MAPlower_final}) is just the area under the MP EXIT curve from $0$ to $\epsilon$, while the second term is the area under the EMP EXIT curve from the MP decoding threshold to infinity. 
Since
\ifonecolumn
\begin{equation} \notag
\int_{0}^1 h^{\rm EMP}(z)\, \diff \epsilon(z) = \int_{0}^{\epsilon(x^*)} L(y(\epsilon'))\, \diff \epsilon' 
+ \int_{\epsilon(x^*)}^{\epsilon(1)} L(y(\epsilon'))\, \diff \epsilon' 
- \int^{0}_{x^{\rm MP}} h^{\rm EMP}(z)\, \diff \epsilon(z),
\end{equation}
\else
\begin{equation} \notag
\begin{split}
&\int_{0}^1 h^{\rm EMP}(z)\, \diff \epsilon(z) = \int_{0}^{\epsilon(x^*)} L(y(\epsilon'))\, \diff \epsilon' \\
&\;\;\;\;+ \int_{\epsilon(x^*)}^{\epsilon(1)} L(y(\epsilon'))\, \diff \epsilon' 
- \int^{0}_{x^{\rm MP}} h^{\rm EMP}(z)\, \diff \epsilon(z),
\end{split}
\end{equation}
\fi
where $x^*$ is taken from Definition~\ref{def:MAP},  it follows that the right-hand side of (\ref{eq:MAPlower_final}) (with $\epsilon=\epsilon(x^*)$) simplifies to
\ifonecolumn
\begin{equation} \notag
\begin{split}
\int_{0}^{\epsilon(x^*)} L(y(\epsilon'))\, \diff \epsilon' + \int^{x^{\rm MP}}_0 h^{\rm EMP}(z)\, \diff \epsilon(z) &=\int_{0}^1 h^{\rm EMP}(z)\, \diff \epsilon(z)  - \int_{\epsilon(x^*)}^{\epsilon(1)} L(y(\epsilon'))\, \diff \epsilon' \\
&\overset{(a)}{=}\int_{0}^1 h^{\rm EMP}(z)\, \diff \epsilon(z)  -  \int_{x^*}^{1} h^{\rm EMP}(z)\, \diff \epsilon(z) \\
&\overset{(b)}{=}0.
\end{split}
\end{equation}
\else
\begin{equation} \notag
\begin{split}
\int_{0}^{\epsilon(x^*)} &L(y(\epsilon'))\, \diff \epsilon' + \int^{x^{\rm MP}}_0 h^{\rm EMP}(z)\, \diff \epsilon(z) \\
&=\int_{0}^1 h^{\rm EMP}(z)\, \diff \epsilon(z)  - \int_{\epsilon(x^*)}^{\epsilon(1)} L(y(\epsilon'))\, \diff \epsilon' \\
&\overset{(a)}{=}\int_{0}^1 h^{\rm EMP}(z)\, \diff \epsilon(z)  -  \int_{x^*}^{1} h^{\rm EMP}(z)\, \diff \epsilon(z) \\
&\overset{(b)}{=}0.
\end{split}
\end{equation}
\fi
Equality $(a)$ follows from
\begin{equation} \notag
\int_{\epsilon(x^*)}^{\epsilon(1)} L(y(\epsilon'))\, \diff \epsilon' = \int_{x^*}^{1} h^{\rm EMP}(z)\, \diff \epsilon(z),
\end{equation}
which again is true because there is no $x'$ in the interval $(x^*,1]$ such that $\epsilon(x')=\epsilon(x)$ (see Definition~\ref{def:MAP}), and $(b)$ follows from Definition~\ref{def:MAP}.
Now, the statement of the theorem follows, since $h^{\rm Maxwell}(\epsilon) \geq 0$ for $\epsilon = \epsilon(x^*)$, which again gives the desired upper bound on the Maxwell decoding threshold since there is no $x'$ in the interval $(x^*,1]$  such that $\epsilon(x')=\epsilon(x^*)$ (which follows from Definition~\ref{def:MAP}).

\section{Proof of \cref{th:maxwell_threshold}}
\label{proof:theorem2}

Run the peeling decoder on the original CB graph until it stops. Then, $g(z)$ for the equivalent residual graph, denoted by $\tilde{g}(z;x)$, is given by $\tilde{g}(z;x)=1-\tilde{\rho}(z;x)$ where
\begin{displaymath}
\tilde{\rho}(z;x) = \frac{\tilde{R}'(z;x)}{\tilde{R}'(1;x)},
\end{displaymath}
and where
%
\begin{align}
\tilde{R}(z;x) &= \frac{\bar{R}(1-x+zx) - \bar{R}(1-x) -zx\bar{R}'(1-x)}{1-\bar{R}(1-x) - x \bar{R}'(1-x)},  \label{eq:residual_R}
\end{align}
$x$ is the largest fixed-point of the DE recursion in (\ref{eq:DECounter2FixedPoint}), and $\bar{R}(z)$ is defined in (\ref{eq:Rbar}).
This can be seen from the analysis of the peeling decoder found in \cite[Sec.~2.10]{mea_thesis}. In particular, the expression for the residual right degree distribution from \cite[Eq.~(3.124)]{modern_coding_theory} is properly adapted to (\ref{eq:residual_R}).

Now, using (\ref{eq:residual_R}), after some calculations, we end up with the following simplifed expression
\begin{equation} \label{eq:barrho}
\tilde{\rho}(z;x) = 1 - \frac{g(x-zx)}{g(x)},
\end{equation}
which immediately leads to the expression given in the theorem for the EMP EXIT curve (defined for the equivalent graph as in (\ref{eq:EBP_EXIT})) for the expected residual CB graph when the peeling decoder stops.

For the second part of the proof, from \cref{th:area}, the area under the EMP EXIT curve (for the equivalent graph using the expected counter node degree distribution in (\ref{eq:residual_R}) with $x = x^*$ (corresponding to the area threshold)) can be written as
\begin{equation} \label{eq:area_ebp_exit_residual}
1-\tilde{\rho}(0;x^*) + k \left( \tilde{\rho}(0;x^*)-\int_0^1 \tilde{\rho}(1-z;x^*)\, \diff z \right),
\end{equation}
where $\tilde{\rho}(z;x^*)$ is given in (\ref{eq:barrho}) with $x=x^*$. Now, 
\ifonecolumn
\begin{equation} \notag
\begin{split}
\int_0^1 \tilde{\rho}(1-z;x^*)\, \diff z &= \int_{0}^1 \tilde{\rho}(z;x^*)\, \diff z 
 = \int_{0}^1 \left(1- \frac{g(x^*-zx^*)}{g(x^*)} \right)\, \diff z \\
&= 1 - \frac{1}{g(x^*)} \int_0^1 g(x^*-zx^*)\, \diff z 
= 1-\frac{1}{x^* g(x^*)} \int_0^{x^*} g(z)\, \diff z.
\end{split}
\end{equation}
\else
\begin{equation} \notag
\begin{split}
\int_0^1 \tilde{\rho}(1-z;x^*)\, \diff z &= \int_{0}^1 \tilde{\rho}(z;x^*)\, \diff z \\
 &= \int_{0}^1 \left(1- \frac{g(x^*-zx^*)}{g(x^*)} \right)\, \diff z \\
&= 1 - \frac{1}{g(x^*)} \int_0^1 g(x^*-zx^*)\, \diff z \\
&= 1-\frac{1}{x^* g(x^*)} \int_0^{x^*} g(z)\, \diff z.
\end{split}
\end{equation}
\fi
From the proof of \cref{th:area_potential_thres_same}, the fixed-point potential $Q(x)$ from \cref{def:fixed_point} for $x=x^*$ is equal to zero. Thus, 
\begin{displaymath}
\int_{0}^{x^*} g(z)\,\diff z = \left(1 -\frac{1}{k} \right) x^* g(x^*),
\end{displaymath}
from which it follows that
\begin{equation} \notag
\int_{0}^1 \tilde{\rho}(1-z;x^*)\, \diff z =1-\frac{1}{x^* g(x^*)} \cdot \left(1 -\frac{1}{k} \right) x^* g(x^*)
=\frac{1}{k}.
\end{equation}
Since $\tilde{\rho}(0;x^*) = 0$, (\ref{eq:area_ebp_exit_residual}) reduces to
\begin{displaymath}
1- 0 + k \left( 0 - \frac{1}{k} \right) = 0
\end{displaymath}
and the result follows.

\section*{Acknowledgment}
The authors are grateful to Henry D.\ Pfister for useful discussions and insightful comments.

\balance

\end{document}